\PassOptionsToPackage{numbers,sort&compress}{natbib}
\documentclass[preprintnumbers,amsmath,amssymb,prd, notitlepage,nofootinbib,twocolumn, superscriptaddress]{revtex4}

\usepackage{amsfonts,amssymb,amsmath}
\usepackage{gensymb}
\usepackage{color}
\usepackage{graphicx}
\usepackage{multirow}
\usepackage[utf8]{inputenc}
\usepackage{aas_macros}
\usepackage{hyperref}

\graphicspath{{./Images/}}
\usepackage[normalem]{ulem}

\newcommand{\greaterthanapprox}{\mathrel{\vcenter{
  \offinterlineskip\halign{\hfil$##$\cr
    >\cr\noalign{\kern2pt}\sim\cr\noalign{\kern-2pt}}}}}
    
    \newcommand{\lessthanapprox}{\mathrel{\vcenter{
  \offinterlineskip\halign{\hfil$##$\cr
    <\cr\noalign{\kern2pt}\sim\cr\noalign{\kern-2pt}}}}}
    
\newcommand{\be}{\begin{equation}}        
\newcommand{\ee}{\end{equation}}

\begin{document}
\title{Large-scale galactic-dust-cleaned cosmic infrared background maps from \textit{Planck} PR4 and HI4PI with \texttt{pyilc} }

\author{Fiona McCarthy}
\email{fmm43@cam.ac.uk}
\affiliation{DAMTP, Centre for Mathematical Sciences, Wilberforce Road, Cambridge CB3 0WA, UK
}
\affiliation{Kavli Institute for Cosmology Cambridge, Madingley Road, Cambridge, CB3 0HA, UK}
\affiliation{Center for Computational Astrophysics, Flatiron Institute, New York, NY, USA 10010}

\date{\today}

\begin{abstract}
The cosmic infrared background (CIB) traces star-forming galaxies throughout cosmic history, with emission peaking at $z\sim1-2$. CIB anisotropies are present at the far-infrared frequencies observed by cosmic microwave background (CMB) experiments such as \textit{Planck}. These contain a lot of astrophysical and cosmological information, but are hard to separate from the dust emission in our own Milky Way galaxy, especially on large scales where the Milky Way contamination severely dominates. This galactic component is often cleaned using information from other galactic tracers such as neutral hydrogen  (HI). In this work we use HI data from the HI4PI survey to clean the 353, 545, and 857 GHz \textit{Planck} NPIPE single-frequency maps using a needlet internal linear combination (NILC) method, with \texttt{pyilc}. This allows us to preserve the CIB anisotropy information on \textit{all} scales, while reducing the variance sourced by the galactic contamination. We also create a NILC CMB map from the  \textit{Planck} NPIPE data, to subtract a CMB template from the 353 GHz map. Our resulting CIB maps are appropriate for cross-correlation studies with cosmological tracers such as CMB lensing maps down to very low $\ell$ ($\ell\sim10$), while achieving similar performance to previous works on intermediate scales. The use of the NPIPE data additionally allows us to achieve lower instrumental noise in the maps than in previous works.  We use our maps, in combination with the \textit{Planck} NPIPE CMB lensing reconstruction, to measure the CIB-CMB lensing cross correlation down to $\ell\sim10$. We make various versions of our maps publicly available to the community for further use in cross-correlation studies, along with a script (and the intermediate data products required) to produce dust-cleaned CIB maps on an arbitrary region of sky. 
\end{abstract}
\maketitle

\section{Introduction}

The cosmic infrared background (CIB) is sourced by thermal emission from dust grains in star-forming galaxies. It traces the star formation history of our Universe, and carries astrophysical information as well as cosmological information through its clustering. We detect it in wavelengths relevant for cosmic \textit{microwave} background (CMB) studies ($\sim$100 GHz and higher), where it can be a dominant foreground to the CMB on small scales, as well as at slightly higher frequencies where it dominates over the CMB on all angular scales. For instance, the 353, 545, and 857  GHz channels of the \textit{Planck}  satellite~\cite{2020A&A...641A...1P} have been used to create signal-dominated CIB anisotropy maps~\cite{2014A&A...571A..30P,2019ApJ...883...75L}; the CIB has also been studied at higher frequencies by other far-infrared instruments such at \textit{Herschel}~\cite{2013ApJ...772...77V}.

The CIB anisotropies have been studied in detail, including through auto power spectra measured from the SPT~\cite{2010ApJ...718..632H}, \textit{Planck}~\cite{2014A&A...571A..30P,2018A&A...614A..39M,2021A&A...645A..40M,2017MNRAS.466..286M} and \textit{Herschel}~\cite{ 2013ApJ...772...77V} data. As a cosmological signal, with the galaxies sourcing the emission tracing the underlying dark matter field, they also exhibit correlation with other cosmological tracers such as CMB lensing and galaxy surveys. Such signals have  been measured using \textit{Planck} CIB maps with high signal-to-noise: see, eg, Refs~\cite{2014A&A...571A..18P,2018A&A...614A..39M,2023PhRvD.108h3522M} for CMB lensing cross correlations and Refs~\cite{2014A&A...570A..98S,2022A&A...665A..52Y,2023MNRAS.520.1895J,2023arXiv231010848Y} for galaxy cross correlations, as well as Ref.~\cite{2023MNRAS.520..583J} for a CIB-galaxy lensing cross correlation study. 

A severe difficulty in interpreting CIB measurements is the foreground dust emission from our own galaxy, which is dominant over the CIB anisotropies on large scales. Due to the similar frequency dependence of the galactic emission to that of the CIB sources, it is not possible to separate with multi-frequency measurements, although techniques such as the generalized needlet internal linear combination (GNILC) have allowed some  separation~\cite{2016A&A...596A.109P}. It is  common, however, to clean the galactic emission using other tracers of the galactic dust, in particular galactic neutral hydrogen 21 cm emission (HI emission). These cleaning techniques make use of the high correlation between hydrogen 21 cm emission and the galactic dust distribution~\cite{1996A&A...312..256B,2011A&A...536A..24P}  to remove the parts of the infrared maps that are sourced in the galaxy. Such techniques have allowed large-area cross correlation measurements~\cite{2014A&A...571A..18P} and produced dust-cleaned maps of the CIB on up to 34\% of the sky, in particular in Ref.~\cite{2019ApJ...883...75L} (hereafter L19), which used the all-sky neutral hydrogen survey HI4PI~\cite{2016A&A...594A.116H} to clean  \textit{Planck} CIB measurements. HI has also been used in a similar way to clean CIB measurements at the power spectrum level~\cite{2017MNRAS.466..286M}.

Generally, the HI-cleaning process works  by inferring from the data one dust-to-hydrogen-gas coefficient in a region of sky, multiplying the hydrogen maps by this coefficient to create a dust map, and subtracting this from the dust+CIB map. The maps of L19 were created by performing an operation like this separately in several regions of the sky, and stitching the results together to create a large scale map. However,  their cleaning also included a ``zero-point offset'', constant in each region of the sky but spatially varying between the separate regions, to account for the dust emission that is not well correlated with the hydrogen gas. Such an operation  removes essentially \textit{all} information in the final CIB map on scales larger than those corresponding to the size over which this zero-point is allowed to spatially vary. This effective high-pass filter resulted in L19 providing CIB maps that are unbiased only on scales $\ell\greaterthanapprox70$.

While it will is difficult to fully remove the large-scale galactic contamination from the CIB, for cosmological studies these large scales are sometimes the most interesting. In particular, Ref.~\cite{2023PhRvD.108h3522M} used the very large scales of the CIB to constrain the local primordial non-Gaussianity parameter $f_{\mathrm{NL}}$, which has the well-known effect of inducing a scale-dependent bias in the clustering of objects with respect to the dark matter clustering~\cite{2008PhRvD..77l3514D}. The CIB has been recognized as a tracer which is highly sensitive to this effect, due to the large bias of the emitting galaxies and their high redshifts~\cite{2016MNRAS.463.2046T}. The analysis of Ref.~\cite{2023PhRvD.108h3522M}  was severely limited by the minimum multipole available in the maps,  that much tighter constraints were forecast to be possible with access to larger scales.

In order to address the lack of large-scale cleaned CIB maps available, in this work we use a needlet internal linear combination (needlet ILC, or NILC)~\cite{2009A&A...493..835D} approach to clean the \textit{Planck} PR4 (NPIPE) full-mission single-frequency maps~\cite{2020A&A...643A..42P} at 353, 545, and 857 GHz using the all-sky HI4PI hydrogen maps. Needlet ILC  allows us to infer dust-to-gas coefficients from the data in a spatially varying manner, as well as in a \textit{scale-dependent} manner such that the analysis can be split between large- and small-scales \textit{without removing large-scale power}. With such a technique, we can create unbiased maps with less dust power than the original single-frequency maps. We note that this does not allow for an estimation of the part of the galactic dust that is uncorrelated with hydrogen, which means that the resulting maps remain contaminated by this component and {{\bf should not be used for large-scale}} ($\ell\lessthanapprox100$) {{\bf auto power spectrum studies}}; however, for cross-correlations with observables that do not contain galactic dust (such as CMB lensing), the effect of this cleaning is to reduce the variance on this measurement while remaining unbiased. On intermediate scales (scales similar to those at which the L19 maps are unbiased), we find similar performance to the methods of L19; on smaller scales, our maps have slightly lower instrumental noise due to our use of the full-mission NPIPE data, c.f. the PR3 data used in L19.

As well as galactic dust, the lower-frequency CIB channels are contaminated by primary CMB anisotropies, which dominate over the CIB at multipoles $\ell\lessthanapprox1000$ at 353 GHz. In order to clean these, we create a CMB template from PR4 data, also using a NILC. We directly subtract this from the 353 GHz map before performing the dust-cleaning at this frequency. These CMB fluctuations, while present, are less of an issue in the 545 and 857 GHz maps, and so we do not subtract the  template in these channels.

Finally, we use our cleaned maps to measure the CIB-CMB lensing cross correlation on several regions down to multipoles as low as $\ell\sim10$. We find that the  measurement can be improved significantly compared to the uncleaned measurement, although we note that  improvements in optimality (in particular with respect to the combination of separate measurements on dusty and clean sky areas) are possible, to be explored in future work.

We use the public software \texttt{pyilc}~\cite{2023arXiv230701043M}\footnote{\url{https://github.com/jcolinhill/pyilc/}} to implement all component separation algorithms. We make 40\%-sky area  CIB maps  at 353, 545, and 857 GHz publicly available at~\url{https://users.flatironinstitute.org/~fmccarthy/CIBmaps\_PlanckNPIPE\_HI4PI\_McCarthy24/}, including full-mission and half-mission maps made on independent splits of the data; we also make available our full-sky PR4 NILC CMB map at~\url{https://users.flatironinstitute.org/~fmccarthy/CMBNILC\_PlanckNPIPE\_McCarthy24/}. Additionally, as cleaning can be slightly more efficient when restricted to a given sky area of interest, and because sometimes larger  sky areas are required to increase statistical power in a cross correlation, we provide an easy-to-use script that can be used  with \texttt{pyilc} to create a cleaned CIB map on an arbitrary region of sky, along with all intermediate data products required.

This paper is organized as follows. In Section~\ref{sec:review_ILC} we review the ILC framework, including NILC. In Section~\ref{sec:data} we summarize the datasets used in this work, including the HI4PI  hydrogen maps, the \textit{Planck} NPIPE single-frequency maps, the CIB maps of L19 which we use for comparison purposes with our own maps, and the \textit{Planck} NPIPE CMB lensing map which we use to measure the CMB lensing - CIB cross correlation with our maps. We discuss the need for CMB template subtraction, and our implementation of this, in Section~\ref{sec:cmbsub}. In Section~\ref{sec:CIBmaps} we explore different choices for the NILC in the creation of our CIB maps at 545 GHz and include explicit comparisons with the maps of L19. In Section~\ref{sec:difffreqs} we present the results for 353 and 857 GHz. In Section~\ref{sec:xpowercmblensing} we present our meausrement of the cross correlation of our maps with the \textit{Planck} CMB lensing convergence map, compare to the uncleaned maps, and quantify the reduction in variance especially at low $\ell$. We conclude in Section~\ref{sec:conclusion}.

\section{Review of ILC}\label{sec:review_ILC}

In this Section, we review the main concepts behind the ILC (and specifically needlet ILC) framework, although we refer the reader to Section III of Ref.~\cite{2023arXiv230701043M} for an in-detail pedagogical review. NILC was first introduced in Ref.~\cite{2009A&A...493..835D}. ILC itself has been used on CMB datasets for decades: see eg~\cite{1992ApJ...396L...7B} for an application on \textit{COBE} data. All ILCs in this work are performed using \texttt{pyilc}~\cite{2023arXiv230701043M}.

\subsection{Internal Linear Combination}

The main assumption behind the ILC algorithm is that a signal of interest contributes to the measured intensity (or temperature) at a given frequency channel (labelled by $\nu$) as follows:
\begin{equation}
T_\nu (\hat n) = a_\nu S(\hat n) + N_\nu (\hat n),
\end{equation}
where $\hat n$ is a unit vector encoding the direction; $T_\nu(\hat n)$ is the temperature at $\hat n$; $S(\hat n)$ refers to the signal of interest; and $N_\nu (\hat n)$ refers to all other contributions to $T_\nu (\hat n)$, including  foreground/background sky components as well as instrumental (or other) noise, and which should be \textit{uncorrelated with $S$}. 
 $a_\nu$, which can be referred to as the spectral energy density (SED) of the component $S$, encodes all of the frequency dependence of the signal, and is \textit{independent of $\hat n$}: the component $S$ must display no inter-frequency decorrelation. If one has temperature measurements at multiple frequencies, any linear combination $ \hat S = \sum_\nu c_\nu T_\nu$ with coefficients $c_\nu$ that obey
\be
\sum _\nu a_\nu c_\nu =1\label{unbiased}
\ee
is ``unbiased to the signal of interest'', in that it is necessarily of the form
\be
\hat S = S + \sum _\nu c_\nu N_\nu.
\ee
The aim of ILC is to create the minimum-variance linear combination that obeys the condition defined in Equation~\eqref{unbiased}. In the case that $a_\nu$ is known (as is the case for the black-body CMB emission~\cite{1965ApJ...142..414D,1996ApJ...473..576F,2009ApJ...707..916F}, the kinetic Sunyaev--Zel'dovich (kSZ)~\cite{1980MNRAS.190..413S} emission which preserves the CMB frequency dependence, and the thermal Sunyaev--Zel'dovich (tSZ) spectral distortion~\cite{1969Ap&SS...4..301Z,1970Ap&SS...7....3S}), the weights $c_\nu$ that define this linear combination can exactly be solved for, given knowledge of the frequency-frequency covariance matrix $\mathcal {C}_{\nu \nu^\prime}$ of the data:
\be
c_\nu = \frac{a_{\nu ^\prime} \left(\mathcal {C} ^{-1}\right)_{\nu \nu^\prime}}{a_\mu  \left(\mathcal {C} ^{-1}\right)_{\mu \mu^\prime}a_{\mu ^\prime}}.\label{ILC_solution}
\ee
In ILC, $\mathcal C$ is estimated \textit{directly from the data} (hence the nomenclature ``internal''). Several choices must be made when performing an ILC, including the \textit{domains on which to calculate and minimize the variance}. In particular, the different scale-dependences of different noise and foreground sources can be accounted for by measuring a scale-dependent covariance matrix ${\mathcal{C}}_\ell$ (known as a ``harmonic ILC''), or the statistical anisotropy of different foreground and noise sources can be exploited by calculating a position-dependent covariance matrix by defining different real-space domains on the sky and calculating $\mathcal C(\hat n)$ separately in each domain. 

Needlet ILC (NILC) combines several of the advantages of harmonic and positionally-dependent real-space ILC by performing the variance minimization on a needlet frame, which allows for both scale-separation as well as spatial localization. In particular, by separating the large-scale and small-scale information, one does not lose the information corresponding to scales larger than that of the patch-size on which one is performing the spatially localized ILC. 

\subsection{Needlet ILC}

In Needlet ILC, the covariance matrix is calculated on a needlet frame. Needlets~\cite{doi:10.1137/040614359} specify a spherical wavelet frame by first specifying a set of harmonic-space window functions (indexed by capital Latin letters $I,J,...$) $h^I_\ell$ which obey the completeness criterion
\be
\sum _I \left(h^I_\ell\right)^2 =1
\ee
at each $\ell$. Each $I$ specifies a different ``needlet scale''. The needlets are further defined by  real-space domains $\mathcal {D}_{\hat n} ^{\mathrm{real},I}$, which can be spatially dependent (ie, different depending on the location of the sphere $\hat n$) and overlapping. One can then project each frequency map onto the harmonic needlet scales by calculating the needlet coefficients $T^I(\hat n)$:
\be
T^I(\hat n) = \sum _{\ell m} h^I_\ell T_{\ell m} Y_{\ell m}(\hat n)
\ee
--- here $ Y_{\ell m}(\hat n)$ are the spherical harmonic functions and  $T_{\ell m} $ are the  spherical harmonic coefficients of $T(\hat n)$---and explicitly measure the frequency-frequency covariance matrices  ${\mathcal  C}^I_{\nu \nu^\prime}(\hat n)$ on the domains $\mathcal {D}_{\hat n} ^{\mathrm{real},I}$. From ${\mathcal C}^I_{\nu \nu^\prime}(\hat n)$, the ILC weights defined by Equation~\eqref{ILC_solution} can be calculated separately for each $I$ and at each $\hat n$, and applied to the needlet coefficients $T^I_\nu (\hat n)$ to create the linear combinations $\hat S^I(\hat n) =\sum _\nu c_\nu^I(\hat n) T^I_\nu(\hat n)$. These define the needlet coefficients of the full ILC solution, which can be co-added to obtain $\hat S(\hat n)$.

Given a fixed choice of the harmonic scales $h^I_\ell$, careful choice of the real-space domains $\mathcal {D}_{\hat n} ^{\mathrm{real},I}$, is important for avoiding a phenomenon known as the ILC bias~\cite{2009A&A...493..835D}. This bias results from the fact that the ILC is not a fully linear operation on the maps, as the weights $c_\nu$ explicitly depend on the data through their dependence on the covariance matrix, which is calculated from the data. If the domains are too small, chance correlations between the noise and the signal of interest can be used to artificially minimize the variance. Fortunately, the size of this effect is well understood and depends on the number of modes used in the calculation of the covariance matrix: 
\be
\frac{b_{\rm{ILC}}}{\left<S^2\right>} = \frac{\left|N_{\mathrm{constraints}}-N_{\mathrm{freq}}\right|}{N_{\mathrm{modes}}}
\ee
where the fractional ILC bias $\frac{b_\mathrm{{ILC}}}{\left<S^2\right>}\sim \frac{\left< S \sum_\nu a_\nu N_\nu\right>}{\left<S^2\right>}$  quantifies the correlation between the signal and the linear combination of noise and foregrounds. The fractional ILC bias depends on the number of frequency channels used in the linear conbination $N_{\mathrm{freq}}$,  the number of constraints the linear combination is forced to obey $N_{\mathrm{constraints}}$, and the total number of modes in the domain on which the covariance matrix is calculated $N_{\mathrm{modes}}$. For the standard, unconstrained ILC the only constraint is that of Equation~\eqref{unbiased} and $N_{\mathrm{constraints}}=1$; for a constrained ILC such as that where a specific component is deprojected~\cite{2011MNRAS.410.2481R}, $N_{\mathrm{constraints}}$ can be higher. 

In general, the real-space domains chosen $\mathcal D_{\hat n}^{\mathrm{real},I}$ should be chosen to be large enough to keep the fractional ILC bias low.

\subsection{ILC with non-temperature tracers}

In the case where one has multiple frequency maps with components with well-defined SEDs, it is easy to calculate the weight vectors with the above approach. One can also use external data with no contribution from the signal of interest to remove further variance from noise or foreground components that are correlated with the external data, by including the external data as an additional ``frequency channel'' where the SED of the component of interest is exactly 0, ie $a_{\nu}=0$ in this channel. This was considered in detail in Ref.~\cite{2023arXiv230308121K}, which explicitly explored the possibility of cleaning future CMB+kSZ component maps using external galaxy-survey data to remove further variance associated with components such as the CIB that are correlated with large-scale-structure. The same idea is true in our case where we use galactic hydrogen data to remove foreground components correlated with our galaxy, where we work under the assumption that there is no cosmological contribution to the hydrogen data at these wavelengths.

\section{Data}\label{sec:data}

We make use of the HI4PI~\cite{2016A&A...594A.116H} all-sky galactic hydrogen survey, and the single-frequency maps from the NPIPE processing of the \textit{Planck} data~\cite{2020A&A...643A..42P}. We describe these datasets in Sections~\ref{sec:HI4PI} and~\ref{sec:PlanckNPIPE} respectively.

We also make use of several further processed data products, for comparison purposes; in particular the CIB maps of~\cite{2019ApJ...883...75L} (L19), which were created using data from the \textit{Planck} PR3  data release~\cite{2020A&A...641A...3P} in combination with the HI4PI data. We describe these below in Section~\ref{sec:LenzCIBmaps}.

Finally, we also use a CMB lensing map to measure its cross correlation with the CIB maps, to explicitly quantify the result of the reduction in variance in the component maps on these statistics. We use the \textit{Planck} NPIPE CMB lensing map~\cite{2022JCAP...09..039C} for this purpose; we describe this in Section~\ref{sec:PlanckNPIPECMB}.

\subsection{HI4PI}\label{sec:HI4PI}

The HI4PI survey~\cite{2016A&A...594A.116H} combines data from the Effelsberg--Bonn HI Survey (EBHIS)~\cite{2011AN....332..637K,2016A&A...585A..41W} (on the northern galactic hemisphere) and the Galactic All-Sky Survey (GASS)~\cite{2009ApJS..181..398M,2010A&A...521A..17K,2015A&A...578A..78K} (on the southern galactic hemisphere). It has an angular resolution of 16.2$^\prime$. It provides spectrally sampled datasets with hydrogen brightness temperature $T_B$ measured for neutral hydrogen with different velocities. The data is provided for velocities (with respect to the local standard of rest $v_{\mathrm{lsr}}$) in the range $-600\, \mathrm{km/s} \le v_{\mathrm{lsr}} \le 600\,  \mathrm{km/s}$ with a velocity channel separation of $\delta v=1.29\, \mathrm{km/s}$. A full-sky integrated map of the hydrogen column density can be obtained by integrating the 3-D spectroscopic brightness temperature data over the whole velocity range:
\be
N_{\mathrm{HI}} (\hat n) [\mathrm{cm}^{-2}]  = 1.823\times 10^{18} \int dv T_B(v,\hat n) [\mathrm{K \, km /s}].\label{velocity_moment_integration}
\ee
 
 The integrated $N_{\mathrm{HI}}$ map  was explicitly calculated on the whole-sky and is provided by the HI4PI collaboration including in HEALPix pixelization\footnote{This is available at \url{http://cade.irap.omp.eu/dokuwiki/doku.php?id=hi4pi}.} As the component separation code we will use, \texttt{pyilc}, explicitly takes and performs calculations on HEALPix maps, we use this map in our calculations. 
 
 To remove degeneracies introduced when the integral over velocities is performed, we also at times use the three-dimensional data to recover more  information. In particular, this three-dimensional data can differentiate between differently spatially-localized HI clouds with different velocities; this approach was also taken in the cleaning of the L19 maps~\cite{2019ApJ...883...75L}. Thus, we also make use of the spectrally binned data from HI4PI.\footnote{This is available at \url{http://cdsarc.u-strasbg.fr/viz-bin/qcat?J/A+A/594/A116\#/browse}.} The spectral resolution is too high for our needs, and so we bin the data  by performing the integral in Equation~\eqref{velocity_moment_integration} over discrete bins in $v_{\mathrm{lsr}}$. This results in several different velocity bins, which we project from the coordinate system at which they are supplied into HEALPix coordinates, which we can input directly into \texttt{pyilc}. We explicitly describe the velocity ranges we use in our binning scheme in Section~\ref{sec:spectral_binning}.

\begin{table*}
\begin{tabular}{|c||c|c|c|}\hline
Frequency (GHz) & Beam FWHM (arcmin) &Noise power spectrum amplitude ($\mathrm{\mu K}$ arcmin) & Noise ($\mathrm{\mu K}^2$)\\\hline\hline
30 & 32.29& 150& 0.00322\\\hline
44 & 27.94& 162&0.00433 \\\hline
70 & 13.08&210 & 0.00335\\\hline
100 & 9.66& 77.4&0.000507 \\\hline
143 & 7.22&33 & 9.21$\times10^{-5}$\\\hline
217 & 4.90&46.8& 0.000185\\\hline
353 & 4.92&154 &0.00200 \\\hline
545 & 4.67&818 &0.0566\\\hline
857 & 4.22 &98064.0&813.71\\\hline
\end{tabular}
\caption{Gaussian beam full width at half maxima (FWHMs) for the \textit{Planck} experiment, and white noise levels. This information comes from Table 4 of~\cite{2020A&A...641A...1P}. The ``noise'' column is related to the ``noise power spectrum amplitude'' column by conversion of the latter to radians (multiplication by $\pi/(180\times60$)) and then squaring it.}
\label{tab:fwhm}
\end{table*}

\subsection{ \textit{Planck} NPIPE intensity data  }\label{sec:PlanckNPIPE}

We use the full-mission single-frequency maps from the NPIPE (PR4) processing  of the \textit{Planck} data~\cite{2020A&A...643A..42P}. These maps are provided at 30, 44, 70, 100, 143, 217, 353, 545, 857 GHz; in our CIB map creation we use only the 353, 545, and 857 GHz maps, as the CIB anisotropies are subdominant to those of the CMB at lower frequencies. Additionally, we use the 30-545 GHz channels in our CMB map NILC estimation. These maps are all provided in $\mu \mathrm{K}_{\rm{CMB}}$. 

As well as full-mission maps, we create maps with independent instrumental noise realizations, in order to allow for auto-power-spectrum estimation that avoids instrumental noise bias; in these cases, we use the half-ring splits from the NPIPE release, with all of the same preprocessing steps as described below.

\subsubsection*{Preprocessing steps}
The kinematic solar dipole anisotropy is the dominant black-body anisotropy in these maps, and so we subtract this from the maps before using them; to do this, we use the solar dipole estimation from the \texttt{Commander} component-separation analysis of the NPIPE maps~\cite{2020A&A...643A..42P}.\footnote{This is available on NERSC at \texttt{\$CFS/cmb/data/planck2020/}  \texttt{all\_data/commander\_dipole\_templates/planck/} \texttt{dipole\_CMB\_n4096\_K.fits}.} 

Additionally, we inpaint point sources using an iterative inpainting routine whereby a point source mask is defined, and edges of the masked regions on the initial maps are filled in with the mean of the surrounding unmasked pixels. To perform this inpainting, we use the \texttt{diffusive\_inpaint} routine included along \texttt{pyilc}. In the 353, 545, and 857 GHz maps used for CIB creation, the mask we use for this inpainting is the \texttt{Planck} point source mask defined at the relevant frequency.

In our creation of the CMB NILC used to subtract the CMB anisotropies from the 353 GHz channel, we use the preprocessed versions of the PR4 30, 44, 70, 100, 143, 217, 353, and 545 GHz single-frequency maps described in Ref.~\cite{2023arXiv230701043M}. The preprocessing involves applying a mask which covers the point sources as well as a small amount of the galactic centre, and diffusively inpainting the regions of the PR4 maps covered by these maps with a diffusive inpainting scheme as described in Ref.~\cite{2023arXiv230701043M}. The preprocessed maps, along with the preprocessing mask, are available at~\url{https://users.flatironinstitute.org/~fmccarthy/ymaps\_PR4\_McCH23}.

Note that these maps are all convolved with the \textit{Planck} instrumental beam (see Table~\ref{tab:fwhm} for estimations of beam sizes). In the creation of the CMB NILC, we reconvolve all maps to a 5 arcminute beam; for the CIB maps, we do not deconvolve any beam.

Where relevant, we convert between $\mathrm{\mu K}_{\rm{CMB}}$ and $\mathrm{Jy\, sr^{-1}}$ using the values in Table~\ref{tab:unit_conversions}.

\begin{table}
\begin{tabular}{|c|c|}\hline
$\nu$ [GHz]& U[ $\mathrm{Jy}\,\mathrm{sr}^{-1} \,\mathrm{\mu K}_{\rm{CMB}}^{-1}$]\\\hline
353 &287.45 \\\hline
545 & 58.04\\\hline
857 &2.27\\\hline
\end{tabular}
\caption{Bandpass-integrated unit conversion for the \textit{Planck} frequency channels, for conversion from $\mathrm{\mu K}_{\rm{CMB}}$ to $\mathrm{Jy}\,\mathrm{sr}^{-1}$ (see eg~\cite{2014A&A...571A...9P}).}\label{tab:unit_conversions}
\end{table}

\subsection{L19 CIB maps}\label{sec:LenzCIBmaps}

We compare our CIB maps explicitly to the large-scale CIB maps of~\cite{2019ApJ...883...75L}; we often refer to these as the L19 maps. These maps were created with the single-frequency PR3 \textit{Planck} maps~\cite{2020A&A...641A...3P} at 353, 545, and 857 GHz. Each frequency channel was cleaned of galactic dust separately by estimating the dust-to-gas ratio at each frequency using the three-dimensional HI4PI data, creating dust maps, and subtracting these from the single-frequency maps. Additional variance sourced by the primary CMB fluctuations was removed by subtracting the \textit{Planck} PR3 SMICA estimation of the black-body CMB~\cite{2020A&A...641A...4P} before performing the dust cleaning. These maps were created on several clean regions of sky, with $f_{\mathrm{sky}}$ increasing from $\sim 10\%$ for the cleaneast region to $\sim 34\%$ for the dustiest regions (the cleaner regions always being subsets of the dustier regions). 

These maps exhibit significant large-scale power loss, due to the subtraction of the large-scale zero-point estimation of the dust-to-gas ratio; Ref.~\cite{2019ApJ...883...75L} explicitly warns that they are only unbiased for multipoles above $\ell\sim70$.

\subsection{NPIPE CMB lensing map}\label{sec:PlanckNPIPECMB}

One aim of this work is to provide CIB maps for CMB lensing cross-correlations. We explicitly explore the reduction in variance achievable in the measurement of the CMB lensing-CIB cross-power spectrum $C_\ell^{\mathrm{CIB}-\kappa}$ by measuring this using CMB lensing maps estimated from \textit{Planck} data.

The CMB lensing potential $\phi$ (or convergence $\kappa$) can be estimated from maps of the CMB temperature and polarization anisotropies by taking advantage of the well-understood non-Gaussianities and statistical anisotropies induced in an underlying Gaussian and statstically isotropic Gaussian CMB map by its (weak) gravitational lensing by intervening matter. This is often done with optimal quadratic estimators~\cite{2003PhRvD..67h3002O,2021PhRvD.103h3524M}.  We use the CMB lensing reconstruction of Ref.~\cite{2022JCAP...09..039C}, created  by applying the ``generalized minimum variance'' (GMV) estimator of Ref.~\cite{2021PhRvD.103h3524M} to the NPIPE CMB data.

\section{CMB subtraction}\label{sec:cmbsub}

In this Section we discuss the subtraction of a CMB template from the CIB maps. In Section~\ref{sec:theory_CMBCIB} we present theory predictions for the CMB and CIB fluctuations, to illustrate on which scales and frequencies the CMB is relevant.  In Section~\ref{sec:CIBcontamination_CMB} we discuss the possibility of oversubtraction of the CIB due to CIB leakage into the CMB template, and assess the expected size of this bias  by creating a NILC CMB template from simulations and assessing the amount of CIB residual in the template.

\subsection{Relative strength of CMB and CIB fluctuations: theory}\label{sec:theory_CMBCIB}

\begin{figure}
\includegraphics[width=\columnwidth]{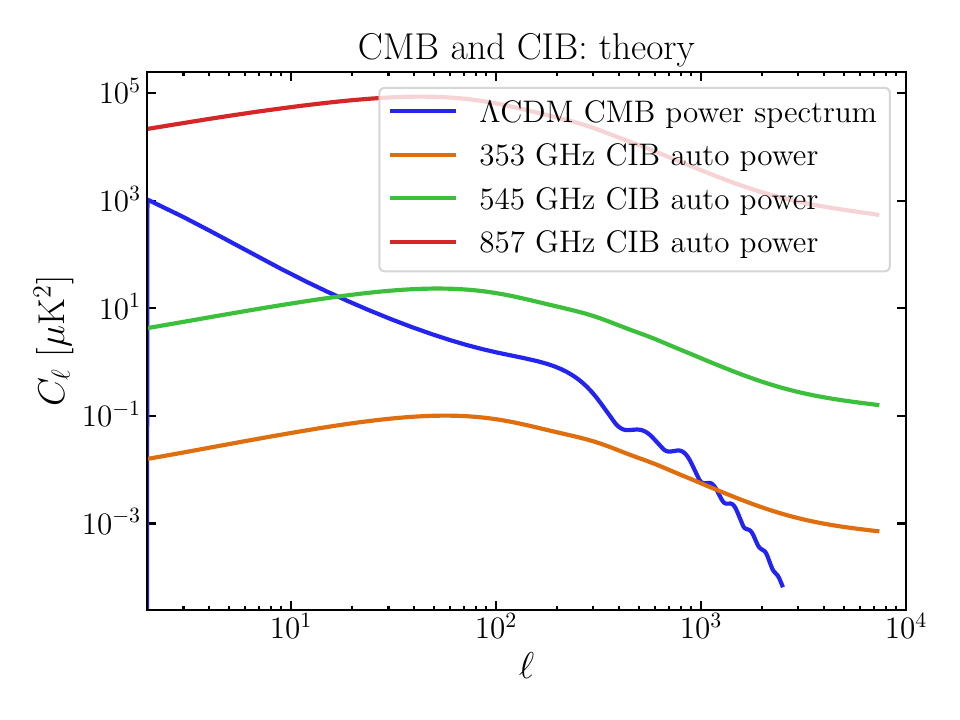}
\caption{Comparison of theory models for the CIB signal at 353, 545, and 857 GHz with the CMB power spectrum. All have been converted to $\mathrm{\mu K}_{\mathrm{CMB}}^2$. We see that the 353 GHz CIB is subdominant to the CMB at $\ell<\sim1000$, while for 545 GHz the crossover is at $\ell<\sim20$. The 857 GHz CIB is always dominant over the CMB.}\label{fig:CMB_CIB_comparison}
\end{figure}

We show in Fig.~\ref{fig:CMB_CIB_comparison}  a comparison of the $\Lambda$CDM CMB power spectrum and some theory predictions for the CIB power spectrum.  We compute the CMB power spectrum with \texttt{class}\footnote{\url{http://class-code.net}}~\cite{2011arXiv1104.2932L,2011JCAP...07..034B}, and the CIB power spectra with \texttt{class\_sz}\footnote{\url{https://github.com/CLASS-SZ/class\_sz}}~\cite{2023JCAP...03..039B,2023arXiv231018482B}, which is a public extension of \texttt{class}. 

In the theory calculations we use the \textit{Planck} 2018 cosmology~\cite{2020A&A...641A...6P}: $\{H_0=67.32 \,\, \mathrm{km/s/Mpc};\,\sigma_8 = 0.812;\, n_s = 0.96605;  \,\Omega_b h^2 = 0.022383;\, {\Omega_{{\mathrm{cdm}}}h^2}=0.12011; \tau=0.0543\}$, where $H_0\equiv100 h$ is the Hubble parameter today; $\sigma_8$ is the amplitude of linear fluctuations on a scale of $8 h^{-1}$ Mpc today (i.e., the linear matter power spectrum integrated over all scales, smoothed with a top-hat window function of $8 h^{-1}$ Mpc); {$n_s$ is the spectral index of the primordial fluctuation power spectrum}; $\Omega_b h^2$ is the physical density of baryons today; $\Omega_{{\mathrm{cdm}}} h^2$ is the physical density of dark matter today; and $\tau$ is the optical depth to the CMB release.

The model used to compute the CIB power spectrum is the parametric halo model (see e.g.~\cite{2002PhR...372....1C} for a review of the halo model)  of Ref.~\cite{2012MNRAS.421.2832S} which was fit to \textit{Planck} CIB auto-spectra in Ref.~\cite{2014A&A...571A..30P}. The implementation in \texttt{class\_sz} follows the description in Ref.~\cite{2021PhRvD.103j3515M}; we refer the reader to this reference for a full description of the model. In our halo model calculations, we use the  halo mass function of~\cite{2008ApJ...688..709T} and the halo bias of~\cite{2010ApJ...724..878T},  assume that halos are spherically symmetric with  Navarro--Frenk--White (NFW) density profiles~\cite{1997ApJ...490..493N}, and use the concentration-mass relation of~\cite{2008MNRAS.390L..64D}.

Inspecting Figure~\ref{fig:CMB_CIB_comparison}, at 353 GHz we see that the CIB anisotropies are subdominant to the CMB anisotropies at $\ell\lessthanapprox1000$; at 545 GHz, the CMB dominates below $\ell\sim20$. Thus, the CMB can add significant variance to the 353 GHz map. We can subtract the CMB fluctuations with an estimation of the CMB; in  L19 this was done by subtracting the \textit{Planck} PR3 SMICA estimation of the primary CMB~\cite{2020A&A...641A...4P}.  We will do this with our own NILC estimation of the CMB from the NPIPE data; we describe our NILC in detail in Appendix~\ref{app:cmbtemplate}.  We only perform this CMB subtraction at 353 GHz, and we do so before performing the HI dust cleaning.

\subsection{CIB contamination in the CMB template: assessment of oversubtraction on simulations}\label{sec:CIBcontamination_CMB}

The CMB template subtraction has the potential to introduce bias to the maps, as the subdominant CIB foregrounds that leak into the NILC estimation will also be subtracted; this is necessarily true for any CMB template subtraction, unless the CIB is explicitly removed, eg by deprojection or clever avoidance of high-frequency channels.\footnote{As CIB deprojection would likely increase the variance on the CMB estimate, we do not explore such options here.} To estimate the size of the bias, we run our NILC pipeline on mm-sky simulations. These simulations include galactic foregrounds from PySM3~\cite{2017MNRAS.469.2821T} and extragalactic foregrounds from Websky~\cite{2020JCAP...10..012S}. We include primary CMB, CIB, kSZ, tSZ, and radio sources~\cite{2022JCAP...08..029L} from Websky, along with the galactic dust, synchrotron radiation, anomalous microwave emission, and free-free components from PySM3 corresponding to the \texttt{d1}, \texttt{s1}, \texttt{a1}, and \texttt{f1}
models respectively (we refer the reader to the PySM3 documentation for descriptions of these models). We also include white noise at levels corresponding to the \textit{Planck} instrument---see Table~\ref{tab:fwhm}---and convolve the simulations with the appropriate beams, also listed in Table~\ref{tab:fwhm}.

\begin{figure}
    \includegraphics[width=\columnwidth]{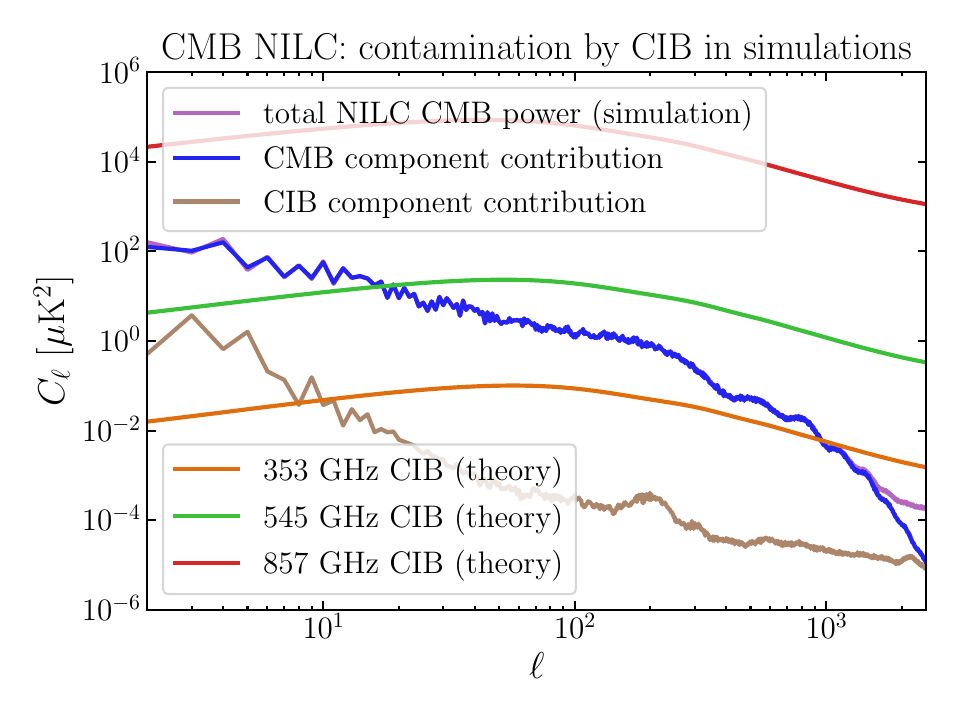}
    \caption{The power spectrum of the CMB NILC map created from the simulations (purple), compared with the power spectrum of the true CMB (blue) and the contribution of the CIB to the NILC (brown) (the purple and blue lines follow each other for most of the plot). We also show the theory  models for the 353, 545, and 857 GHz CIB fluctuations. The subtracted CIB is comparable to the signal at 353 GHz only for $\ell<\sim10$.}
    \label{fig:tracing_components_CMBNILC}
\end{figure}

We run our NILC pipeline on these simulations, save the resulting ILC weights, and apply them to each component separately to understand how much foreground bias remains in the final estimation of the CMB. The results are shown in Fig.~\ref{fig:tracing_components_CMBNILC}, where we compute the power spectrum of several of the linearly-combined components that contribute to the final NILC map. In this plot we have computed the power just on the \textit{Planck} galactic plane mask which leaves 20\% of the sky available, apodized with an apodization scale of 2 degrees, to get an idea of the CIB contamination in in the (galactically) cleaner regions of the CMB map that we are interested in. We see that the CIB contamination is always less than 0.1\% of the total power spectrum at $\ell\greaterthanapprox100$; this contamination is comparable to the intrinsic CMB fluctuations at 353 GHz at $\ell\lessthanapprox10$ and always less than the intrinsic CMB fluctuations at 545 GHz, as indicated in the figure. For $\ell>50$ the spurious CIB subtraction is less than 1\% of the signal at 353 GHz. 

While such conclusions are model dependent, this is certainly a reassuring result; exploration of the results for different CIB models is beyond the scope of this work.

\section{CIB maps: dust cleaning at 545 GHz}\label{sec:CIBmaps}

In this Section we explicitly describe the choices we make in the creation of our CIB maps, explore various configurations, and compare the auto spectra of various configurations of our 545 GHz maps to those of the L19 maps (and to the auto spectra of the uncleaned single-frequency map). 

\begin{figure*}[t]
\includegraphics[width=0.32\textwidth]{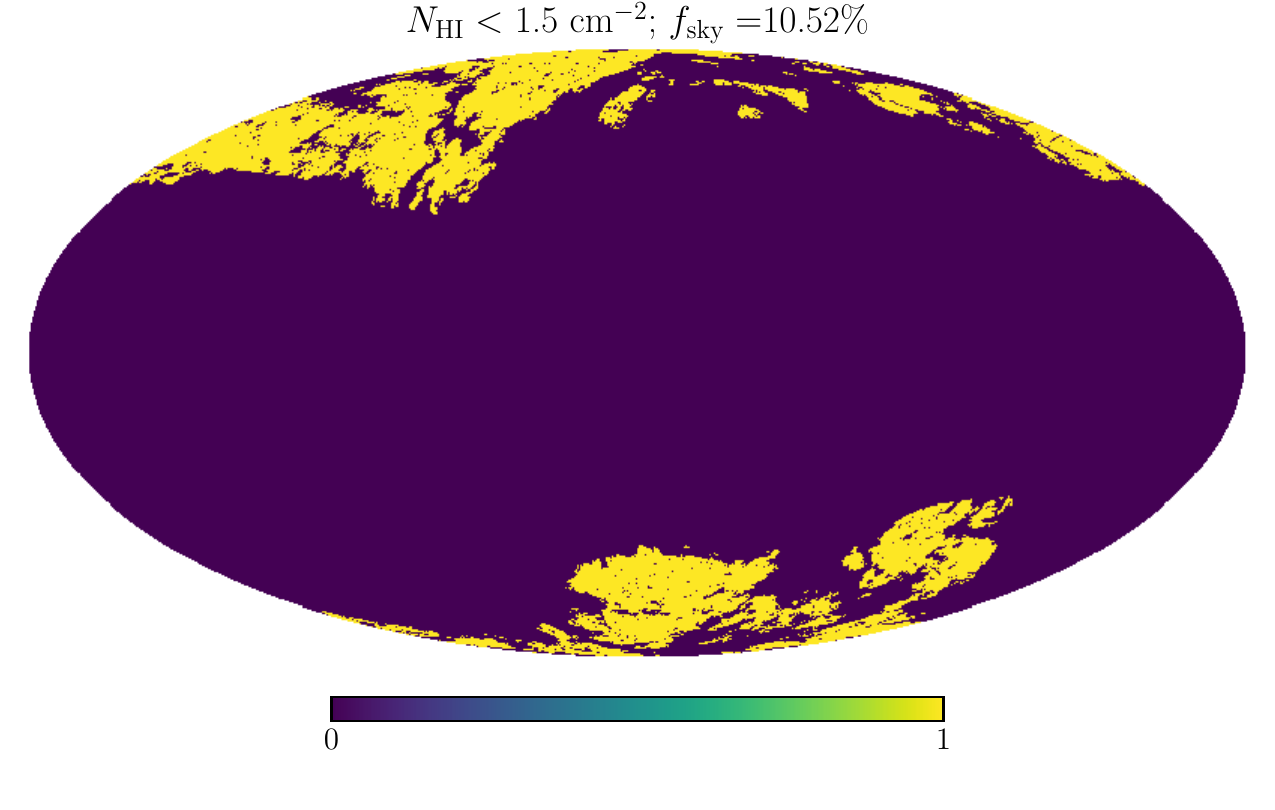}
\includegraphics[width=0.32\textwidth]{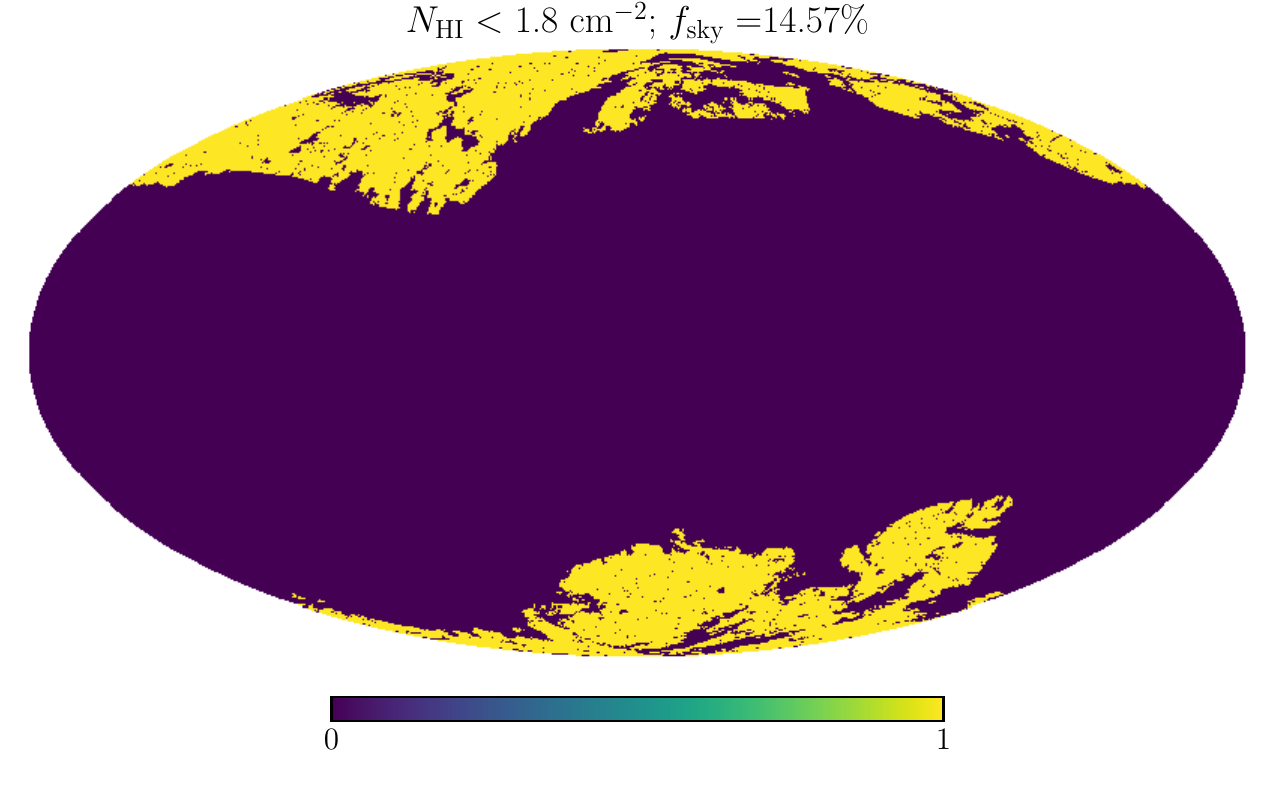}
\includegraphics[width=0.32\textwidth]{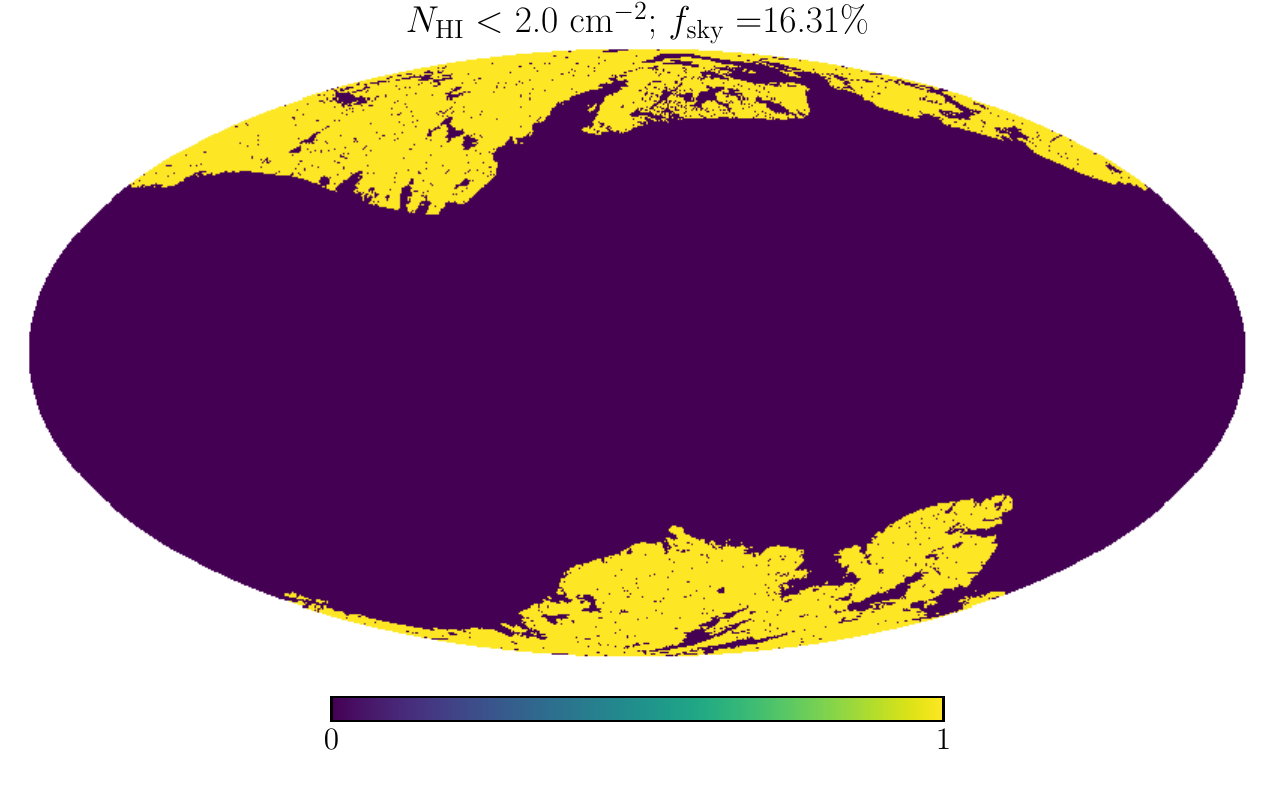}
\includegraphics[width=0.32\textwidth]{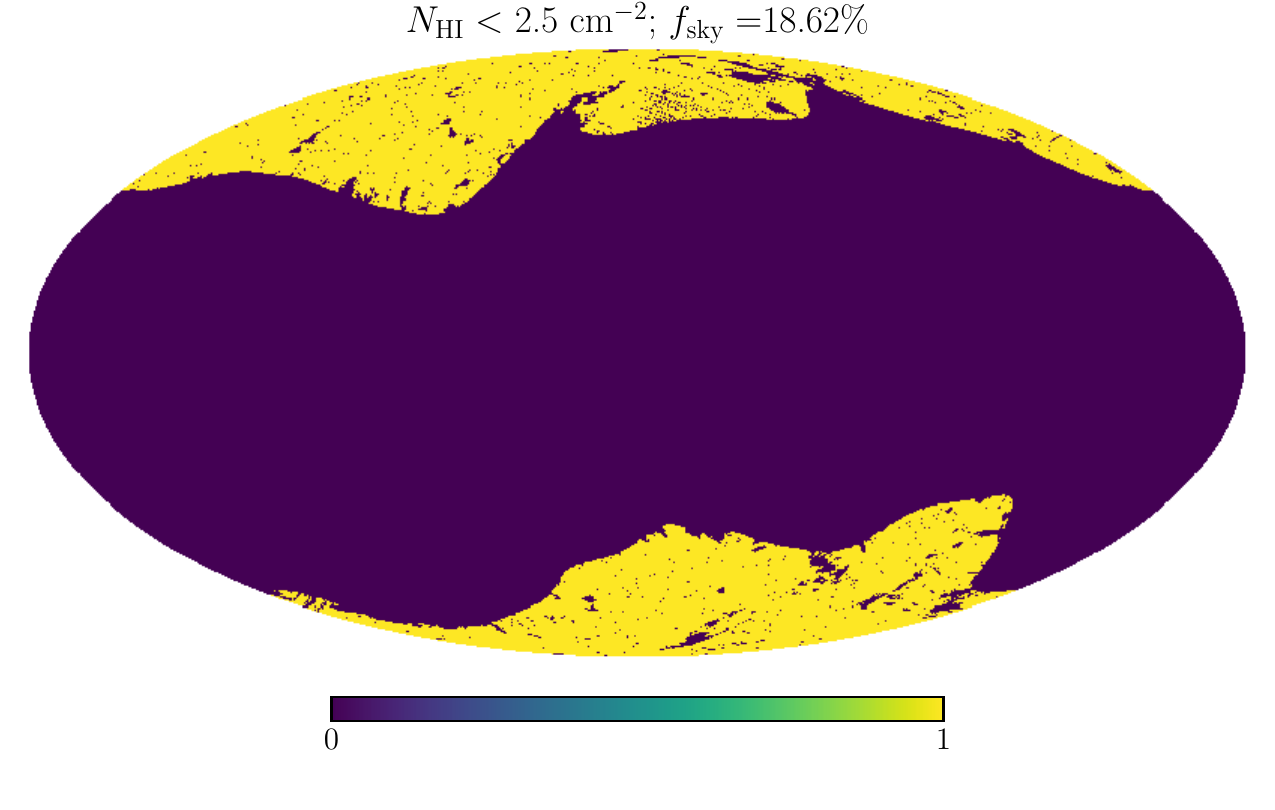}
\includegraphics[width=0.32\textwidth]{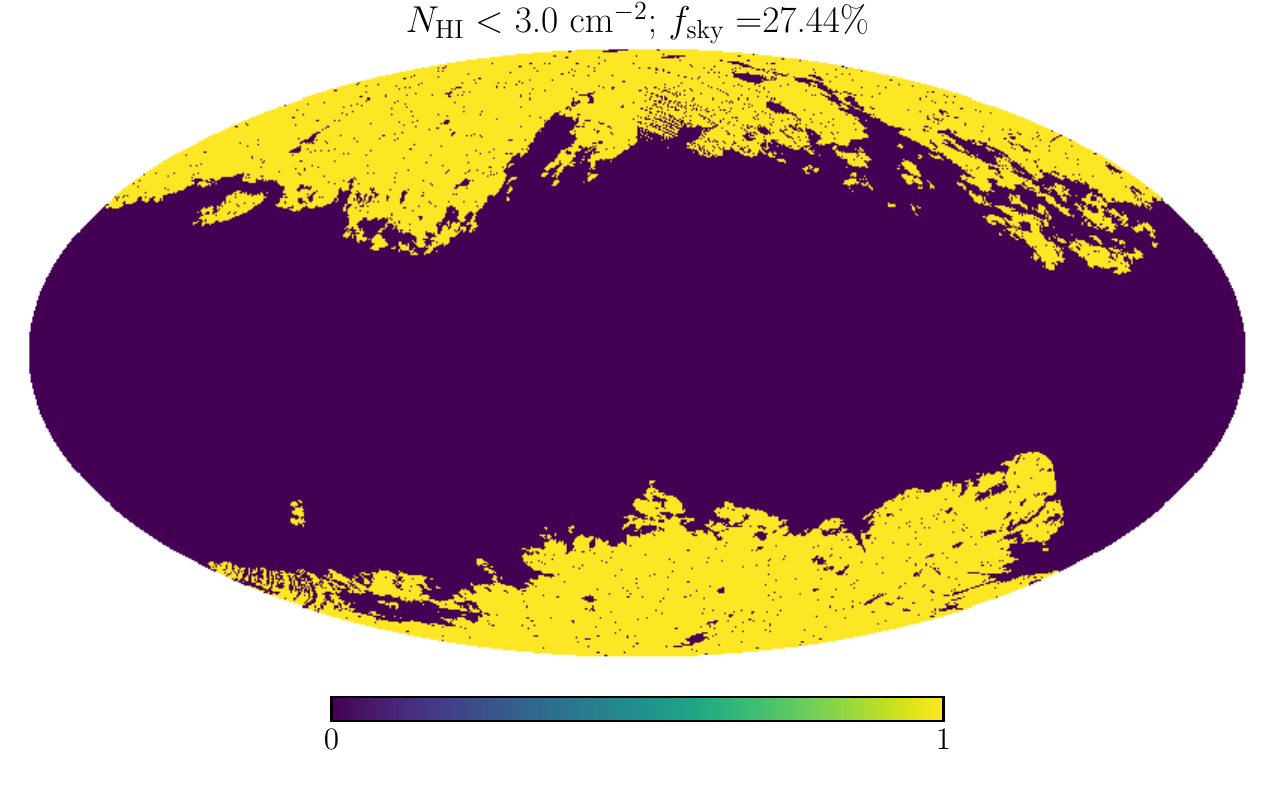}
\includegraphics[width=0.32\textwidth]{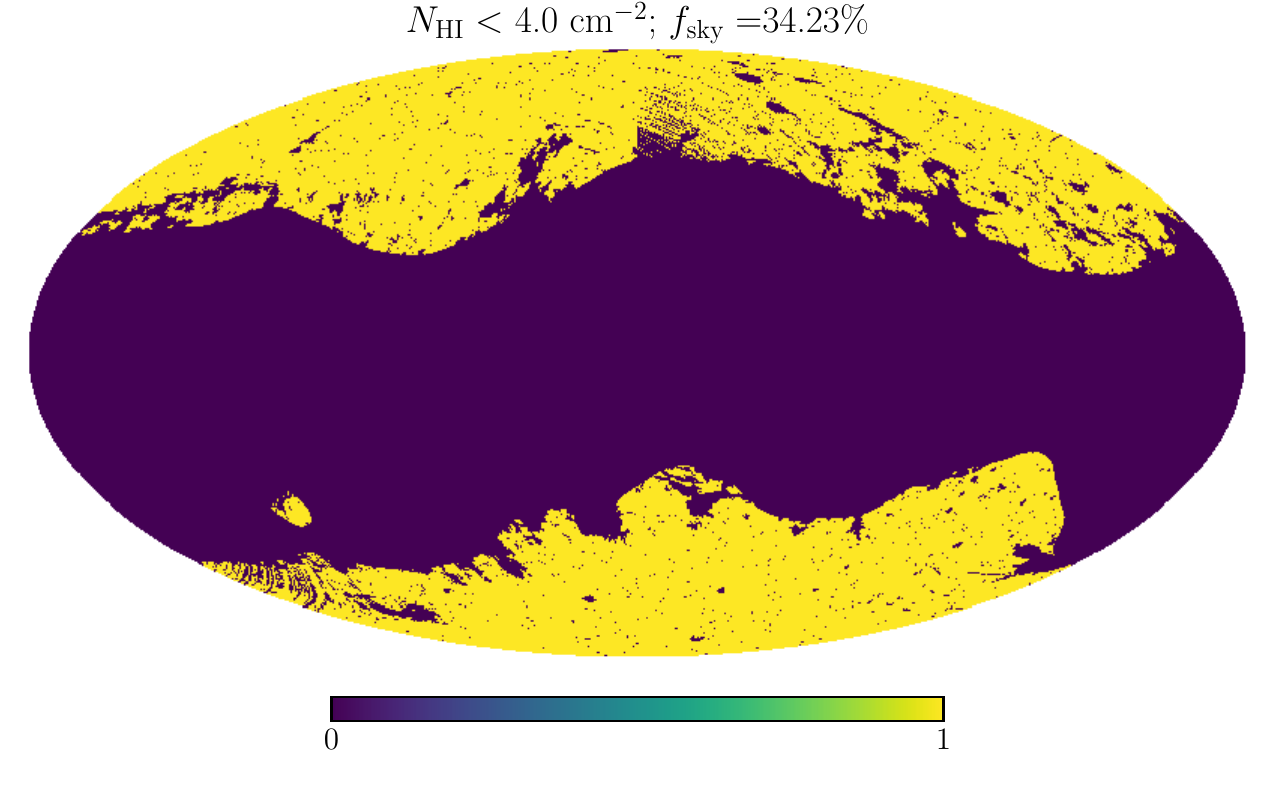}
\caption{The sky regions on which the 545 GHz maps of Ref.~\cite{2019ApJ...883...75L}  were released.}\label{fig:lenzmasks}
\end{figure*}

We start in Section~\ref{sec:realspaceILC} by performing an analagous operation to what was done in the creation of the L19 maps, where we perform a spatially-varying real-space ILC on subdomains of the map, and subtract a large-scale ``zero-point'', on spatial regions of similar size to those used for the L19 maps. 

In Section~\ref{sec:largerdomainsize} we proceed to restore the large scale information by simply increasing the size of the domain on which we perform the real-space ILC (ie, by allowing for less large-scale spatial variation). In Section~\ref{sec:NILC_cleaning}  we then explore using a NILC in which the large scales are cleaned separately to the smaller scales.

Throughout this Section, we only consider for comparison purposes the 545 GHz map; we will present several results for 353 and 857 GHz in Section~\ref{sec:difffreqs} without such a thorough exploration.

\subsection{Comparison with the L19 maps: real-space ILC}\label{sec:realspaceILC}

In this Subsection, we use \texttt{pyilc} to create real-space ILC maps similar to the L19 maps, and compare our results to the L19 maps. 

In Section~\ref{sec:L19_algorithm} we briefly summarize the L19 algorithm.  In Section~\ref{sec:realspaceILC_with_pyilc} we describe how we use \texttt{pyilc} to implement a real-space ILC with the large-scale mean removed. In Section~\ref{sec:power_suppression} we discuss the large-scale power suppression introduced by the large-scale mean subtraction. In Section~\ref{sec:spectral_binning} we present our hydrogen velocity-binning scheme.  In Section~\ref{sec:autopower_realspaceILC} we explicitly calculate the auto power spectra of our resulting CIB maps, and compare to the L19 maps as well as to the uncleaned maps. Finally, in Section~\ref{sec:validation_realspace} we present the results of a pipeline validation where we inject and recover a known signal, in order to assess the bias of our cleaning algorithm.

\begin{figure}
\includegraphics[width=\columnwidth]{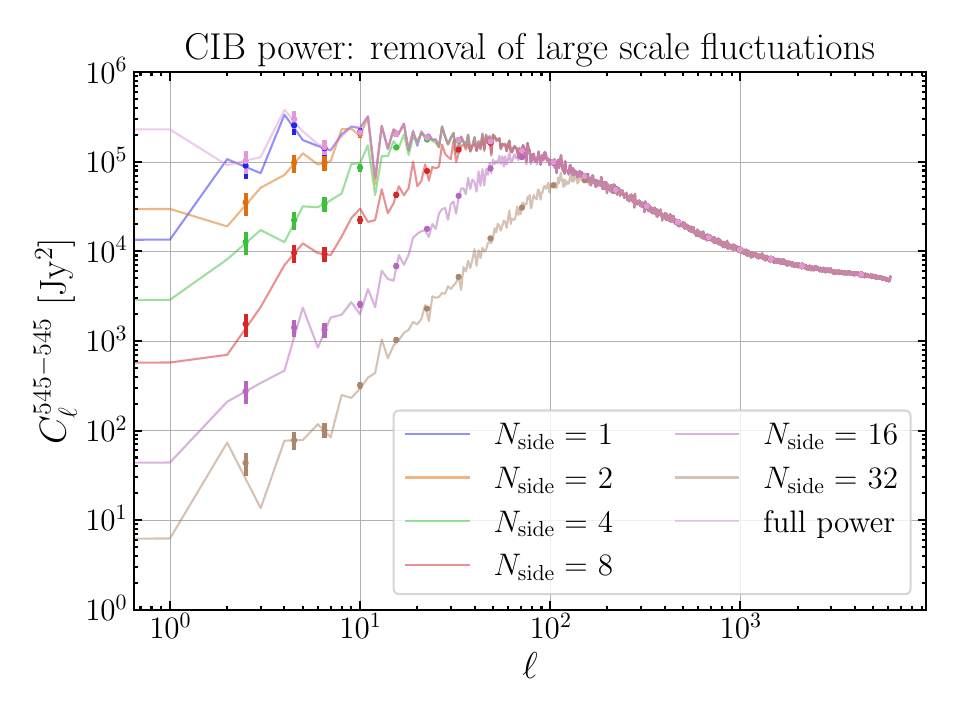}
\includegraphics[width=\columnwidth]{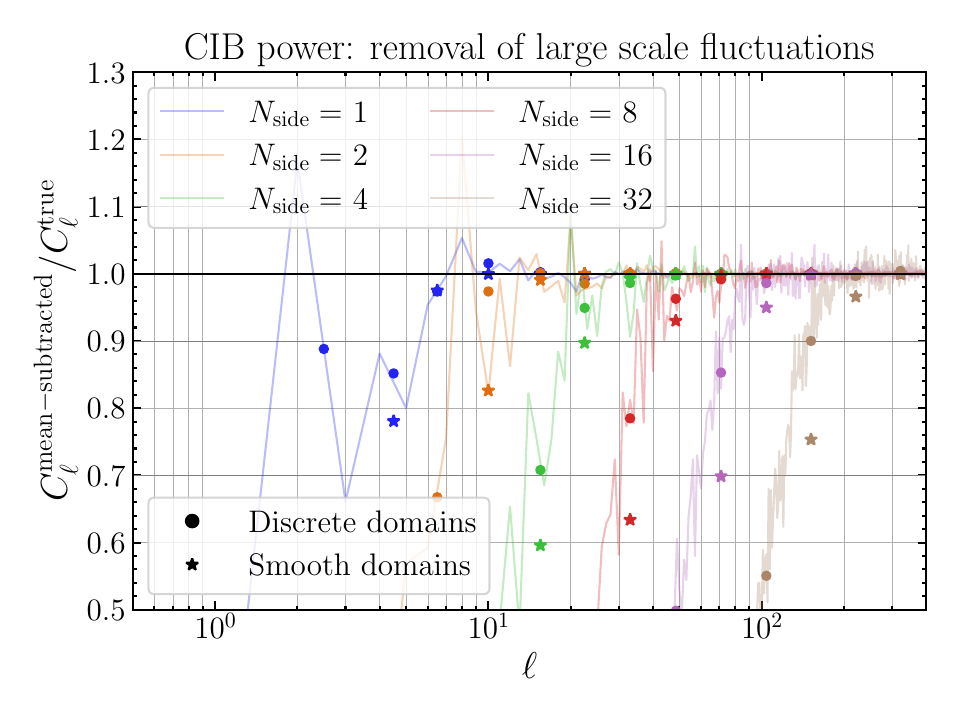}
\caption{Effect on the measured (full-sky) power spectrum of the Webksy CIB map, when the fluctuations above a given $N_{\mathrm{side}}$ have been removed. We show the power spectra on top and the ratio of the power spectra to the truth on the bottom. For $N_{\mathrm{side}}=1,2,4,8,16,32$ this ratio is with $\sim3\%$ of 1 for $\ell>\sim7,15, 24,40,90,200$ respectively, and within $\sim10\%$ of 1 for $\ell>\sim6,7,20,40,15, 20,30,70,150$ respectively (although of course there is considerable sample variance at low $\ell$. Note that we have removed the errorbars from the bottom plot as the sample variance (which dominates the errorbars) should cancel in this ratio. We indicate this ratio both when the large scale fluctuations are subtracted in a discrete manner (by downgrading the map to the given $N_{\mathrm{side}}$, upgrading again to the full  $N_{\mathrm{side}}$, and subtracting; shown by dots), and in a continuous manner (by subtracting the mean smoothed with a Gaussian kernel with FWHM scale corresponding to the resolution of the given $N_{\mathrm{side}}$; shown by asterisks).}\label{fig:suppression_nside}
\end{figure}

\subsubsection{The L19 dust-cleaning algorithm}\label{sec:L19_algorithm}
\begin{figure*}
\includegraphics[width=0.32\textwidth]{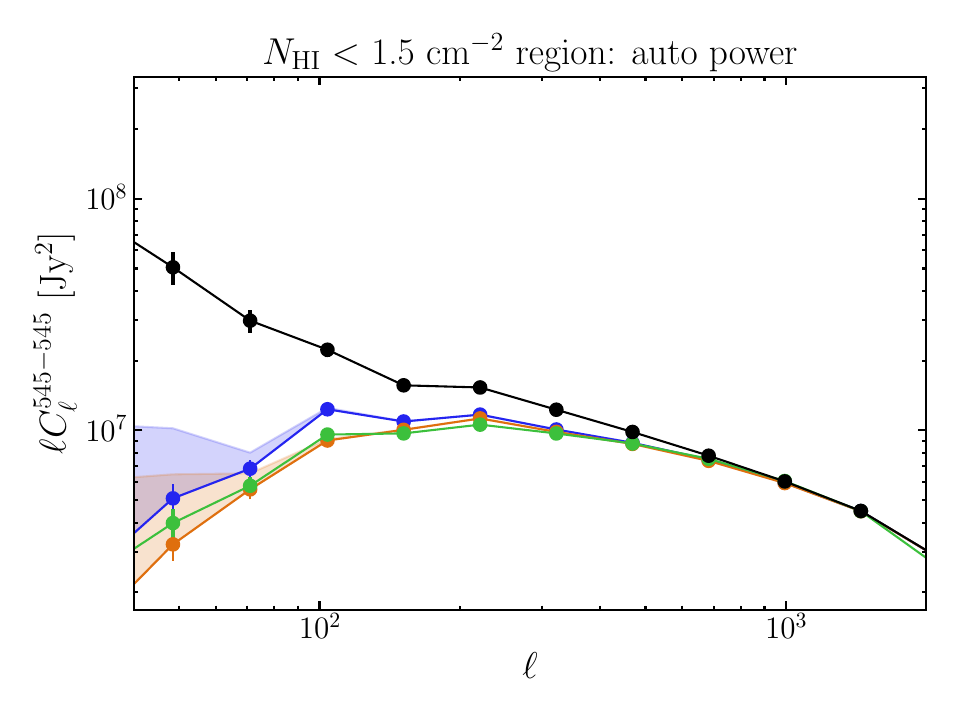}
\includegraphics[width=0.32\textwidth]{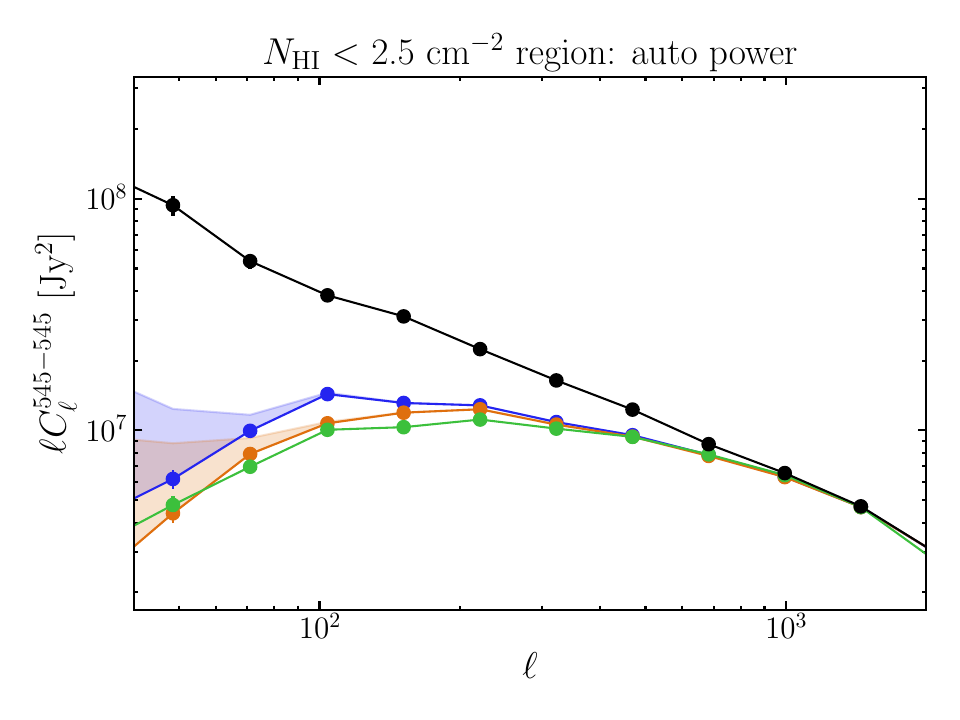}
\includegraphics[width=0.32\textwidth]{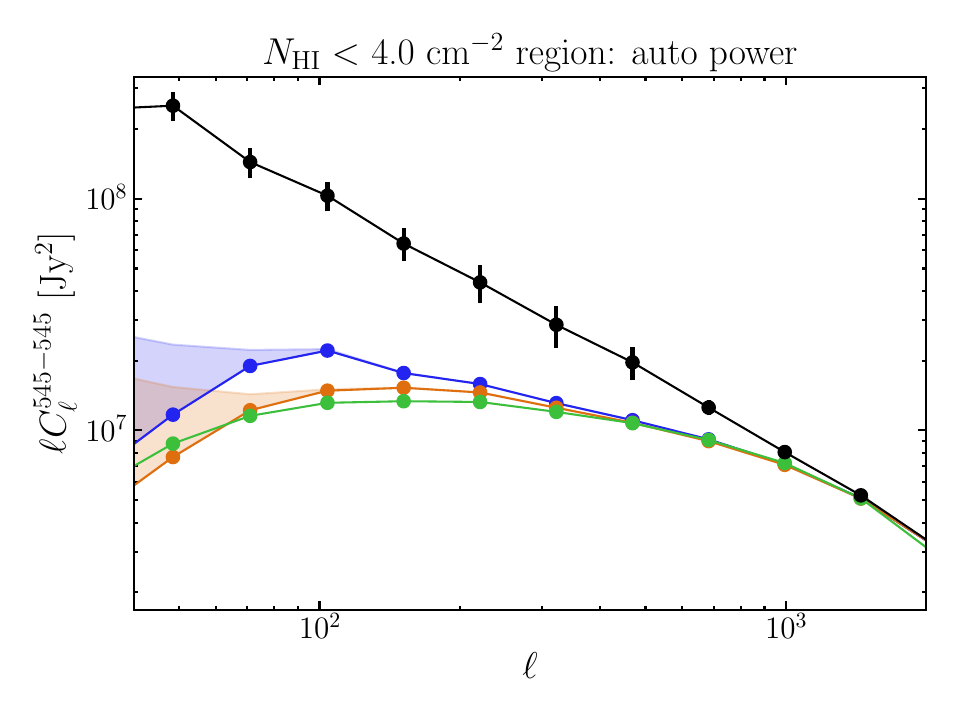}\\
\includegraphics[width=1.2\columnwidth]{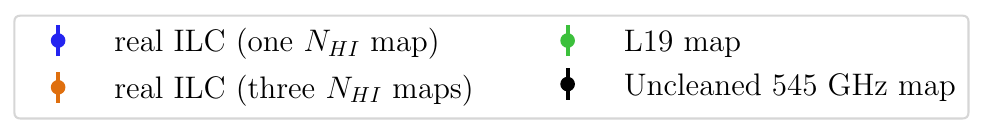}
\caption{The auto power spectra of the uncleaned 545 GHz map (black) and the L19 CIB maps, on several of the sky regions on which they were released, along with several real-space ILC \texttt{pyilc} maps with two different binning schemes for the HI emission (``1 map'' is the full column density map, whereas ``3 maps'' refers to our spectral binning scheme with three bins in the hydrogen velocity space). The shaded area indicates the loss of power due to the transfer function due to removing the large-scale power (see Fig.~\ref{fig:suppression_nside}); the maps of Ref.~\cite{2019ApJ...883...75L} have a similar transfer function, although we do not indicate this on the plot. }\label{fig:lenz_comparison_auto}
\end{figure*}

Ref.~\cite{2019ApJ...883...75L} provided large-scale CIB maps at 353, 545, and 857 GHz, created from PR3 \textit{Planck} data and cleaned via the subtraction of a dust map estimated by minimizing a $\chi^2$ based on the HI4PI data and the  \textit{Planck}  data. At each frequency, they  dust emission in a given spatial region (labelled by $A$) is modelled as 
\be
I^{A}_\nu(\hat n)^{\mathrm{dust}} = \sum_{\mathrm{ch}} \epsilon _\nu^{\mathrm{ch},A} T_{B}^{\mathrm{ch},A}(\hat n)+ \beta ^A_\nu,\label{d2g}
\ee
where $\epsilon_\nu^{\mathrm{ch},A}$ are the dust-to-gas ratios (which are parameters to be fit), $\beta^A_\nu$ is the zero-point-offset in the dust-to-gas ratio (also a parameter to be fit), and $T_{B}^{\mathrm{ch},A}(\hat n)$ is the HI column density in a given velocity bin (with the velocity bins labeled by ``$\mathrm{ch}$'').

A data-informed dust model is then created by minimizing a loss function defined as 
\be
\mathcal L_\nu^A =\sum_{\hat n}\left( \hat I^A_\nu(\hat n) - I^A_\nu(\hat n)^{\mathrm{dust}}\right)^2 + \gamma \sum_{\mathrm{ch}}||\epsilon_\nu^{\mathrm{ch}}||_1\label{loss}
\ee
where $ \hat I^A_\nu(\hat n)$ is the measured intensity of the dust+CIB at frequency $\nu$ (ie, the \textit{Planck} map) in the region $A$;  $\gamma$ is a regularization parameter (the inclusion of the regularization term proportional to $\gamma$ is intended to prevent over-fitting); and $||\cdot||_1$ indicates the $L^1$-norm. 

The minimization of~\eqref{loss} is performed independently in the different spatial regions $A$ (and also independently for the different frequencies). The spatial regions chosen are  super-pixels defined by the HEALPix pixels with $N_{\mathrm{side}}=16$. This allows them to find the best-fit parameters for the dust model defined in Eq.~\eqref{d2g}, which they can then explicitly subtract from the \textit{Planck} intensity map $\hat I_\nu(\hat n)$ to create the CIB map.

We note that the L19 maps are provided on several different sky areas (see Fig.~\ref{fig:lenzmasks}). The masking is always done \textit{prior} to the minimization of Equation~\eqref{loss}, such that the masked pixels do not inform the dust model.

\subsubsection{Real-space ILC CIB maps with \texttt{pyilc}}\label{sec:realspaceILC_with_pyilc}

The L19 algorithm is, in spirit, similar to a real-space ILC (without the regularization term in the loss function, it is exactly a real-space ILC). Indeed, we can reproduce results similar to those of L19 by running \texttt{pyilc} in ``real-space ILC mode'', whereby we:
\begin{enumerate}

\item  set the harmonic needlet filters to a top-hat containing all $\ell$s (removing the harmonic localization)
\item Define the real-space domains to be Gaussian kernels with FWHMs corresponding to the resolution of the superpixels used in the creation of the L19 maps (this is $\sim 3.66$ degrees);
\item subtract the spatially-varying mean of the maps calculated over these domains from the maps before performing the ILC;
\item perform the ILC as standard.
\end{enumerate}

Step 3. plays the role of the $\beta_\nu$ parameter, the parameter describing the large-scale ``zero-point offset'' in the dust-to-gas-ratio in Equation~\eqref{d2g}. In the context of creating a  \textit{galactic dust} map, it assumes that all fluctuations on scales larger than the domain size are galactic dust fluctuations; in the context of CIB estimation, it severely suppresses all fluctuations greater than this scale. 

Note that, in order to compare properly with the maps of L19, the same masks are be applied to the data \textit{before} calculating the covariance matrices (note that this means that the CIB estimations on smaller sky regions are not simply subsets of the CIB estimations on larger sky regions, as they are  cleaned differently due to this more tailored estimation of the covariance matrices).

\subsubsection{Large-scale power suppression}\label{sec:power_suppression}
\begin{table}[h!]
\begin{tabular}{|c||c|}\hline
$N_{\mathrm{side}}$&FWHM (degrees)\\\hline\hline
 1   &58.6\\\hline
 2   &29.3\\\hline
 4   &14.7\\\hline
 8   &7.33\\\hline
 16   &3.66\\\hline
  32  &1.83\\\hline
\end{tabular}
\caption{The FWHMs of the Gaussian real-space domains appropriate for the corresponding $N_{\mathrm{side}}$ values (calculated with \texttt{healpy.nside2resol()}).}\label{tab:nsideresol}
\end{table}
We illustrate explicitly the large-scale power suppression caused by the large-scale mean subtraction in Fig.~\ref{fig:suppression_nside}.
Here, we have computed the power spectrum of the CIB of the Websky~\cite{2020JCAP...10..012S} simulations after first downgrading to a given $N_{\mathrm{side}}$, upgrading back to the original $N_{\mathrm{side}}$, and subtracting this low-pass-filtered map from the original map. We see that the effective ``transfer function'' on $C_\ell^{\nu\nu}$ is within 3\% of 1 ($>0.97$) for  $\ell>\sim7,15, 24,40,90,200$
and within 10\% of 1 ($>0.9$) for r $\ell>\sim6,7,20,30,15, 20,40,70,150$ respectively for $N_{\mathrm{side}}=1,2,4,8,16,32$. We note that Ref.~\cite{2019ApJ...883...75L}, who use an $N_{\mathrm{side}}=16$ scheme for their superpixels, claim unbiasedness at $\ell>\sim70$, which is consistent with our transfer function being within 10\% of 1, although neglects a small power loss at $70<\ell<100$.

In practice, \texttt{pyilc} computes a covariance matrix at every pixel which varies smoothly across the map (as opposed to in a small number of discrete superpixels), making use of \texttt{healpy}'s \texttt{smoothing} function, which smooths the map with a Gaussian beam. For the FWHM of this beam, we use the resolution corresponding to the given $N_{\mathrm{side}}$ we are interested in (see Table~\ref{tab:nsideresol}). In the bottom panel of Fig.~\ref{fig:suppression_nside}, we indicate the ratio of the power as measured for the discrete version described above (smooth lines), as well as binned bandpowers for two versions of the large-scale removal: subtraction of a map defined by the original map, downgraded to a given $N_{\mathrm{side}}$ and upgraded back to the higher $N_{\mathrm{side}}$ (dots); and also subtraction of the map smoothed with the Gaussian kernel (stars). In the latter case (which is the relevant one for maps produced with \texttt{pyilc}), we see that the power goes to zero slightly faster than in the former case.

\subsubsection{Spectral binning of the \textit{HI4PI} data; resolution and regularization}\label{sec:spectral_binning}

We note that Ref.~\cite{2019ApJ...883...75L} used many HI tracers in their spectral binning scheme ($\sim13$, from their Fig. 1). Using so many tracers in an ILC results in an intolerable ILC bias and allows CIB fluctuations to be spuriously removed by chance correlations between the data and the signal; to control this, the $\gamma$-regularization procedure of Equation~\eqref{loss} was used by Ref.~\cite{2019ApJ...883...75L}, as well as a masking scheme which removed regions with low HI emission before the minimization of the likelihood.  Instead of explicit regularization, we  simply use fewer tracers, with broader spectral bins. Explicitly, we compare one case  with just the fully-integrated HI column density map, and also a case with 3 spectral bins defined by $-25 \mathrm{km/s}<v_{\mathrm{lsr}}-12.5\mathrm{km/s}$;  $-12.5 \mathrm{km/s}<v_{\mathrm{lsr}}<0\mathrm{km/s}$; and  $0 \mathrm{km/s}<v_{\mathrm{lsr}}-25\mathrm{km/s}$.
\begin{figure}
\includegraphics[width=\columnwidth]{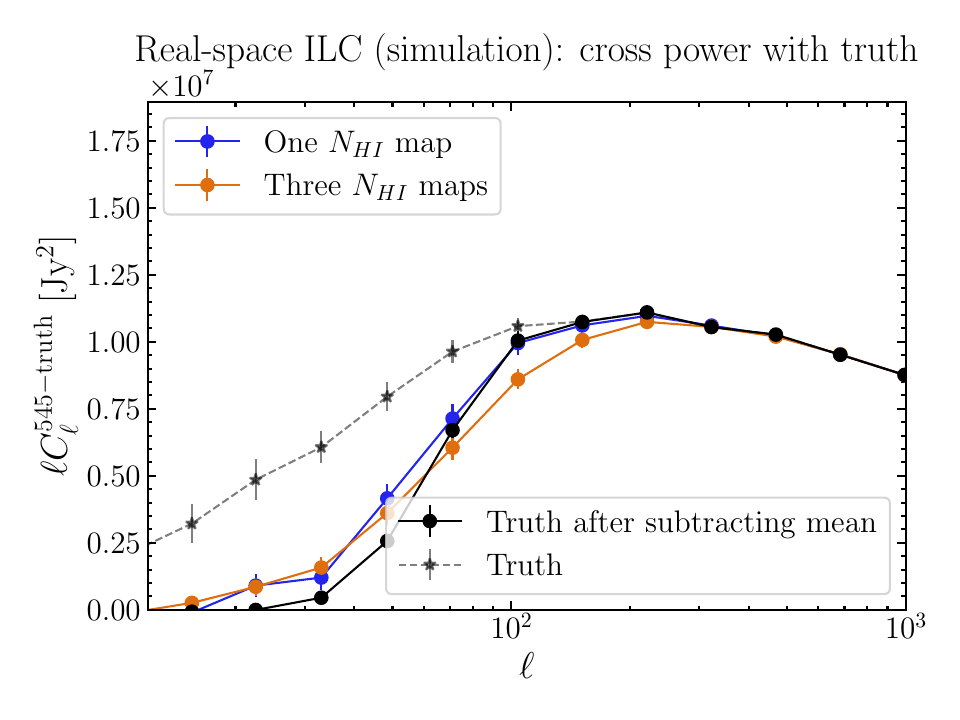}
\caption{Validation of the ``real-space'' ILC \texttt{pyilc} maps shown in Fig~\ref{fig:lenz_comparison_auto}. We have artificially add the websky CIB to the data, run our pipeline, and measure the cross-correlation of the output with the truth. For an unbiased map, we require this cross-correlation to be exactly equal to true power, which is indicated in black (after performing the large-scale mean subtraction), and dashed gray (before performing the large-scale mean subtraction). We see that, for the case when we use only one hydrogen tracer, the ILC bias is under control and we have not over-subtracted any CIB other than what we expect from the large-scale mean removal; however, there is a bias at $\ell<\sim300$ when we have introduced the spectral binning. We show this on the $N_{\mathrm{HI}}<4.0\,\mathrm{cm}^{-2}$ region;  the results are similar on the other regions.}\label{fig:test_ILCbias_realspaceILC}
\end{figure}

\begin{figure*}[t]
\includegraphics[width=0.32\textwidth]{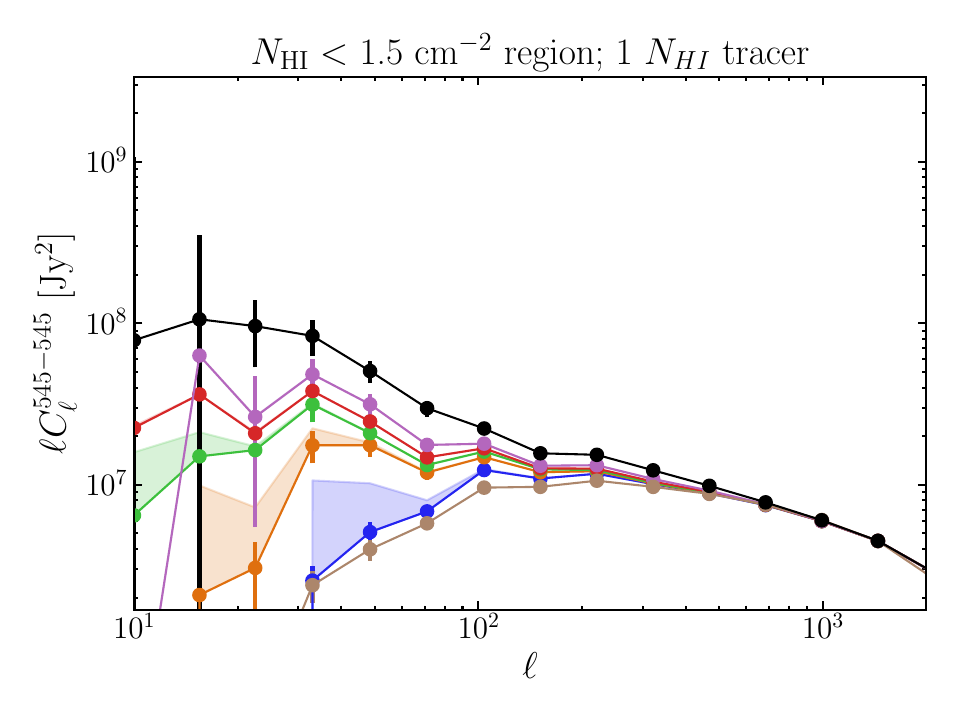}
\includegraphics[width=0.32\textwidth]{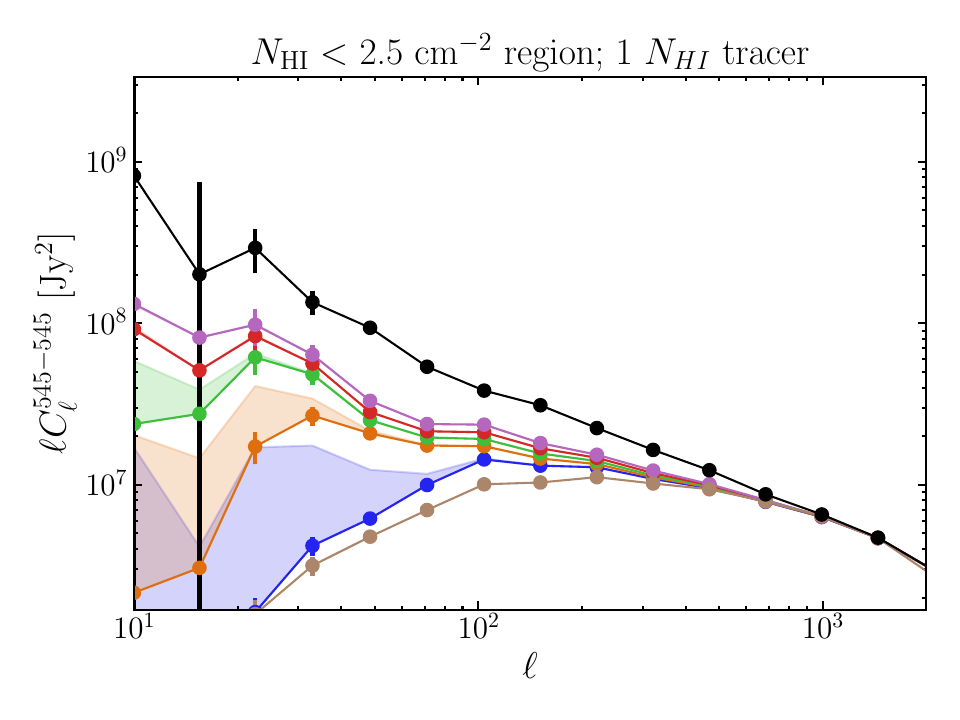}
\includegraphics[width=0.32\textwidth]{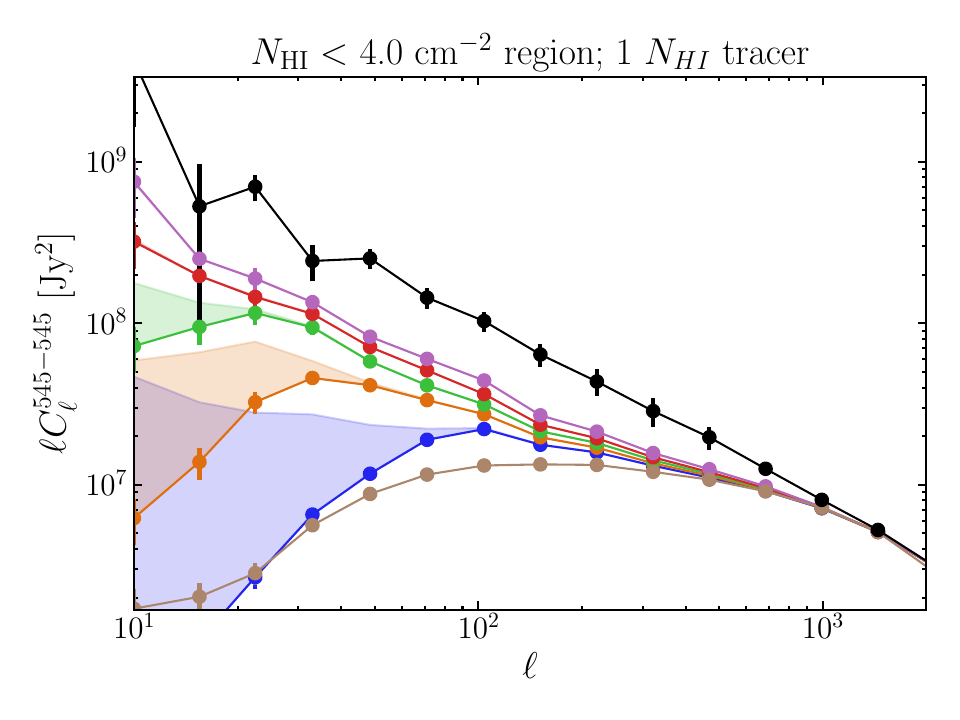}
\includegraphics[width=0.75\textwidth]{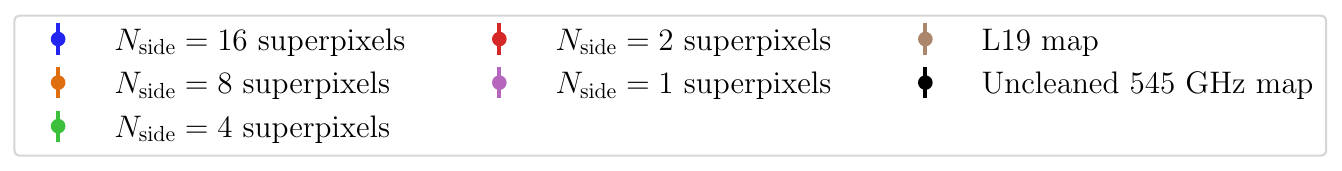}

\includegraphics[width=0.32\textwidth]{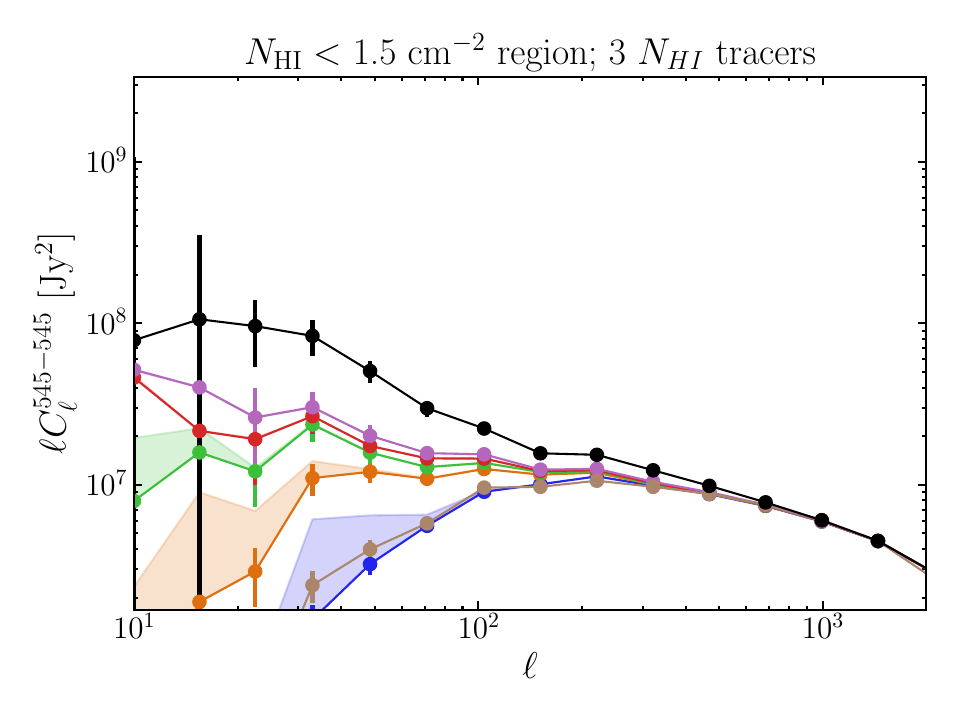}
\includegraphics[width=0.32\textwidth]{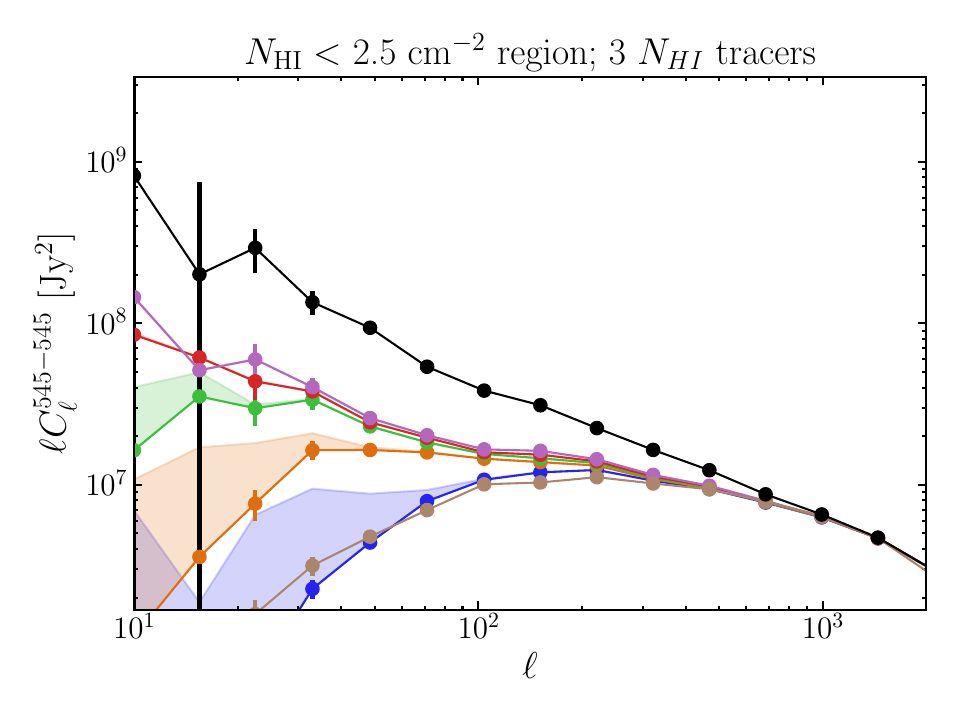}
\includegraphics[width=0.32\textwidth]{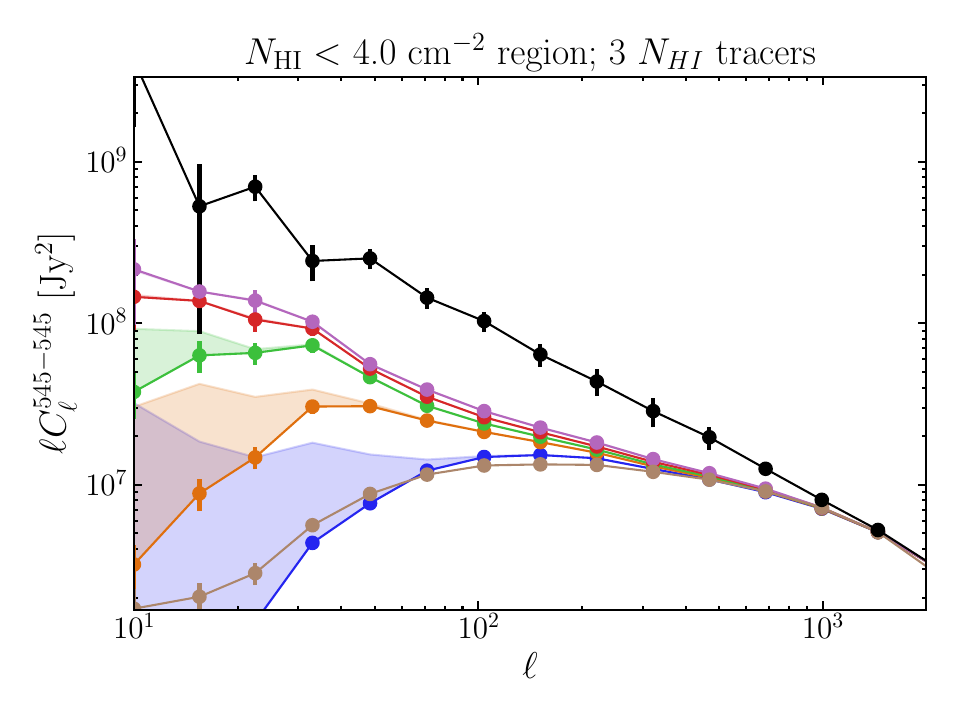}

\caption{The auto power-spectra after performing real-space ILC on super-pixels corresponding to different $N_{\mathrm{side}}$, as indicated, using only the fully-integrated hydrogen column density map to clean (\textit{top}), and with 3 spectrally binned $N_{\rm{HI}}$ maps (\textit{bottom}). For the high-$N_{\mathrm{side}}$ super-pixels, there is a large transfer function as indicated in Fig.~\ref{fig:suppression_nside}; we indicate the size of this by a shaded region (the maps of  Ref.~\cite{2019ApJ...883...75L} have the same transfer function as the $N_{\mathrm{side}}=16$ maps, although we do not indicate this on the plot).}\label{fig:explore_nside_realspaceILC}

\end{figure*}
\subsubsection{Auto power spectra of the CIB maps}\label{sec:autopower_realspaceILC}

We perform the real-space ILC described above  on the sky regions on which the L19 maps were released. The power spectra of the resulting 545 GHz CIB maps (for three of the sky regions) are shown in Fig.~\ref{fig:lenz_comparison_auto}, along with the power spectra of the L19 map on the same sky region as well as the power spectra of the uncleaned single-frequency map measured on the same sky region. 

In all cases we measure the mask-decoupled power spectra with \texttt{namaster}~\cite{2002ApJ...567....2H,2019MNRAS.484.4127A}, and assign errorbars by computing the mask-decoupled covariance appropriate for a Gaussian field with their measured power spectra, although we stress that these are not appropriate covariance estimates for the auto power spectra of maps containing the highly non-Gaussian dust component.

From inspecting Fig.~\ref{fig:lenz_comparison_auto}, it is clear from comparing the black lines (which indicate the power spectra of the uncleaned maps) to any of the other, cleaned power spectra, that we remove a significant amount of dust power, on all of the sky areas. We also reproduce the power spectra of the L19 maps  at $\ell\greaterthanapprox400$, although we have an excess of power on large scales when we use only the unbinned $N_{\mathrm{HI}}$ map; much of this is removed when we use 3 spectral bins as our tracers. 

At high $\ell$ ($\ell\greaterthanapprox1000$), the HI data (which have a resolution of 16.2$^\prime$) have no information and cannot be used to clean the galactic dust; thus the power spectra of the cleaned maps asymptote to those of the uncleaned map at these scales.

\subsubsection{Validation}\label{sec:validation_realspace}

To validate that our method is unbiased (or to quantify the bias of our method), it would be ideal to have an ensemble of simulations on which we can run our component separation algorithms. The dust and hydrogen data are highly non-Gaussian and we do not have access to this. Instead, we artificially add a known signal to the data before cleaning it, and verify that we recover it. 

To do this, we add the Websky 545 GHz CIB to the 545 GHz \textit{Planck} map, perform the component separation, and validate whether the resulting cleaned map has the appropriate correlation with the true Websky map. 

We show these validation results in Fig.~\ref{fig:test_ILCbias_realspaceILC} for the ``real-space ILC'' \texttt{pyilc} maps. We see that the maps which were cleaned using the fullly-integrated $N_{\mathrm{HI}}$ map are unbiased (within what is expected given the large-scale removal), as the blue lines overlap with the black true-power lines. However, there is a hint of a small non-zero ILC bias introduced at $\ell<\sim300$ when we spectrally bin the $N_{\mathrm{HI}}$ maps into 3 velocity bins. While we show this plot only one one specific sky region, the conclusions for the other sky regions are similar.

\subsection{Restoring the large scales: larger domains for the real-space ILC}\label{sec:largerdomainsize}

One way to access larger scales is by using larger super-pixels. Such choices were explored in Ref.~\cite{2019ApJ...883...75L} but due to the increase in the dust contribution to the auto power spectrum upon the use of $N_{\mathrm{side}}=8$ superpixels, the $N_{\mathrm{side}}=16$ superpixels were chosen for the final maps. However, even if there is an increase in variance (ie, increased autopower), the larger super-pixels can allow us to achieve unbiased cross-correlation measurements at lower $\ell$.

In Fig.~\ref{fig:explore_nside_realspaceILC} we explore how the auto-spectra of the maps change when we choose real-space domains corresponding to different values of $N_{\mathrm{side}}$  (see Table~\ref{tab:nsideresol} for the FWHM of the Gaussian kernels we use). In the top row of Fig.~\ref{fig:explore_nside_realspaceILC}, we use only the fully-integrated $N_{\mathrm{HI}}$ map in the ILC; in the bottom row we use the 3 spectrally binned $N_{\mathrm{HI}}$ measurements as described in Section~\ref{sec:spectral_binning}. Note that, while there was an ILC bias in this case for the $N_{\mathrm{side}}=16$ super-pixel case, there are more modes available when we use larger super-pixels so the ILC bias will decrease. In all cases the power is decreased when we use three spectral bins compared to the case when we use only one.

It is clear from Fig.~\ref{fig:explore_nside_realspaceILC} that we can reduce the errorbar on $C_\ell^{\nu\kappa}$ compared to the uncleaned map down to $\ell\sim20$. 

\begin{figure}
\includegraphics[width=\columnwidth]{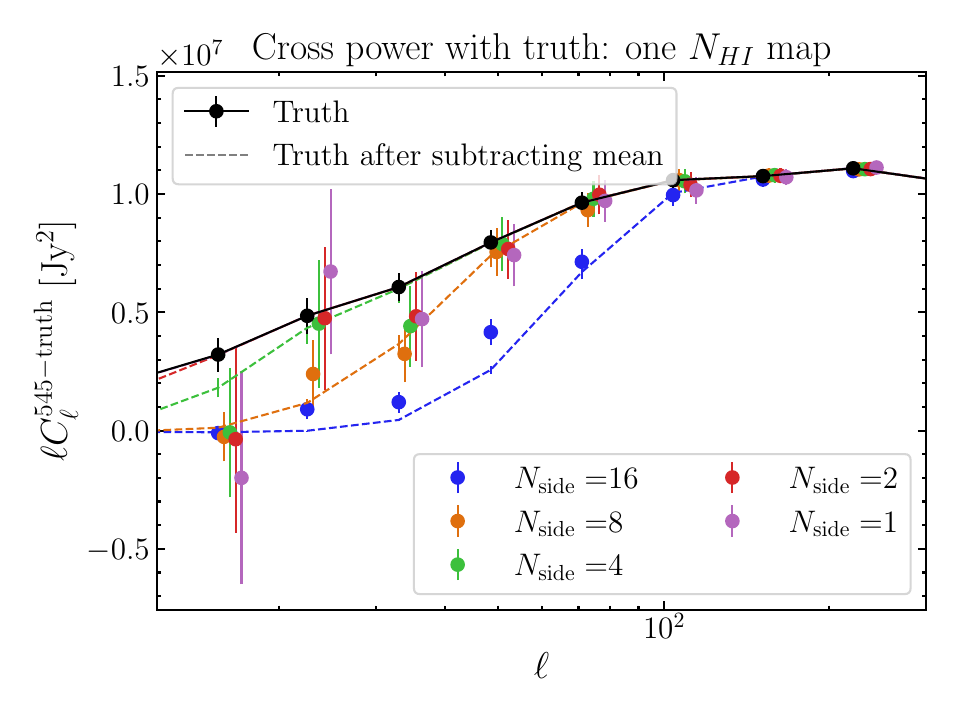}
\includegraphics[width=\columnwidth]{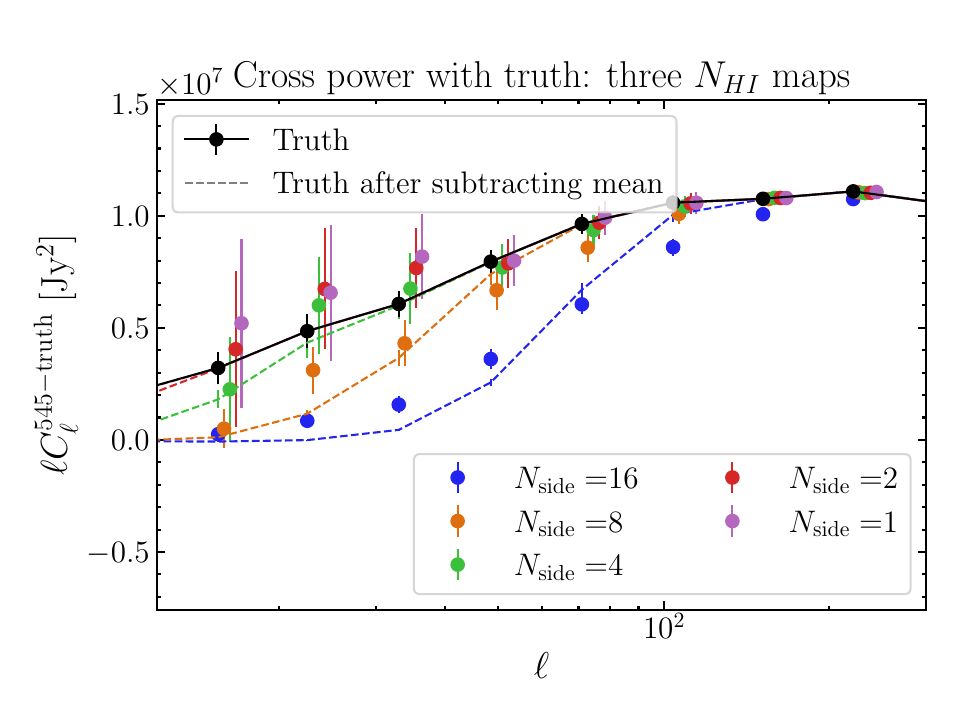}
\caption{Validation of the ``real-space'' ILC \texttt{pyilc} maps shown in Fig~\ref{fig:explore_nside_realspaceILC}, which are shown for the three-$N_{\mathrm{HI}}$-spectral-bin case. We see very little bias in the one-$N_{\mathrm{HI}}$-tracer case (above). For the three-$N_{\mathrm{HI}}$-tracer case (below), we again see the slight bias at $\ell\sim100$ for the $N_{\mathrm{side}}=16$ case, but no significant bias in the other cases: the areas on which we compute the covariances are large enough to avoid this.}\label{fig:ILC_websky_compare_nside}
\end{figure}

We show in Fig.~\ref{fig:ILC_websky_compare_nside} the ``validation checks'' performed: the result of injecting a known signal (again the Websky CIB signal) into the data, running the pipeline, and measuring the cross-correlation of the output with the truth. We see that most of the  configurations are essentially unbiased by this metric, except for the previously noted ILC bias in the $N_{\mathrm{side}}=16$ case when three spectral bins are used, and a small bias in the $N_{\mathrm{side}}=8$ case as well (evident at $\ell\sim80$).

\subsection{Keeping the mean information with NILC: cleaning large scales separately}\label{sec:NILC_cleaning}

Using larger superpixels for the cleaning reduces the ability of the ILC to clean the intermediate scales.  Thus,  to attempt to achieve achieve a better performance we perform a needlet ILC, whereby the small-scale modes ($\ell\lessthanapprox100$) are cleaned separately to the larger scales ($\ell\greaterthanapprox100$). In this Section we explore such a procedure. 

We describe the harmonic filters used in our NILC in Section~\ref{sec:harmonic_filters}, and the real-space filters in Section~\ref{sec:realspace_filters}. We show the resulting auto power spectra of the 545 GHz map in Section~\ref{sec:autospectra_NILC}, and present the validation results upon injection of a known signal in Section~\ref{sec:validation_NILC}.

\subsubsection{Harmonic filters}\label{sec:harmonic_filters}
\begin{figure}
\includegraphics[width=\columnwidth]{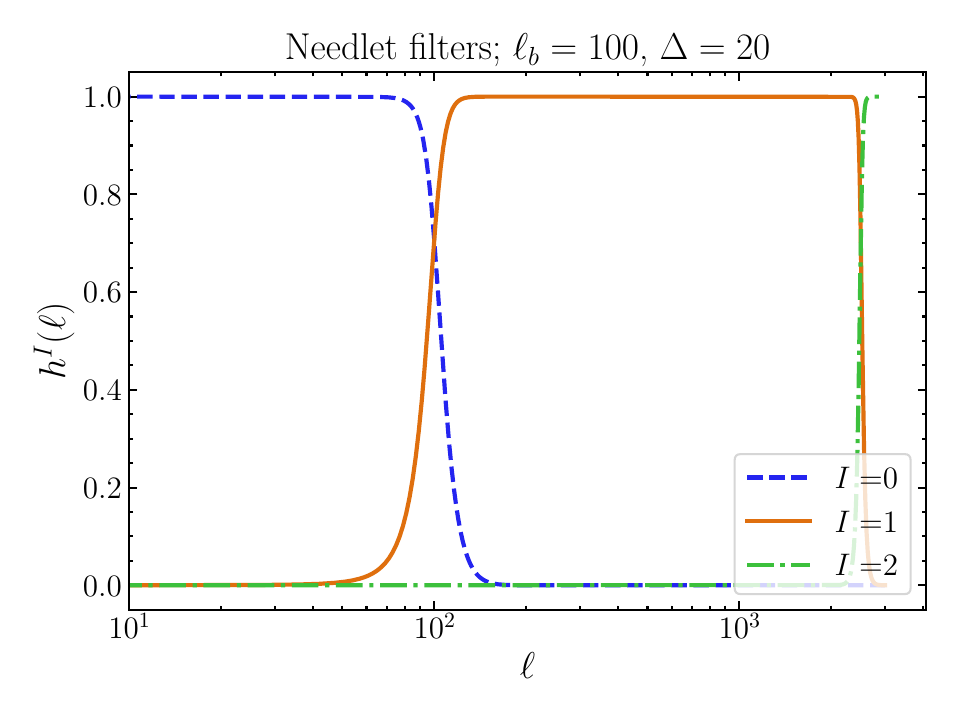}
\caption{The needlet filters used with a boundary $\ell_b=100$, and a taper width of $\Delta=20$. These filters effectively separate the large and small scales at $\ell_b$. Note that there as an additional very small-scale filter ($I=3$) with \{$\ell_b=2500$, $\Delta=100$\}, which we use to ensure no HI data is used at $\ell>2500$, which helps with numerical stability.}\label{fig:needlets}
\end{figure}

In every case we use two needlet filters, in order to separate large and small scales, with some boundary multipole, $\ell_{b}$. The wavelet filters are as follows:
\begin{equation}
h^I(\ell)^2 = \begin{cases}
    1-0.5 \tanh\left(1+\frac{2}{\Delta}\left(\ell-\ell_b\right)\right)&\ell<\ell_b\\
 1+0.5 \tanh\left(1+\frac{2}{\Delta}\left(\ell-\ell_b\right)\right)&\ell>\ell_b,
\end{cases}
\end{equation}
where $\Delta=20$ sets the width of the transition region between the two needlet filters. The needlet filters for $\ell_b=100$ are shown in Fig.~\ref{fig:needlets}.

Note that in practice we find that when the needlet coefficients are computed on the full-sky (this is the operation in which the real-space maps are transformed to harmonic space, filtered by the harmonic needlet scales, and transformed back to real space), the large anisotropic power on the large scales makes it difficult to isolate locally the large-scale modes alone; thus we apply the boolean mask relevant to the region we are interested in to the maps before performing all computations (ie, \textit{before computing the wavelet coefficients of each map}).

\subsubsection{Real-space filters}\label{sec:realspace_filters}

For the large scale ($I=0$) needlet coefficients, we choose the 
real-space kernels to be Gaussian filters  corresponding to $N_{\mathrm{side}}=1$ superpixels.
 We continue to subtract the means of the maps calculated on these regions before creating the final map, so we are insensitive to modes on larger scales than this (ie, modes with $\ell\lessthanapprox10$). This corresponds to a FWHM of 58.6 degrees (see Table~\ref{tab:nsideresol}).
 
 For the small-scale modes ($I=1$ needlet coefficients), we explore using Gaussian filters corresponding to $N_{\mathrm{side}}=8,16$ superpixels; again see Table~\ref{tab:nsideresol} for the FWHMs. We do not subtract any means on these needlet scales.

\subsubsection{Autopower spectra}\label{sec:autospectra_NILC}

\begin{figure*}
\includegraphics[width=0.32\textwidth]{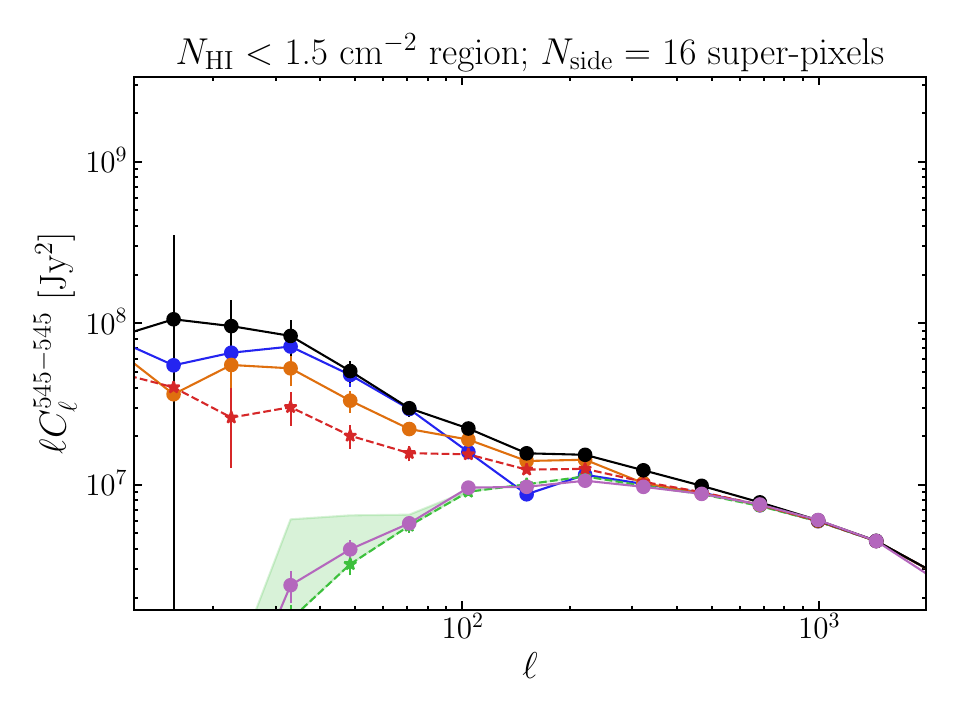}
\includegraphics[width=0.32\textwidth]{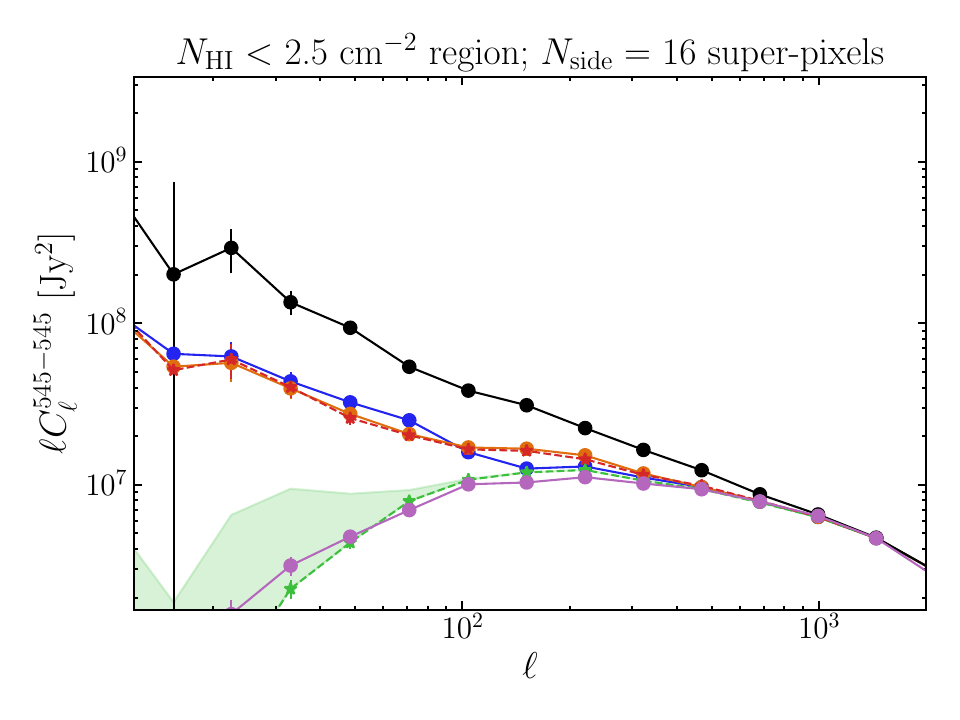}
\includegraphics[width=0.32\textwidth]{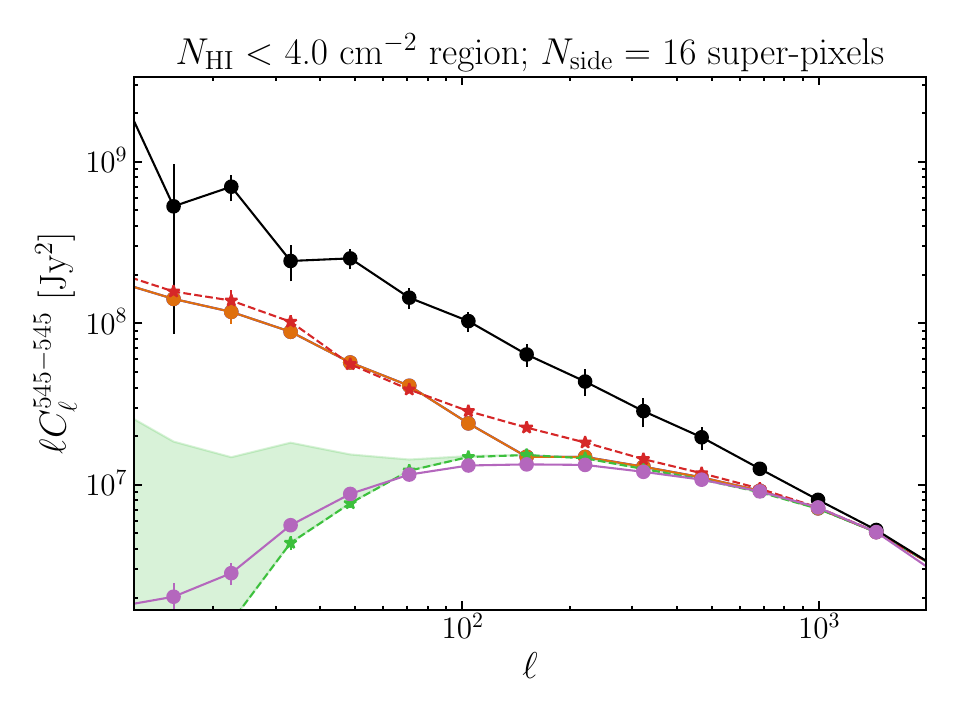}
\includegraphics[width=0.6\textwidth]{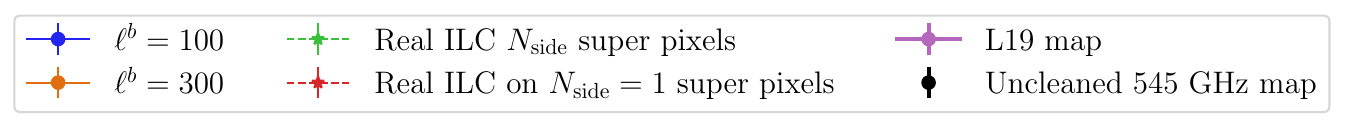}\\
\includegraphics[width=0.32\textwidth]{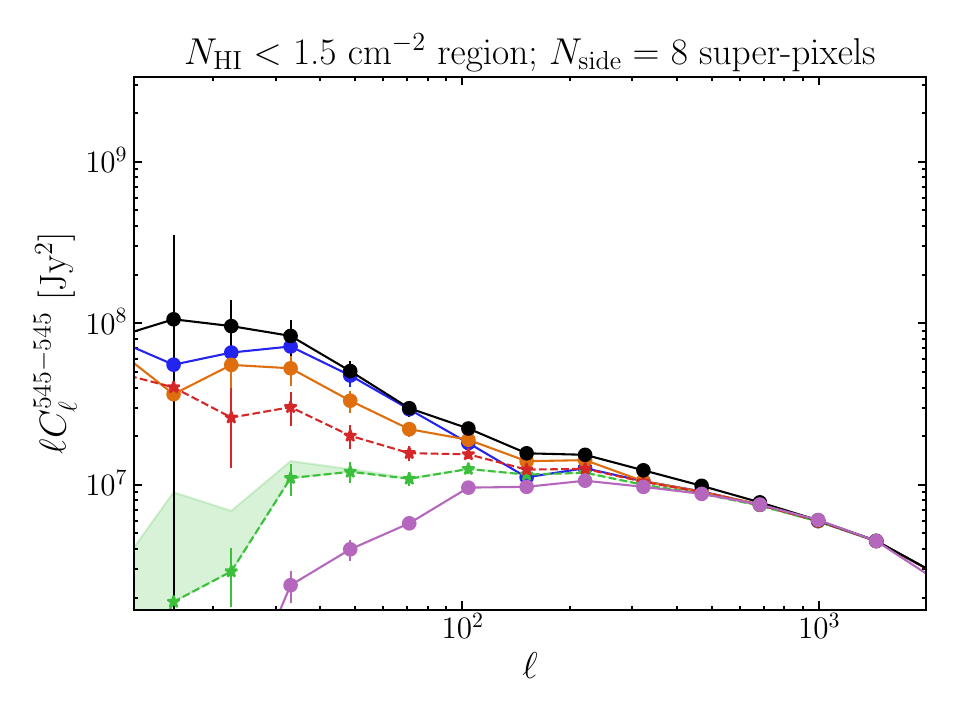}
\includegraphics[width=0.32\textwidth]{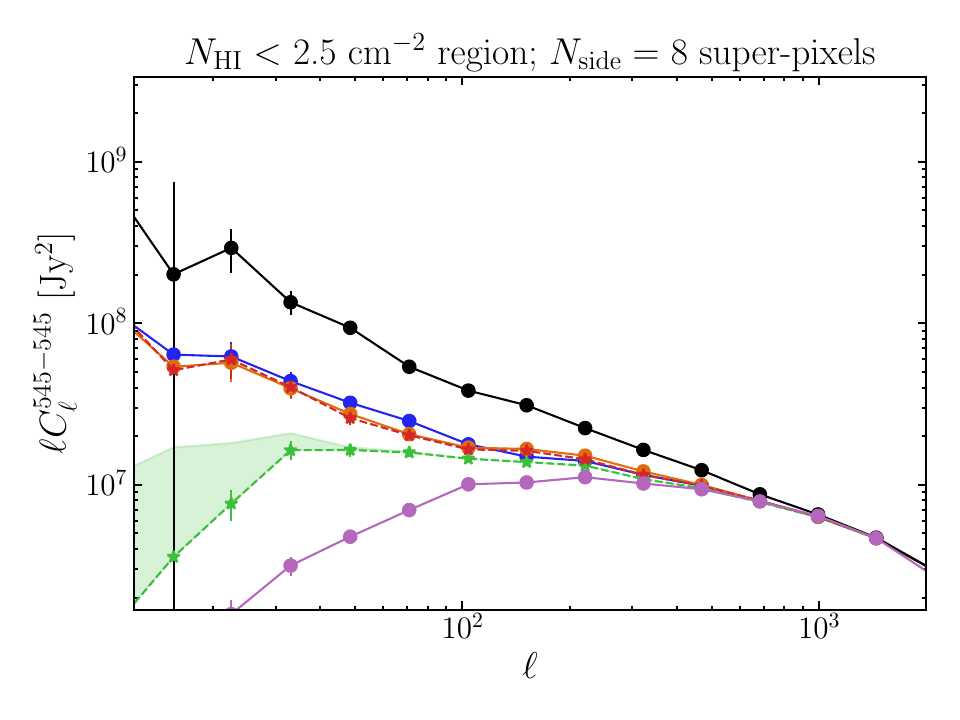}
\includegraphics[width=0.32\textwidth]{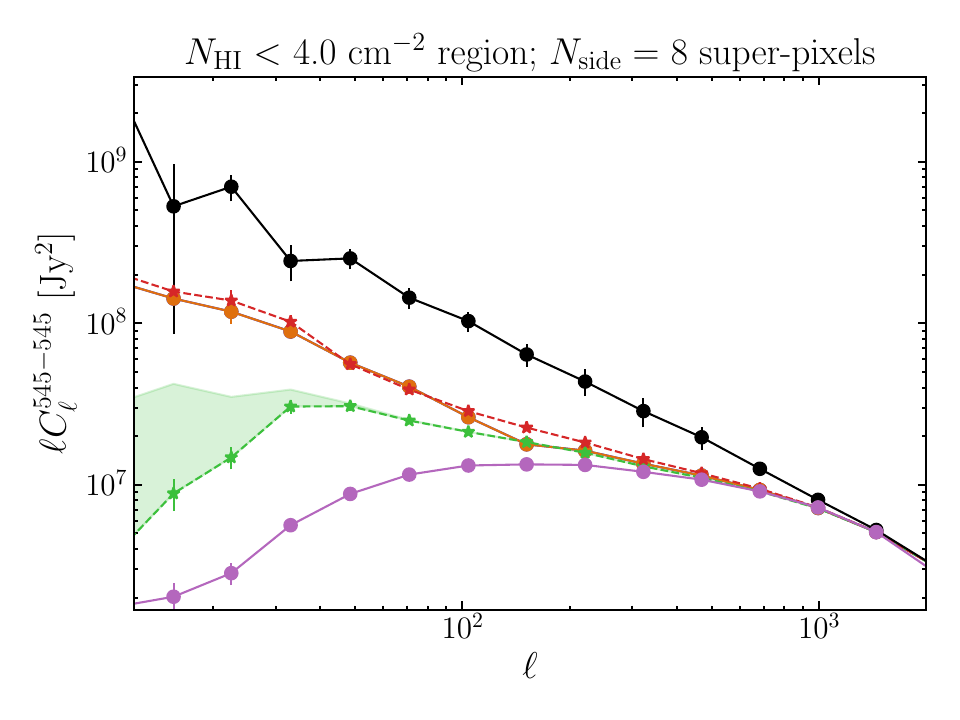}
\caption{The effect of performing needlet ILC with two (tapered) top-hat $\ell$ bins, with boundaries at $\ell^b=100,300$. For $\ell>\ell^b$ we perform ILC with real-space kernels corresponding to $N_{\mathrm{side}}=16$ (top plots) or $N_{\mathrm{side}}=8$ (bottom plots) pixels, and for $\ell<\ell^b$ we perform ILC with real-space kernels corresponding to $N_{\mathrm{side}}=1$ pixels, and subtract the mean as computed on these domains (thus we are insensitive to modes with $\ell<\sim 10$, as indicated in Fig.~\ref{fig:suppression_nside}). All of these maps have been cleaned with 3 spectrally binned $N_{\rm{HI}}$ tracers.}\label{fig:diff_ell}
\end{figure*}

We show in Fig.~\ref{fig:diff_ell} the auto power spectra of the resulting maps, including two different choices of $\ell_b$. We find that it is difficult to do much better than the standard $N_{\mathrm{side}}=1$ real-space ILC, but that there some $\ell$s around the boundary ($\ell\sim100-300$) where we can improve on this; however, this depends slightly on the region on which we perform the component separation.  We will also see in  below  that this is at the cost of some ILC bias.

\subsubsection{Validation and ILC bias}\label{sec:validation_NILC}
\begin{figure}
\includegraphics[width=\columnwidth]{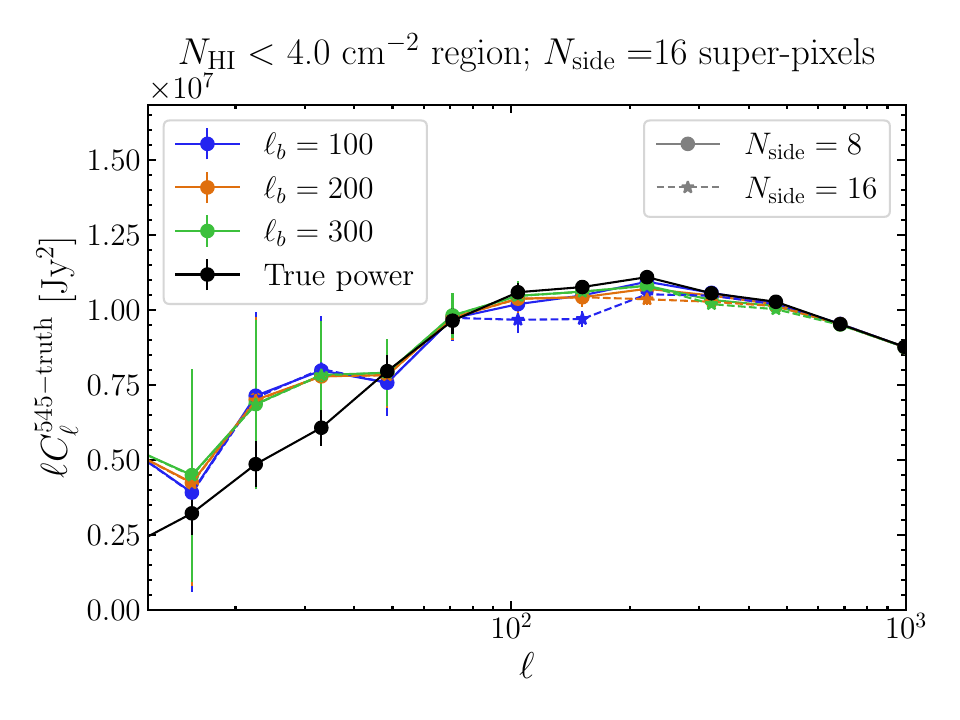}
\caption{The cross-power of the recovered signal with the truth, when we inject the Websky CIB into the data and perform NILC with an $\ell$ at the boundary as indicated by the colours, for $N_{\mathrm{side}}=16$ and $N_{\mathrm{side}}=8$ superpixels for the large $\ell$s; the auto power of such maps are shown in Fig.~\ref{fig:diff_ell}. Overall we see very little ILC bias (with the low-$\ell$ increase in power consistent with a fluctuation), although there is a hint of a bias for \{$\ell_b=100$,  $N_{\mathrm{side}}=16$ superpixels\}.} \label{fig:ILCbias_splitell}
\end{figure}

It is important to assess whether we have introduced any ILC bias, especially on the large scales where there are very few modes available to calculate the covariance matrices; we indicate this in Fig~\ref{fig:ILCbias_splitell} where we inject the known websky signal as before and very that it can be recovered. We see a very low bias in the recovered signal, except for the case when $\ell_b=100$ and the superpixels correspond to $N_{\mathrm{side}}=16$, where there is a small amount of power loss around $\ell\sim100-200$. This is the one reason why one may choose to use the $N_{\mathrm{side}}=8$ superpixel maps instead, despite their slightly increased variance.

\subsection{Sky area for cleaning}\label{sec:skyarea}

\begin{figure*}
\includegraphics[width=0.3\textwidth]{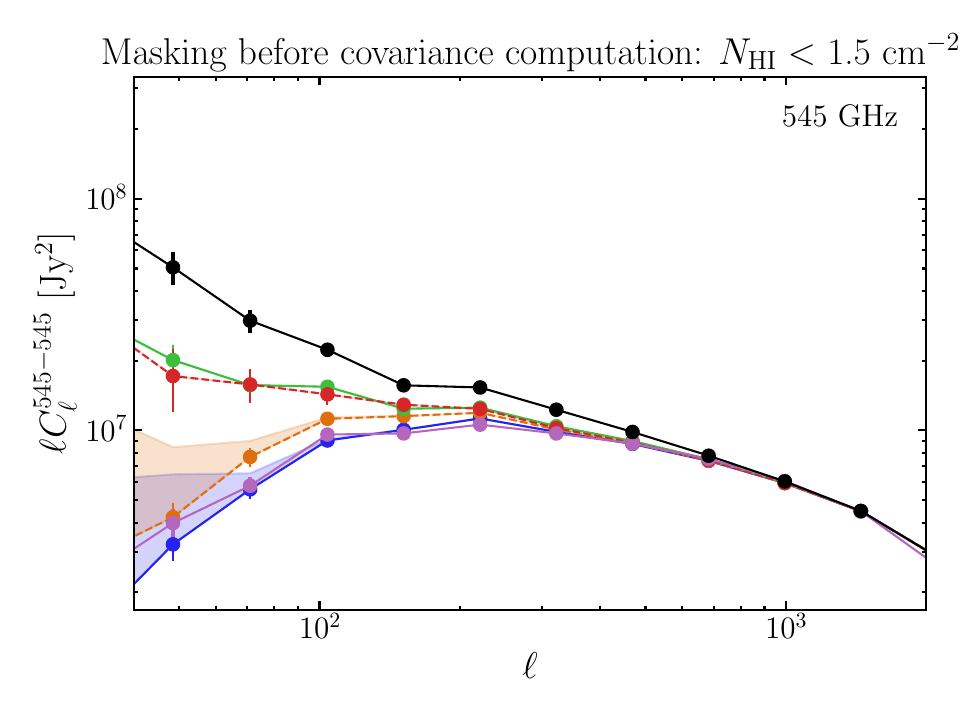}
\includegraphics[width=0.3\textwidth]{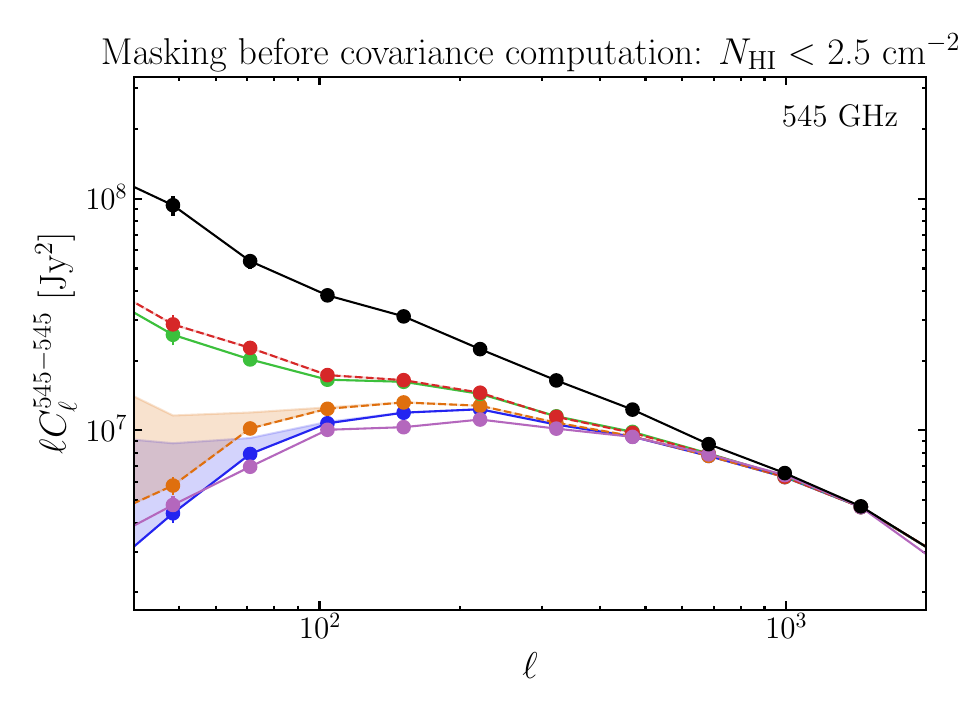}
\includegraphics[width=0.3\textwidth]{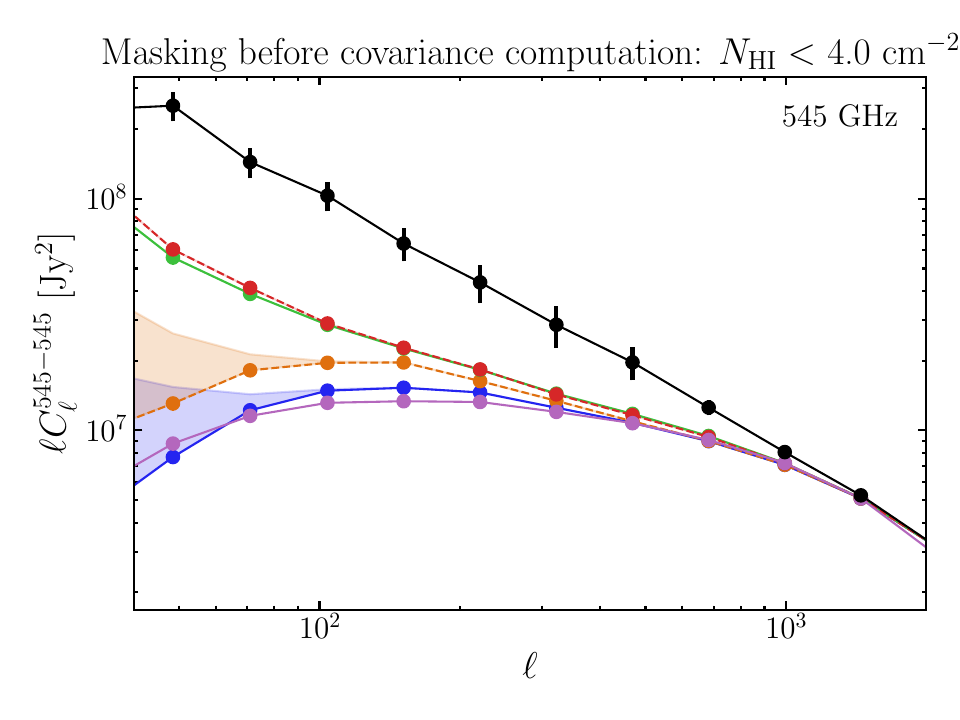}

\includegraphics[width=0.8\textwidth]{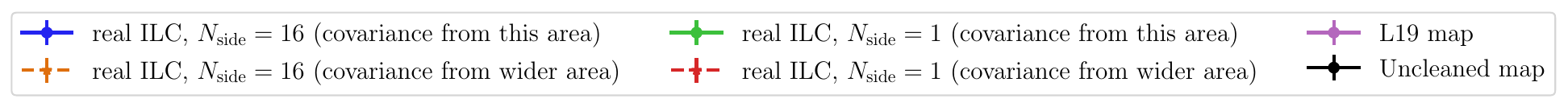}
    \caption{Effects of applying a mask before or after covariance computation. There are at times a small penalty to computing the covariance on a wider sky area than that on which we are interested in computing the power spectrum; this is especially true for the $N_{\mathrm{side}}=16$ real-space ILC case. }\label{fig:compare_skyarea}
\end{figure*}

Our previous results have been shown on sky areas corresponding to the masks of Ref.~\cite{2019ApJ...883...75L}, moving from smaller-area, low-dust sky regions to wider-area, dustier regions. In all cases we have been computing the covariance required for the ILC weights \textit{after} applying the masks, such that the covariance matrices are appropriately estimated. However, the penalty for computing the covariance matrix on a wider area and then applying the masks \textit{after performing the component separation} is sometimes quite small; see Figure~\ref{fig:compare_skyarea}. Note that on the ``wider area'' plots, we have still applied the \textit{Planck} 40\% galactic plane mask  (the mask which leaves 40\% of the sky unmasked) before computing the covariance matrices.

As the difference is small, we only make the CIB maps with covariance computed on the  40\% galactic plane mask publicly available, although all plots in this work are shown for maps with covariances computed on the relevant area. However, we also provide the necessary dataproducts and a sample script that can use \texttt{pyilc} to perform the component separation on an arbitrary area of sky, such that users of the maps can perform a tailored component separation on their area of interest.

\section{CIB maps at 353 and 857 GHz}\label{sec:difffreqs}

\subsection{Autopower spectra}

In Section~\ref{sec:CIBmaps} we compared different maps with the maps of L19 at 545 GHz. In this Section we look also at the 353 and 857 GHz maps, created with similar pipelines. In all cases we use  3 spectral bins for the HI data as described in Section~\ref{sec:spectral_binning}. For the case of 353 GHz, we always subtract a CMB NILC estimate to remove the contribution from the CMB; we do this before performing the NILC with the single-frequency data and the hydrogen data. The CMB map comes with an additional preprocessing mask; thus we always multiply the masks on which we calculate the final power spectra by this mask (apodized with an apodization scale of 2 degrees, using the ``C1'' apodization routine of \texttt{namaster}). 

\begin{figure*}
\includegraphics[width=0.32\textwidth]{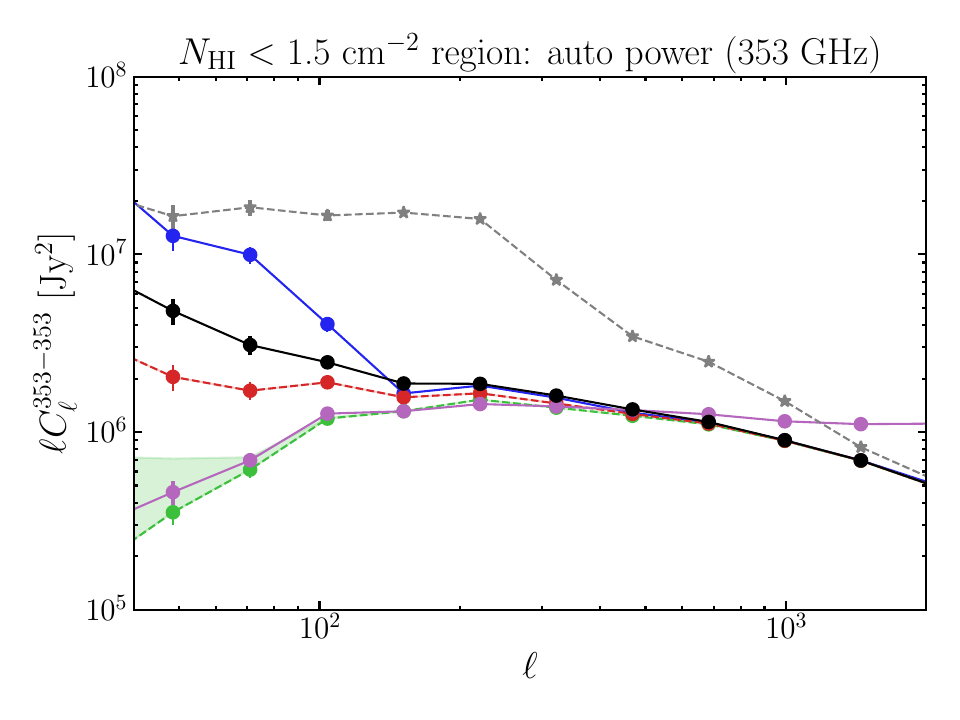}
\includegraphics[width=0.32\textwidth]{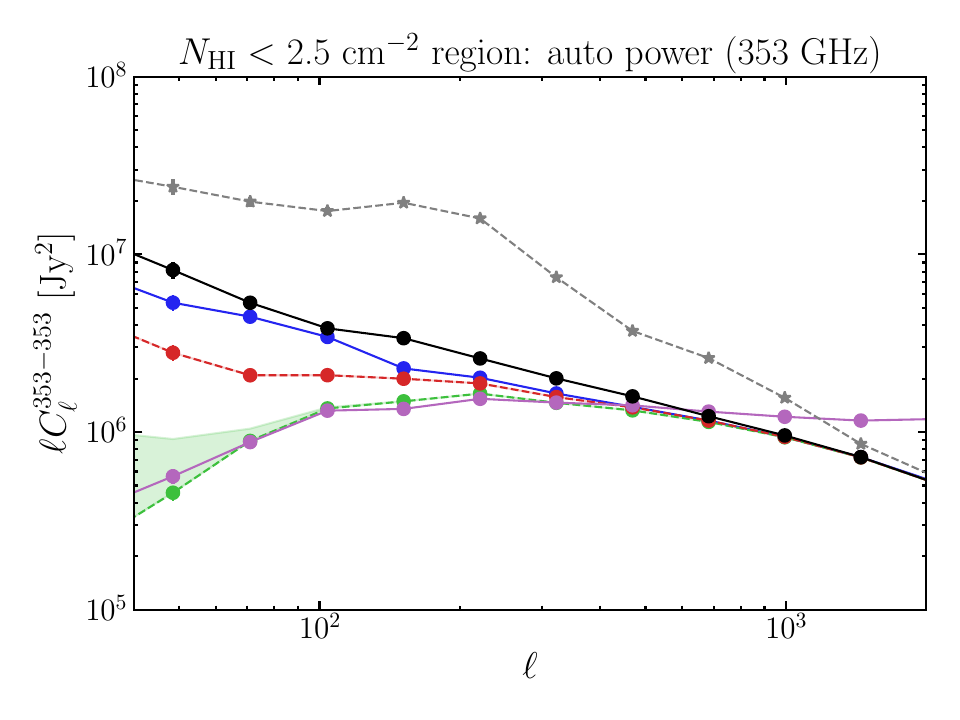}
\includegraphics[width=0.32\textwidth]{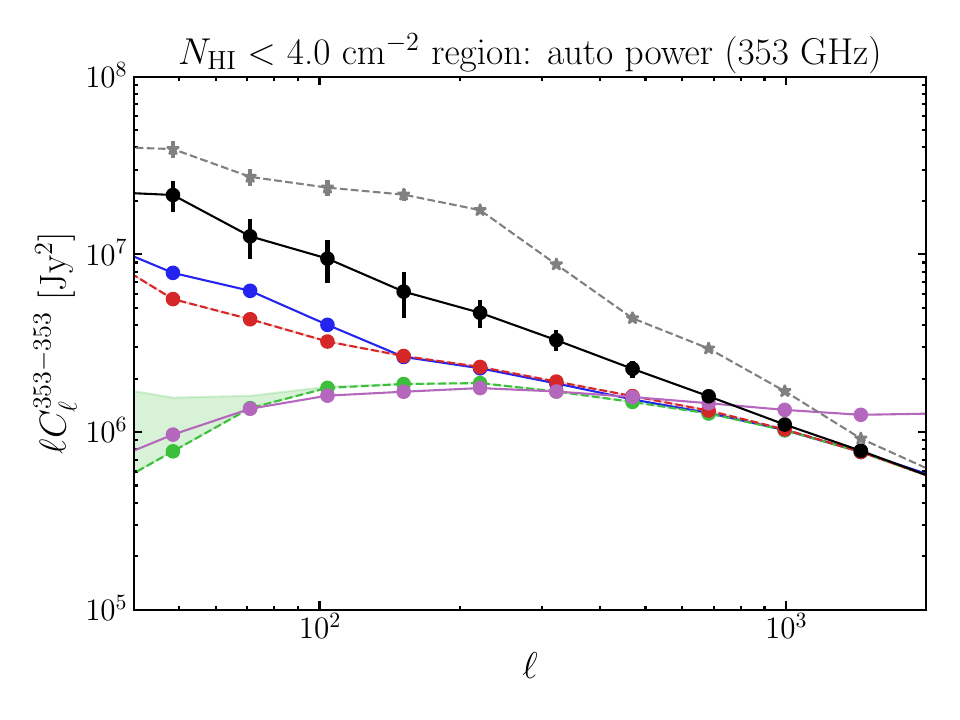}
\includegraphics[width=0.8\textwidth]{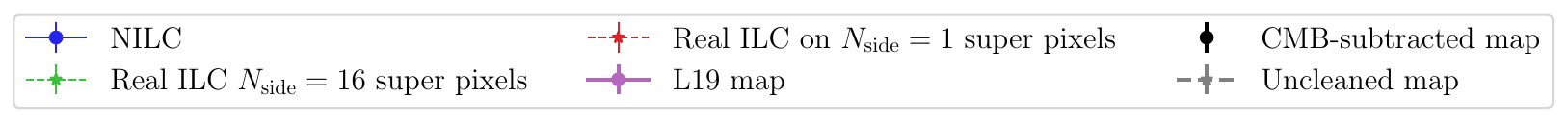}\\
\includegraphics[width=0.32\textwidth]{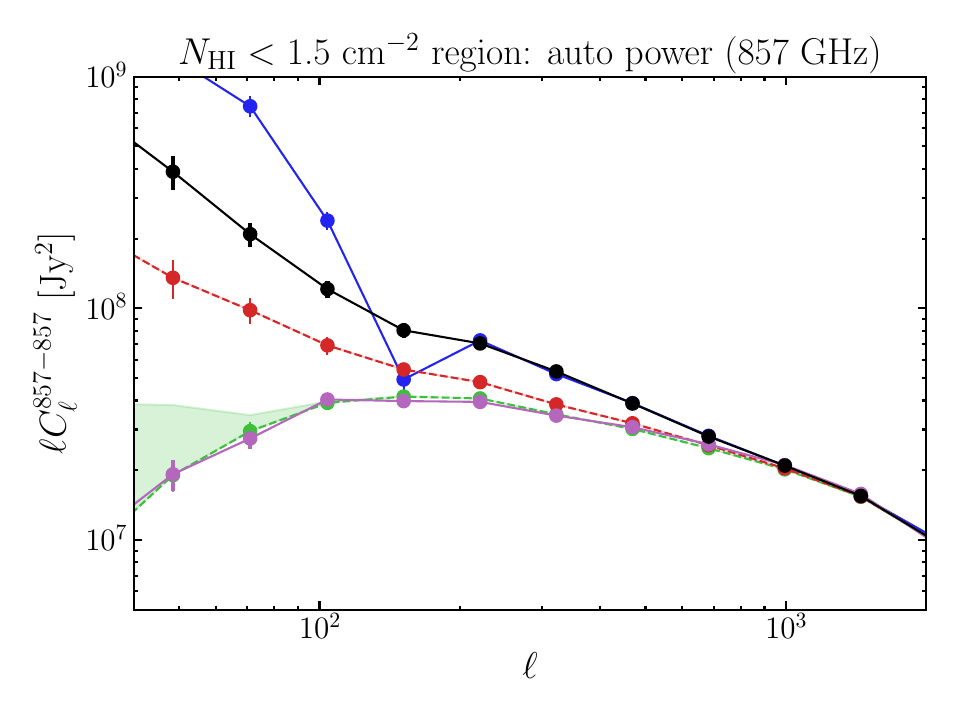}
\includegraphics[width=0.32\textwidth]{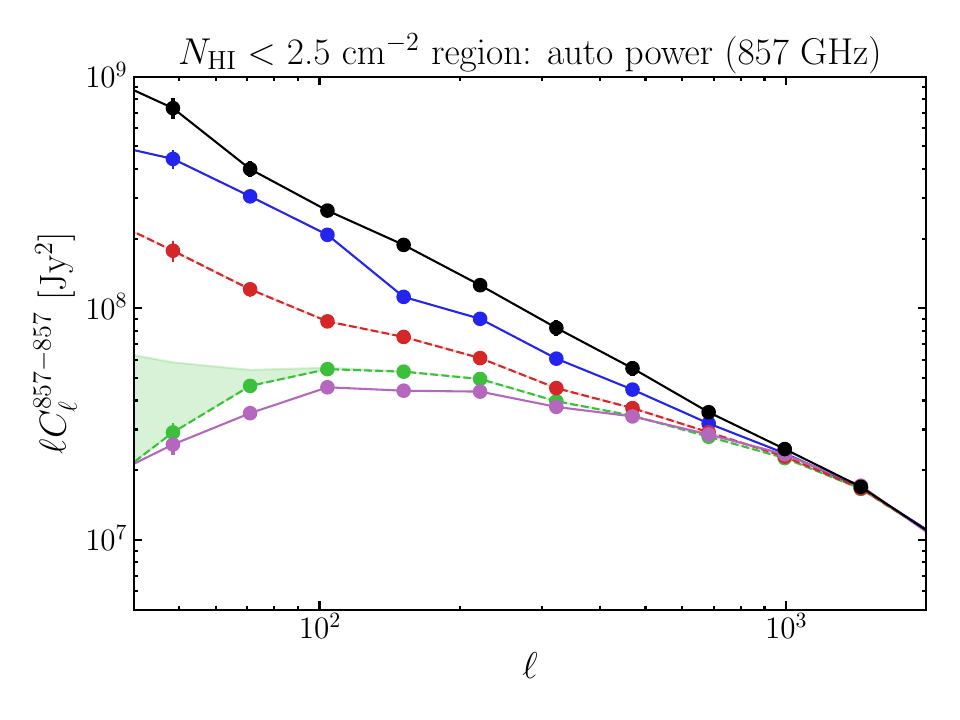}
\includegraphics[width=0.32\textwidth]{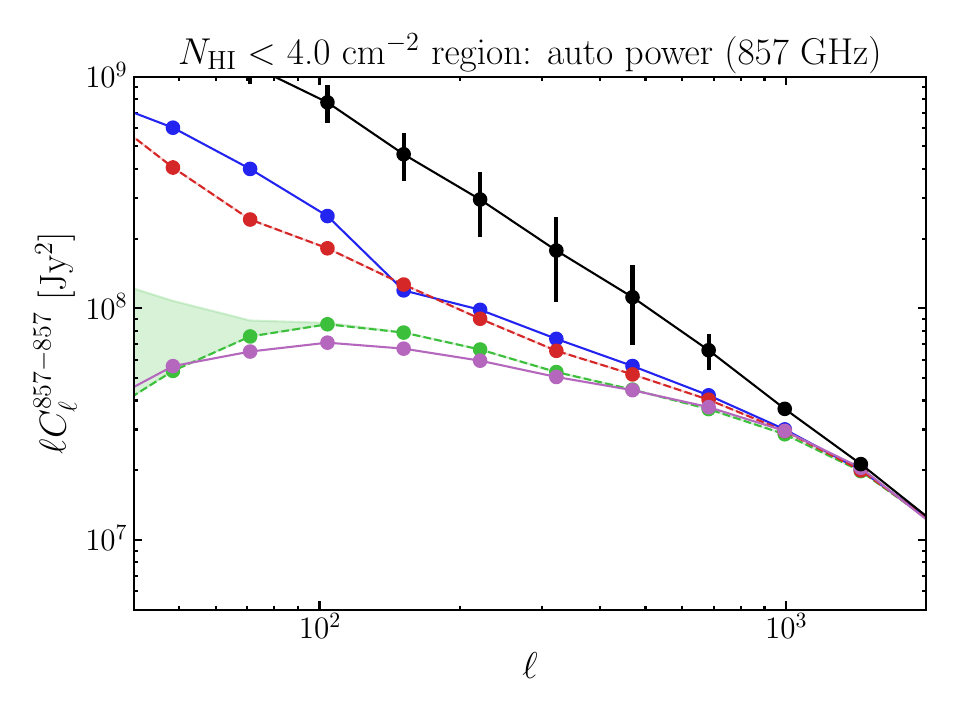}

    \caption{Autopower of the 353 GHz CIB maps (\textit{top}) and the 857 GHz maps (\textit{bottom}). We indicate the raw power of the single-frequency map (dashed gray), as well as the CMB-subtracted single-frequency map (solid black), as well as the power spectra after various cleaning pipelines. In particular, in blue we have a NILC with $\ell_b=100$, and in dashed red and green we have real-space ILC with $N_{\rm{side}}=1,16$ respectively. For the NILC, we used real-space kernels appropriate for $N_{\rm{side}}=16$ for the intermediate scales. We also show the power of the L19 maps on the same region. As before, the transfer function of our real-space ILC maps is indicated with a shaded region. All of these maps have been cleaned with 3 spectrally binned $N_{\rm{HI}}$ tracers.}\label{fig:353_power}
\end{figure*}

The power spectra of the 353 GHz and 857 GHz maps, for various choices of NILC settings, on the same regions as the L19 maps, are shown in Fig.~\ref{fig:353_power}. These plots should be compared with Fig.~\ref{fig:diff_ell} for the 545 GHz case. 

Generally, in these cases our NILC algorithm does not in general do better than the real-space ILC with $N_{\mathrm{side}}=1$ pixels. However, in general we can improve upon the uncleaned maps and achieve similar performance to the maps of L19 at intermediate and small scales. 

\begin{figure*}
    \includegraphics[width=0.32\textwidth]{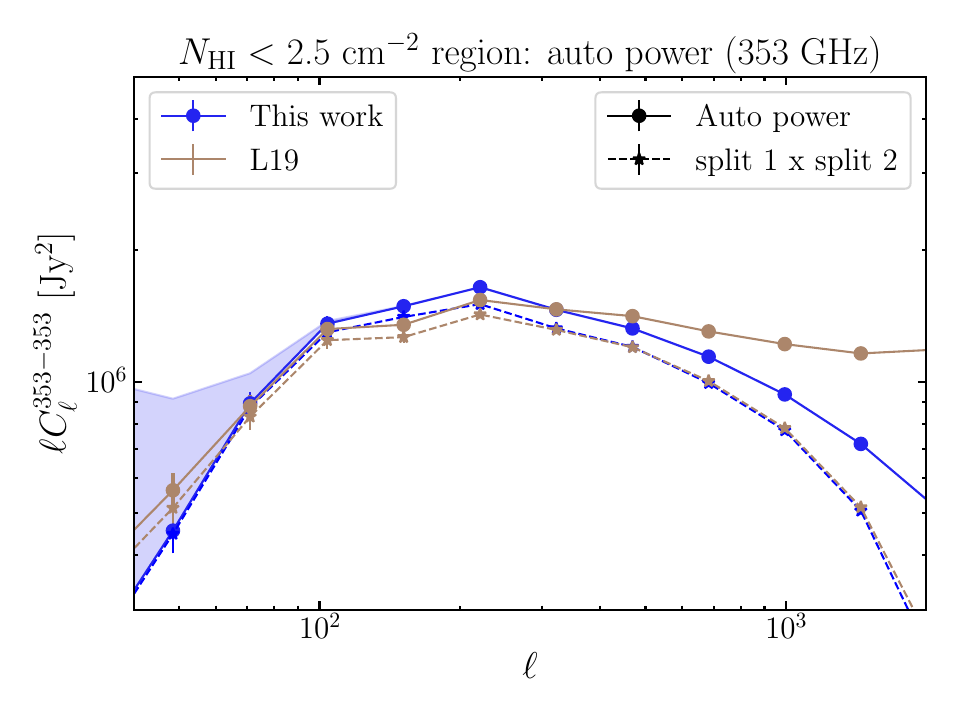}
    \includegraphics[width=0.32\textwidth]{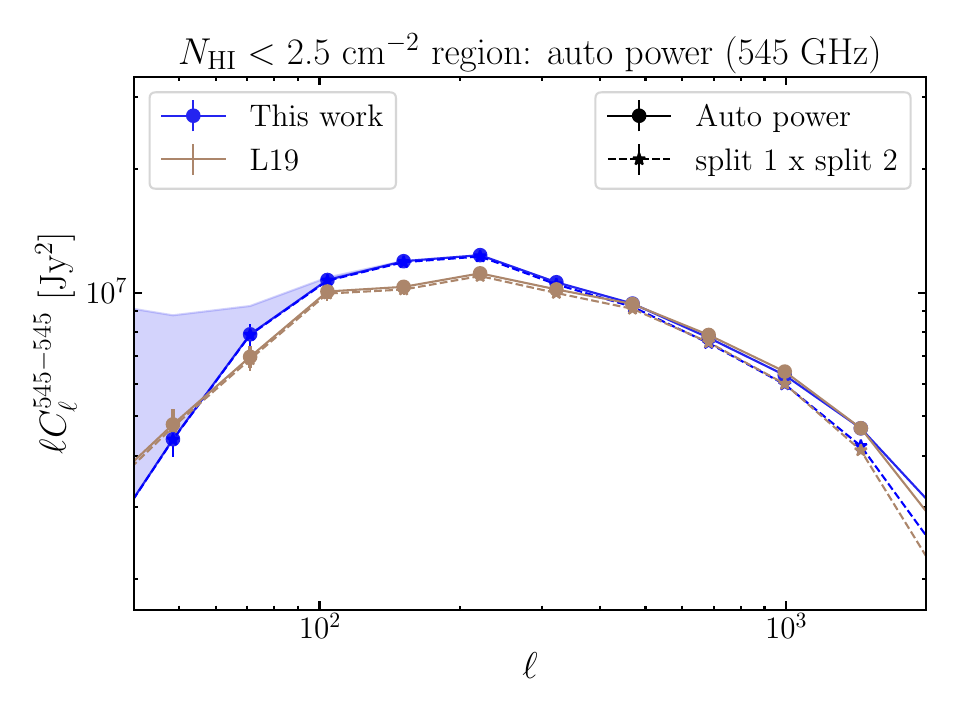}
    \includegraphics[width=0.32\textwidth]{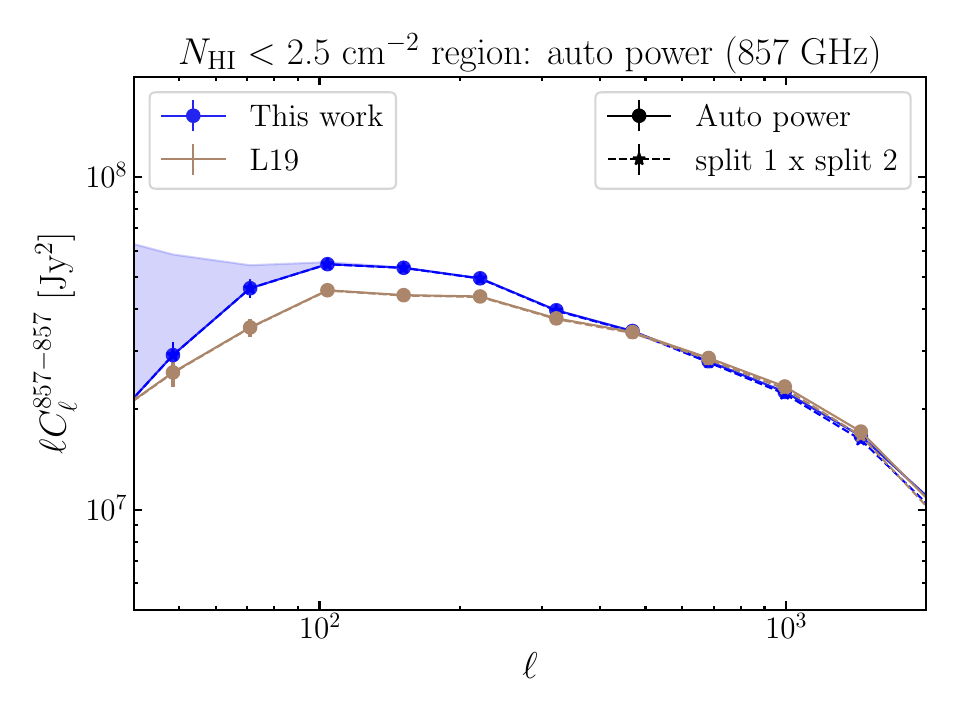}
    \caption{Auto power spectra of the CIB maps, measured on independent splits of the data. We show the measurements both on our maps and on the L19 maps. A clear improvement is seen in 353 GHz where the noise bias is significantly lower in our PR4 maps than those of L19, which were made with PR3 data; there are no significant improvements at the higher frequencies. In all cases the maps we use from this work are $N_{\mathrm{side}}=16$ real-space ILC maps, with 3 tracers, to correspond best to the L19 maps.}
    \label{fig:splits}
\end{figure*}

Finally, this is the first case where we have seen an improvement compared to L19 due to the lower instrumental noise; this is evidenced by the fact that our auto power-spectra are lower than that of L19 at $\ell\greaterthanapprox600$. To verify that this is indeed due to the lower noise properties of the PR4 data cf the PR3 data, we can measure the cross power-spectra of maps created from two independent splits of the data; see Fig.~\ref{fig:splits}. In this case, we have created one map with the ringhalf-1 split and the other with the ringhalf-2 split provided by the NPIPE analysis; in order to achieve this we also created two ringhalf NILC CMB maps from these splits as well, such that the CMB templates we subtract from the 353 GHz channel also have independent instrumental noise realizations. 

We see in Fig.~\ref{fig:splits} that indeed the cross-power spectra of the two split L19 maps is equal to the cross power spectra of our split maps, indicating that indeed the reduction in the auto power spectrum is due to lower instrumental noise. At 545 and 857 GHz any improvements in instrumental noise are not visible by-eye.

\section{Cross power with CMB lensing}\label{sec:xpowercmblensing}

\subsection{CIB-CMB lensing cross correlation measurement}

As the CMB, which is sourced at redshift $z\sim1100$, travels through the Universe, it interacts gravitationally with all of the intervening matter, leading to a phenomenon known as gravitational lensing. This  can be detected statistically through its well-understood effects on the CMB, in particular its non-Gaussian imprints, which can be exploited to create a map integrated matter that caused the lensing. CMB lensing is an excellent probe of the dark matter field in our Universe, has been detected with increasing signal to noise in many CMB datasets, and is a well-established cosmological probe~\cite{2007PhRvD..76d3510S,2011PhRvL.107b1301D,2012ApJ...756..142V,2020ApJ...888..119B,2020A&A...641A...8P,2021MNRAS.500.2250D,2022JCAP...09..039C,2023arXiv230405202Q,2023arXiv230405203M,2023PhRvD.108l2005P}. For a review of CMB lensing, see~\cite{2006PhR...429....1L}.

The CMB is lensed by matter at all redshifts, withmost efficiency at $z\sim2$, close to where the peak CIB intensity is. Thus, it has long been recognised that the CIB should be highly correlated with the CMB lensing distribution~\cite{2003ApJ...590..664S}. This correlation has been detected with high significance many times (see, e.g.~\cite{2014A&A...571A..18P,2015ApJ...802...64B}). As the CMB lensing maps are created from CMB frequency channels that do not contain a large amount of galactic dust, the dust contamination in the CIB maps does not cause a significant bias to the CMB lensing-CIB cross correlation measurement, instead just contributing variance to the measurement; thus the removal of dust can reduce the errorbar on the measurements, but residual dust should not cause an incorrect measurement. Therefore, we can use our maps to measure this signal even on large scales where there remains dust contamination. We ensure that the measurements are consistent before and after cleaning, and quantify the reduction in variance caused by our cleaning.

To measure this cross correlation, we use the NPIPE CMB lensing maps. When we  make the measurements, we multiply the previous masks (the L19 masks, multiplied by the apodized CMB pre-processing mask for the case of 353 GHz) by the NPIPE CMB lensing reconstruction mask, apodized with an apodization scale of half a degree; the CMB lensing mask only removes a small additional region of sky in every case.

\begin{figure*}
\includegraphics[width=0.32\textwidth]{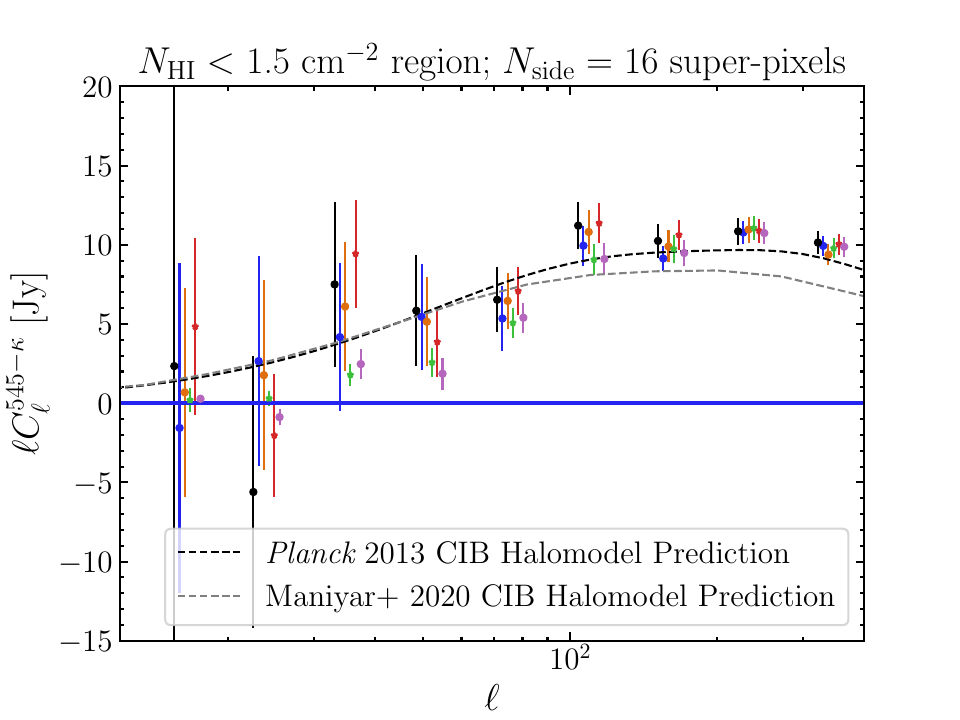}
\includegraphics[width=0.32\textwidth]{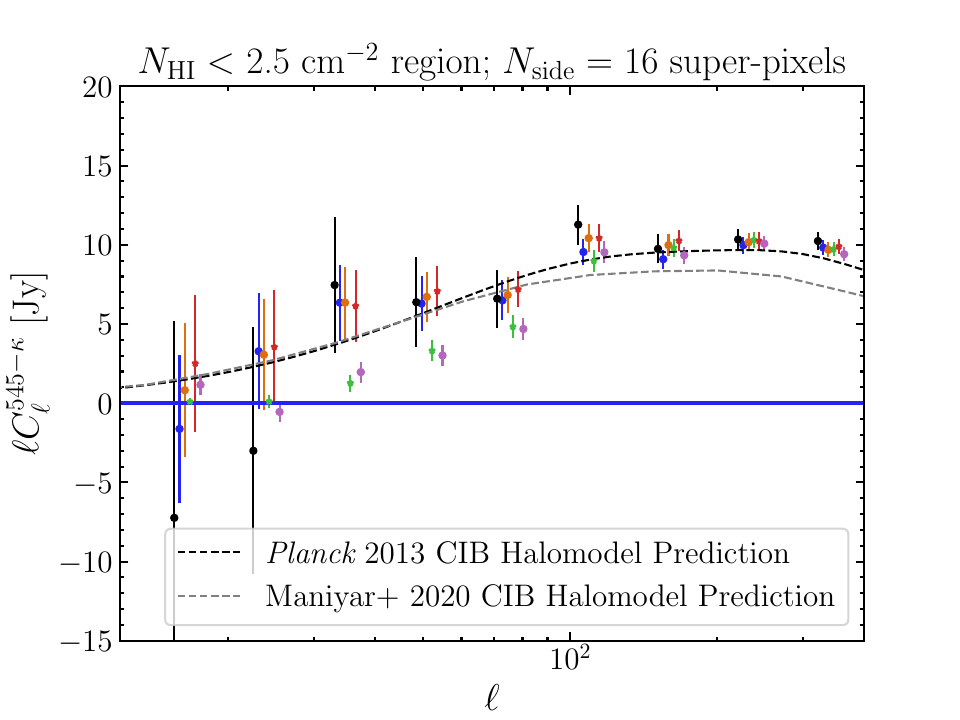}
\includegraphics[width=0.32\textwidth]{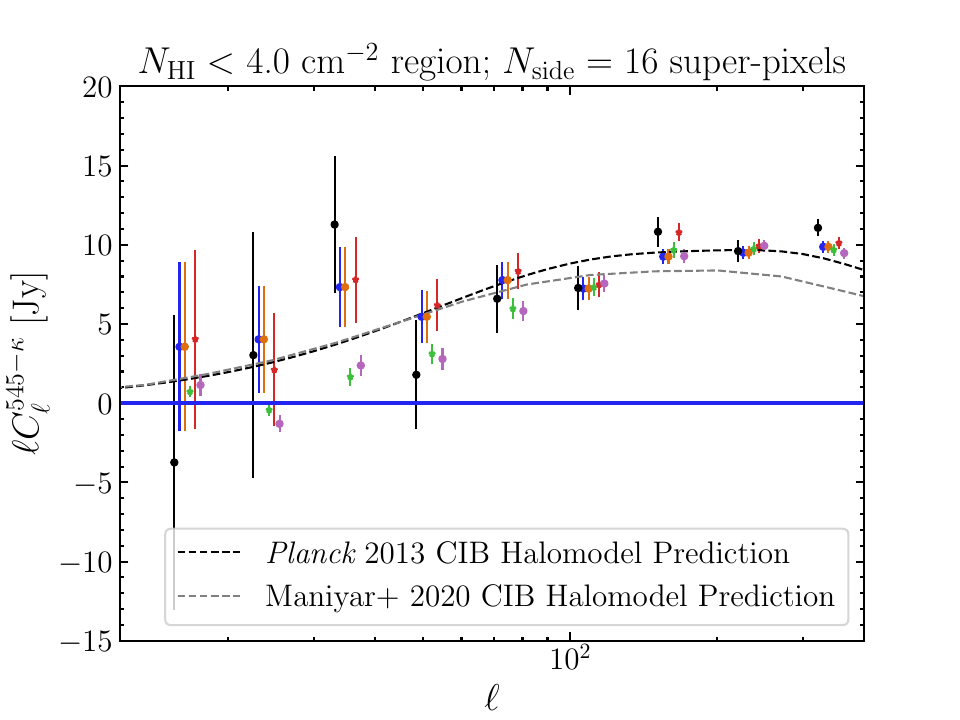}
\includegraphics[width=0.6\textwidth]{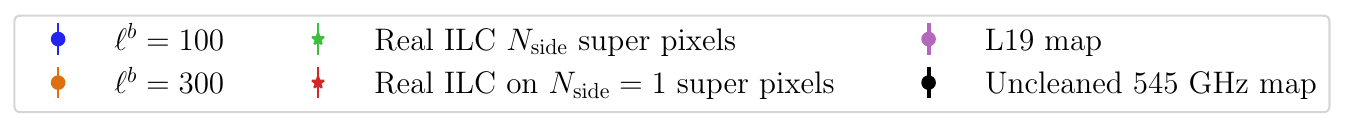}\\
\includegraphics[width=0.32\textwidth]{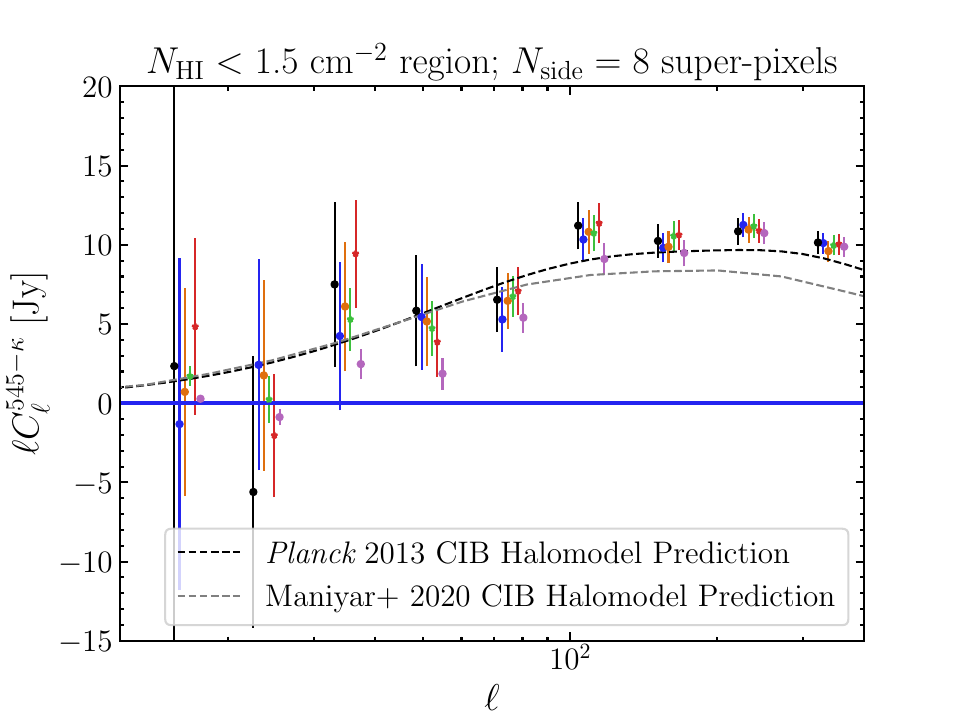}
\includegraphics[width=0.32\textwidth]{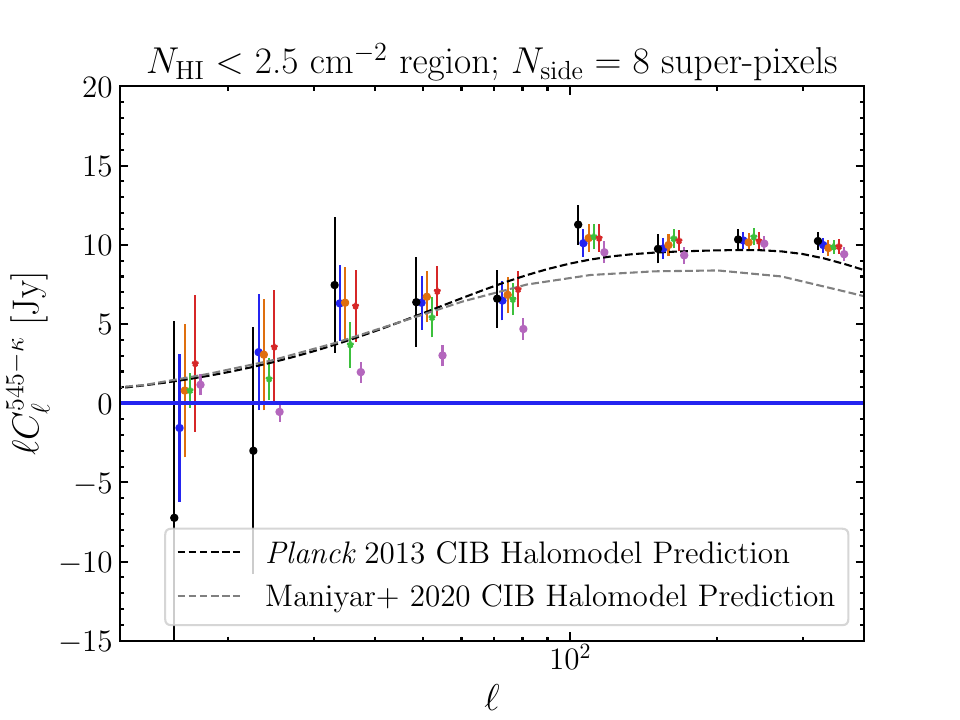}
\includegraphics[width=0.32\textwidth]{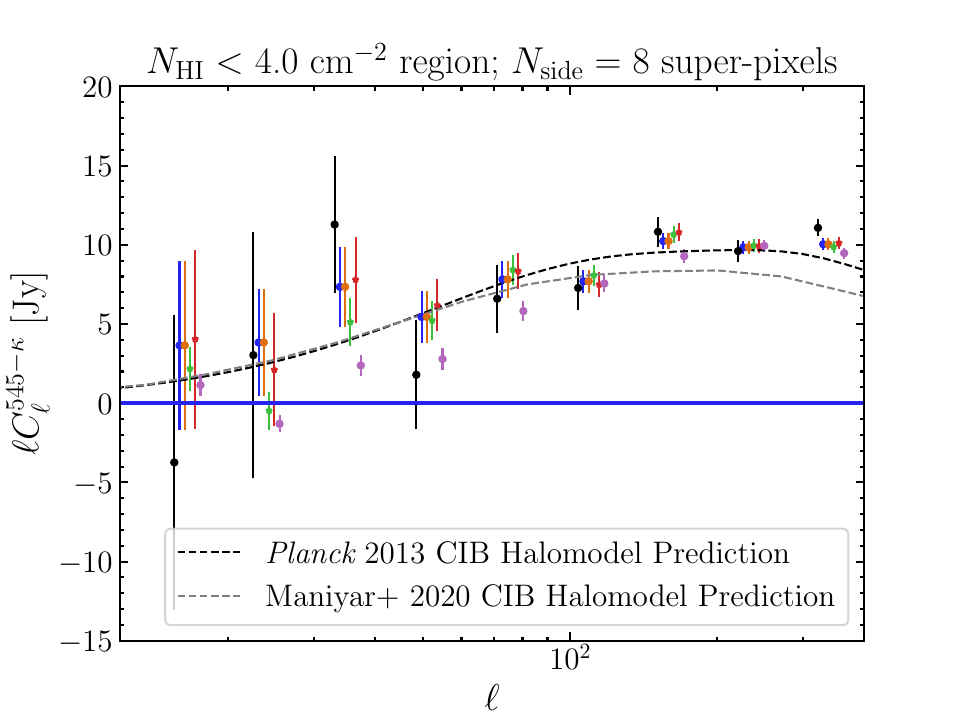}
\caption{The cross power with the NPIPE CMB lensing reconstruction of the maps whose auto spectra are shown in Fig.~\ref{fig:diff_ell}. Theory predictions from the CIB halomodels of~\cite{2014A&A...571A..30P} (``\textit{Planck} 2013'')and~\cite{2021A&A...645A..40M} (``Maniyar 2020'') are indicated in dashed lines. These were computed with \texttt{class\_sz}.}\label{fig:Xcorr_lensing}
\end{figure*}

\begin{figure*}
\includegraphics[width=0.32\textwidth]{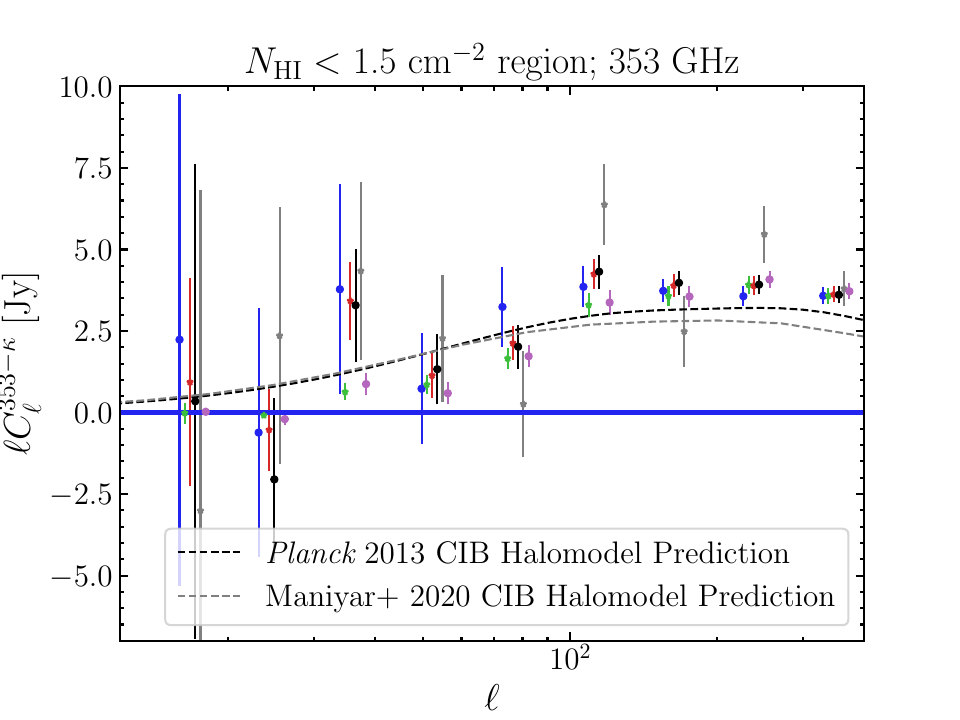}
\includegraphics[width=0.32\textwidth]{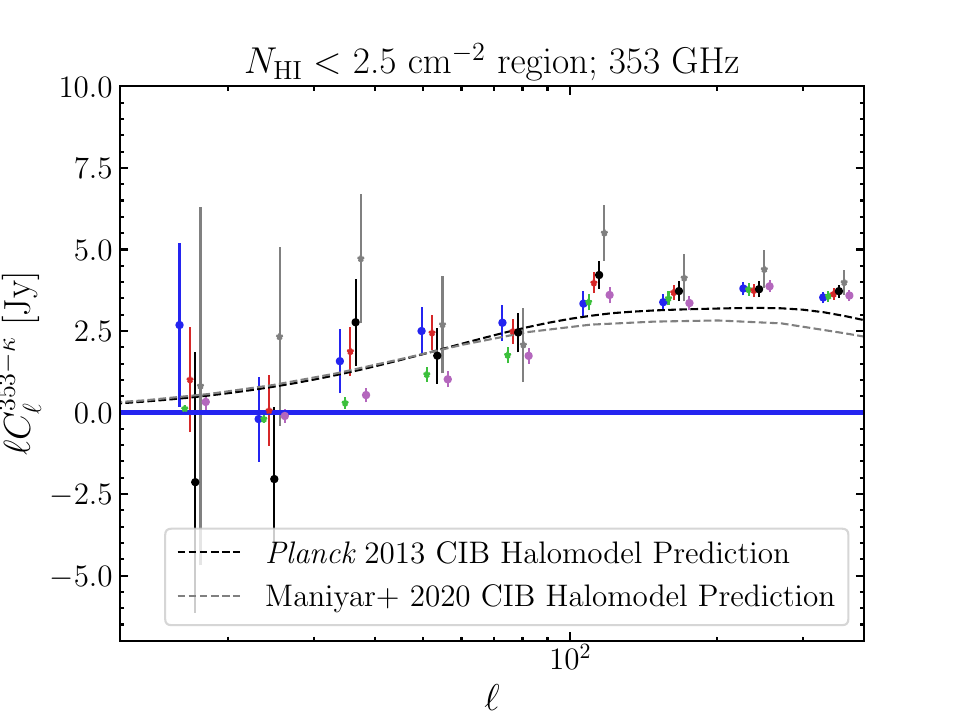}
\includegraphics[width=0.32\textwidth]{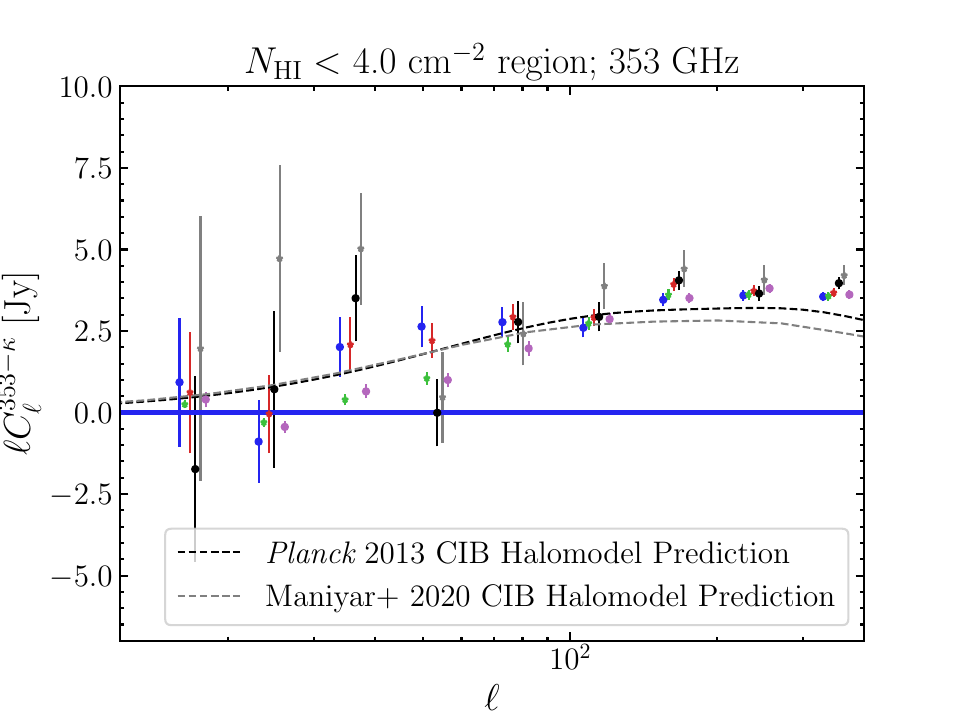}
\includegraphics[width=0.8\textwidth]{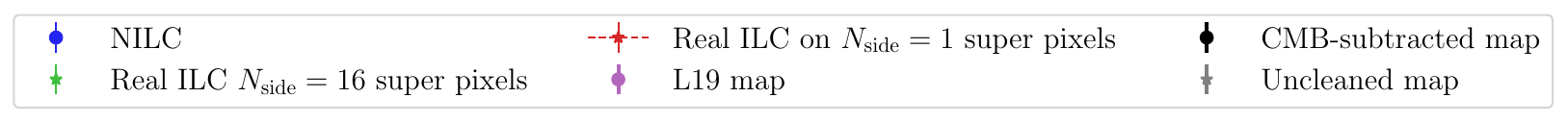}\\
\includegraphics[width=0.32\textwidth]{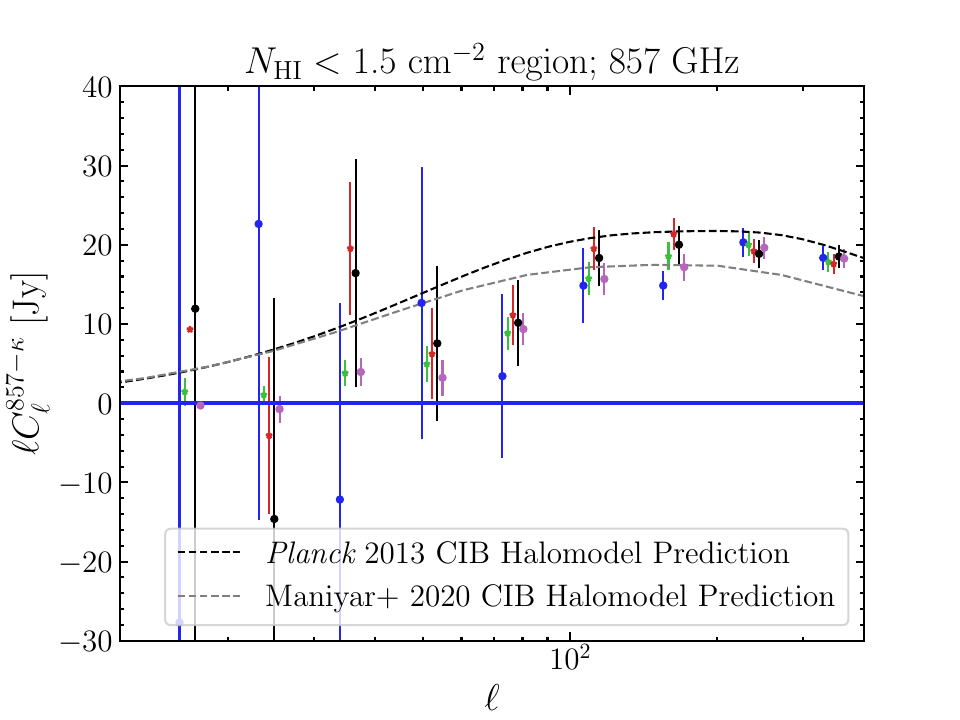}
\includegraphics[width=0.32\textwidth]{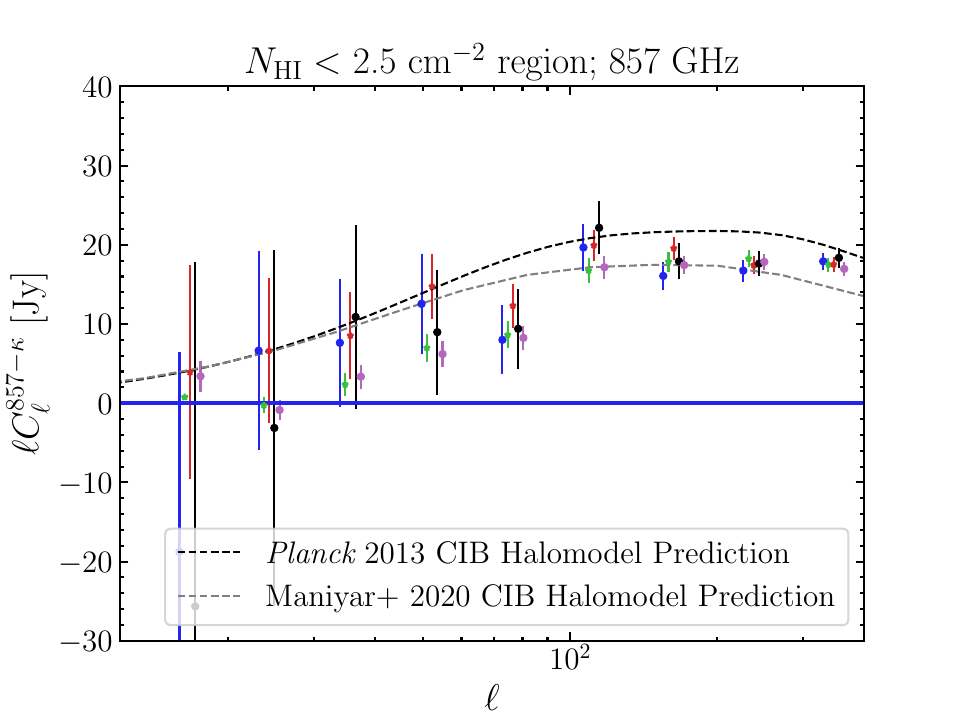}
\includegraphics[width=0.32\textwidth]{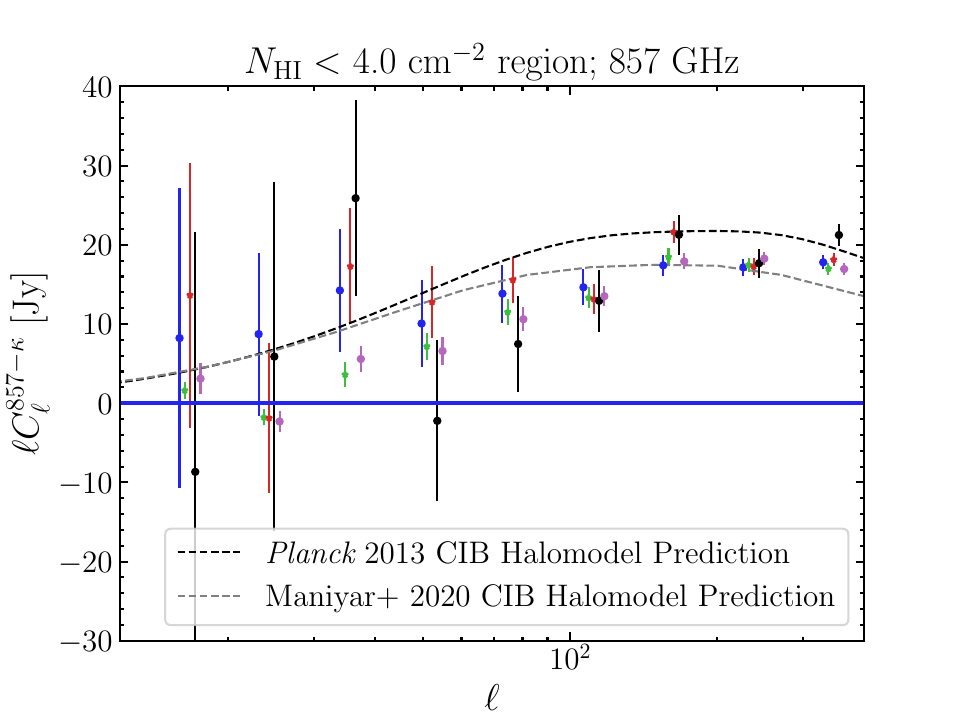}
\caption{The cross power of the NPIPE CMB lensing reconstruction with the 353 GHz CIB (\textit{top}), and the 545 GHz CIB(\textit{bottom}). The autospectra of these maps are  shown in Fig.~\ref{fig:diff_ell}. Theory predictions from the CIB halomodels of~\cite{2014A&A...571A..30P} (``\textit{Planck} 2013'')and~\cite{2021A&A...645A..40M} (``Maniyar 2020'') are indicated in dashed lines. These were computed with \texttt{class\_sz}.}\label{fig:Xcorr_lensing353}
\end{figure*}

Some results for different choices of cleaning pipelines for 545 GHz are shown in Fig.~\ref{fig:Xcorr_lensing}. We show similar results (for fewer choices of cleaning pipelines) for 353 and 857 GHz in Fig.~\ref{fig:Xcorr_lensing353}.

\begin{figure*}
\includegraphics[width=0.32\textwidth]{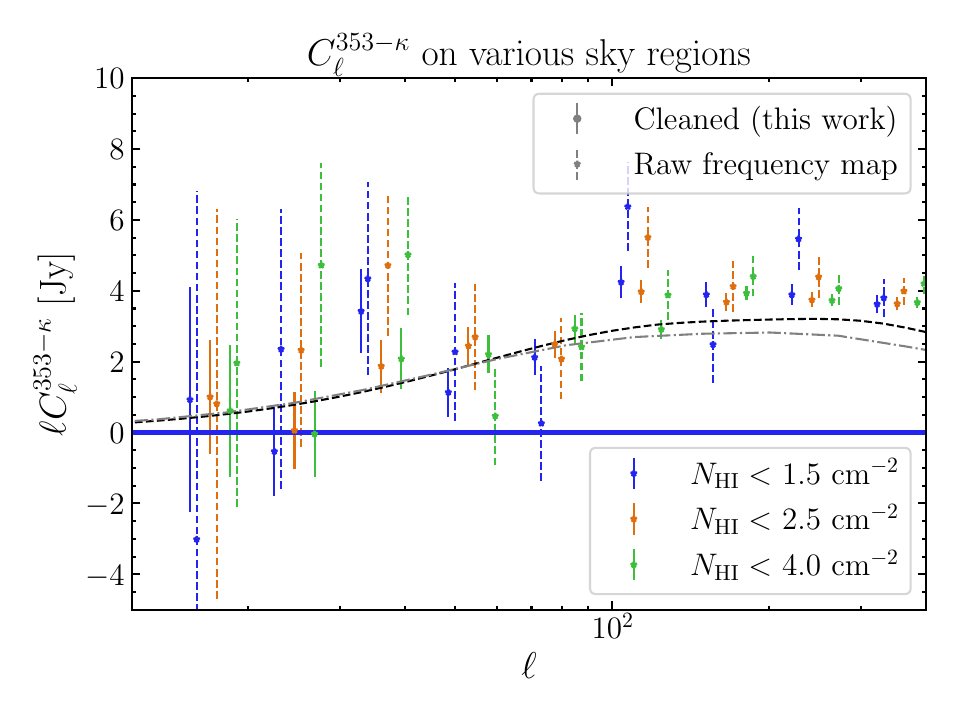}
\includegraphics[width=0.32\textwidth]{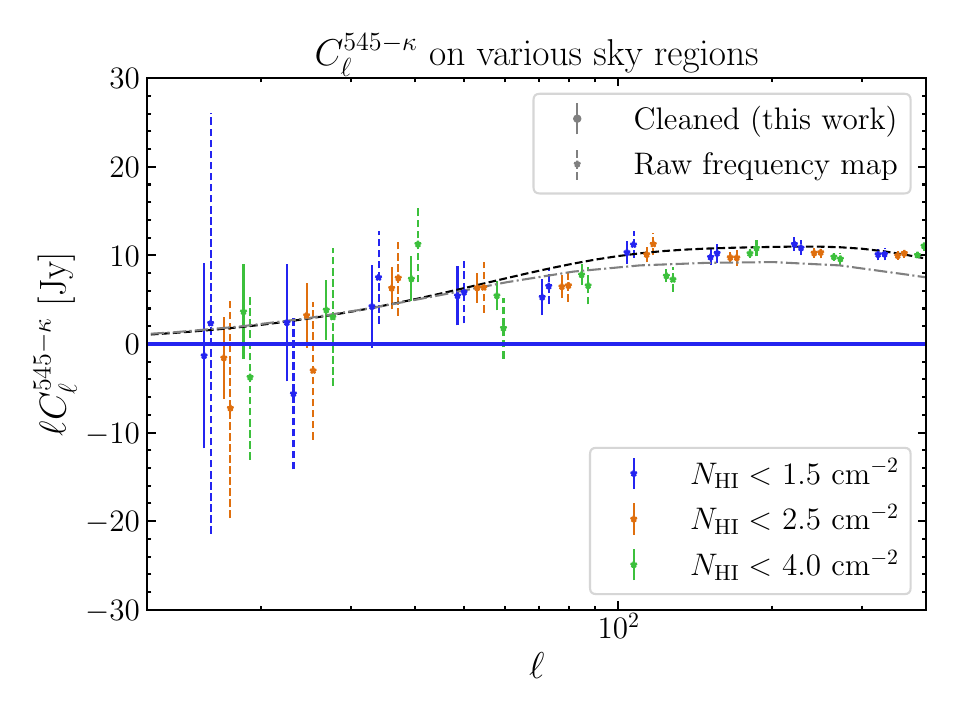}
\includegraphics[width=0.32\textwidth]{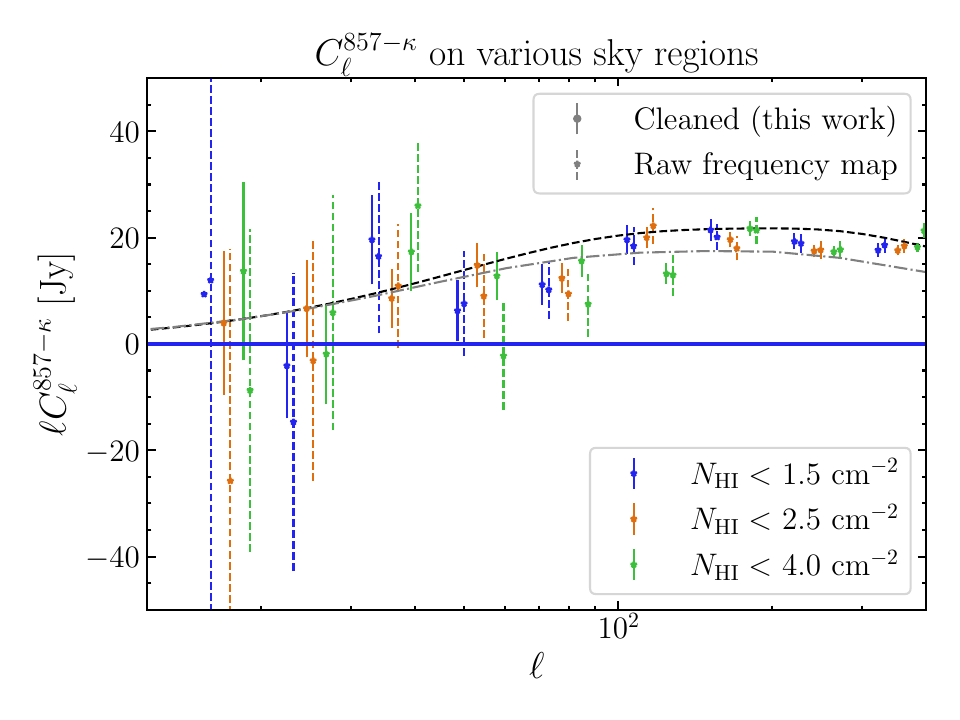}
\includegraphics[width=0.8\textwidth]{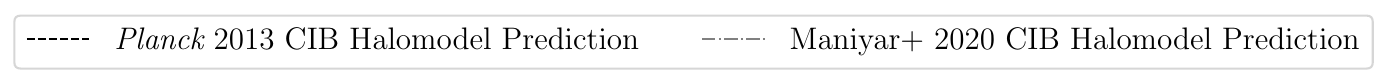}

\caption{The cross-power spectrum $C_\ell^{\nu\kappa}$ measured on the different sky regions for the raw frequency map and the NILC-cleaned 545 GHz CIB map (cleaned using 3 $N_{\mathrm{HI}}$ maps with $N_{\mathrm{side}}=8$ real-space pixels with a boundary of $\ell_b=100$).}\label{fig:allregions_crosspower}
\end{figure*}

We also indicate theory predictions for the signal, from the CIB halomodels of~\cite{2014A&A...571A..30P}  and~\cite{2021A&A...645A..40M}, which were fit to CIB auto-power spectra data. 
The theory curves were computed using \texttt{class\_sz}, using the same choices for the halomodel and cosmological parameters as we did for the auto power spectra calculation, as described in Section~\ref{sec:theory_CMBCIB}.  We illustrate directly the consistency across different sky regions in Fig.~\ref{fig:allregions_crosspower}; no significant scatter (outside of the errorbars) is seen by-eye for the points measured on dustier or cleaner regions. The reduction in variance compared to the raw frequency map is clear.

\subsection{Impact of sky region}

The consistency of the measurements when we move from smaller to larger sky area is reassuring, as it is indicative that we do not pick up a bias when we increase the amount of dust in the maps.

However, we note that the increase in signal-to-noise when we move from smaller to larger sky areas is much smaller than would be expected from a naive $f_{\mathrm{sky}}$ scaling, due to the larger variance present on the larger sky areas due to increased amounts of dust. In principle, the data could be combined more optimally by measuring the power spectra separately on non-overlapping regions of sky, accounting for the covariance between the different sky areas in a signal-to-noise calculation. While we do not take such an approach in this work, such an operation could be interesting in further analyses of the signal.

\section{Conclusion}\label{sec:conclusion}

In this work we have applied NILC cleaning to HI data and single-frequency \textit{Planck} maps, to make large-scale CIB maps that have reduced dust variance compared to the single-frequency maps, and which are unbiased on all multipoles $\ell\greaterthanapprox10$. We have quantified the reduction in variance by explicitly measuring the cross correlation between the CIB and CMB lensing potential on large scales. We release our maps publicly for use in further cross correlation studies, as well as auto power spectrum studies at intermediate-to-high $\ell$.

Our cleaning algorithm achieves similar performance to the maps of~\cite{2019ApJ...883...75L} on intermediate-to-high scales, as well as significantly reducing the contribution to the variance of the CIB maps on large ($\ell\lessthanapprox100$) scales.

The CIB-CMB lensing cross correlations we have made are ideal for further studies of the scale-dependent bias of the CIB, such as done by Ref~\cite{2023PhRvD.108h3522M} at $\ell>70$; we leave such a study to future work. Also of interest would be cross-correlation measurements with new CMB lensing datasets such as the large-scale CMB lensing map released by the Atacama Cosmology Telescope (ACT)~\cite{2023arXiv230405203M}.

The maps we make public are cleaned on the 40\% of the sky left uncovered by the \textit{Planck} 40\% galactic-plane mask. We note, however, that larger sky areas may be of interest for cross correlations with experiments on different footprints (or for the detection of signals that require a large sky area~\cite{2020PhRvD.102d3520M}), and maps cleaned on custom regions may be required to maximize cleaning efficiency; we encourage users to create their own maps on the region of sky they are interested in, using the \texttt{pyilc} scripts and intermediate data products available at~\url{https://users.flatironinstitute.org/~fmccarthy/CIBmaps_PlanckNPIPE_HI4PI_McCarthy24/pyilc_examples/}.

\begin{acknowledgements}

We are especially grateful to Colin Hill for many discussions and collaboration on \texttt{pyilc}. We also thank
Boris Bolliet, Will Coulton, Yogesh Mehta, Blake Sherwin, and Alex van Engelen for useful discussions, and. We thank the Scientific Computing Core staff at the Flatiron Institute for computational support. The Flatiron Institute is a division of the Simons Foundation. We acknowledge support from the European Research Council (ERC) under the European Union’s Horizon 2020 research and innovation programme (Grant agreement No. 851274).

We perform all map manipulations using the Python package \texttt{healpy}\footnote{\url{https://healpy.readthedocs.io/en/latest/}}~\cite{Zonca2019}, a python implementation of \texttt{HEALPix}~\cite{2005ApJ...622..759G}. We also made extensive use of the \texttt{numpy} package~\cite{harris2020array} and the \texttt{matplotlib} package~\cite{Hunter:2007}. All (cross-)power spectrum calculations  were performed using \texttt{namaster}.\footnote{\url{https://namaster.readthedocs.io}}~\cite{2019MNRAS.484.4127A}

\end{acknowledgements}

\bibliography{references}
\appendix
\section{PR4 NILC CMB map}\label{app:cmbtemplate}

\begin{table*}
\begin{tabular}{|c||c|c|c|c|c|c|c|c|c|c|c|c|c|}\hline
{Needlet scale number $I$}& 0 & 1 & 2 &  3&  4&  5&  6&  7& 8 & 9&10&11&12\\\hline\hline
{$\ell^{\mathrm{peak}}$}& 0&100& 200& 300& 400& 600& 800& 1000& 1250& 1400& 1800& 2200& 4097\\\hline
{Real-space FWHMs (degrees)} & 91.8& 35.6& 25.2& 20.6& 14.1&
       10.3&  8.26&  6.31&  6.11&  4.74&
        3.55&  1.34&  1.28\\\hline
\end{tabular}
\caption{The details of the NILC, including the values of $\ell^{\mathrm{peak}}$ and the FWHMs of the relevant real-space filters.}\label{tab:fwhm_cosinefiltersCMB}
\end{table*}
In this Appendix we describe the NILC CMB map we create using \texttt{pyilc} from \textit{Planck} NPIPE data.
We create the map with a resolution of 5$^\prime$. Characterization of this map (such as power spectrum measurements and in particular comparisons to other CMB maps) is beyond the scope of this work, although we make the map public at~\url{https://users.flatironinstitute.org/~fmccarthy/CMBNILC\_PlanckNPIPE\_McCarthy24/}.

We describe the data used to create the map in Section~\ref{sec:nilcdata}, then describe our choices for the needlet ILC algorithm in Section~\ref{sec:needletILCdetailsCMB}, including the harmonic and real-space needlet filters and the range of frequencies included at each scale.

\subsection{Data: \textit{Planck} NPIPE Single frequency maps}\label{sec:nilcdata}
The single-frequency maps we use are the preprocessed PR4 single-frequency maps described in Ref.~\cite{2023arXiv230701043M}, which were used in that reference to create PR4 NILC $y$ maps. These maps have had some preprocessing steps applied, as described in that reference. In particular,  preprocessing mask which masks point sources and  a very small ($\sim 3\%$ of the sky) region in the galactic centre was applied, and the regions of the maps covered by these masks were iteratively inpainted with the publicly available diffusive inpainting code (which is available in the \texttt{pyilc} repository).\footnote{These preprocessed maps are available at~\url{https://users.flatironinstitute.org/~fmccarthy/ymaps\_PR4\_McCH23/inpainted\_input\_maps/}} We create the map including  data from the \textit{Planck} 30, 44, 70, 100, 143, 217, 353, and 545 GHz channels. We deconvolve the beams of each maps by approximating them as Gaussian with FWHMs as listed in Table~\ref{tab:fwhm}, and reconvolve them all with Gaussian beams of FWHM 5$^\prime$.

\begin{figure}
\includegraphics[width=\columnwidth]{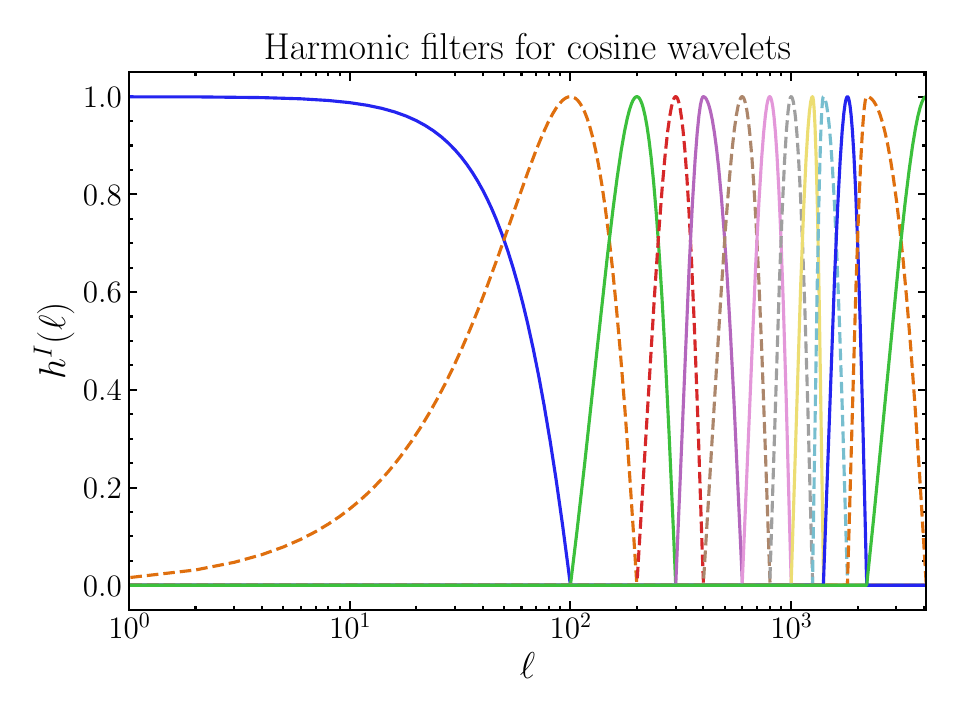}
\caption{The needlet filters for the cosine needlets. The different colours and linestyles are used to aid in differentiation between the different scales.}\label{fig:harmonicfilters_cosine}
\end{figure}

\subsection{Needlet ILC choices}\label{sec:needletILCdetailsCMB}
\subsubsection{Harmonic space needlet filters}
We use cosine needlets in as the harmonic filters for the needlet basis. These are given by
\be
h^I(\ell) = \begin{cases}
\cos\left(\frac{\pi}{2}\frac{\ell^I_{\mathrm{peak}}-\ell}{\ell^I_{\mathrm{peak}}-\ell^{I-1}_{\mathrm{peak}}}\right)&\ell^{I-1}_{\mathrm{peak}}\le\ell<\ell^{I}_{\mathrm{peak}}\\
\cos\left(\frac{\pi}{2}\frac{\ell-\ell^I_{\mathrm{peak}}}{\ell^{I+1}_{\mathrm{peak}}-\ell^I_{\mathrm{peak}}}\right)&\ell^{I}_{\mathrm{peak}}\le\ell<\ell^{I+1}_{\mathrm{peak}}\\
0 & \mathrm{otherwise}.
\end{cases}
\ee
We use 13 harmonic needlet scales with the $\ell_{\mathrm{peak}}$s given by \{0,100, 200, 300, 400, 600, 800, 1000, 1250, 1400, 1800, 2200, 4097\} (following Ref.~\cite{2023arXiv230701258C}). The filters are shown in Fig.~\ref{fig:harmonicfilters_cosine}.

\subsubsection{Real-space needlet filters}

We use Gaussian window functions for our real-space filters. The FWHMs of these Gaussians are chosen by demanding that the fractional ILC bias is always less than 0.01. These are given in Table~\ref{tab:fwhm_cosinefiltersCMB}.

\subsubsection{Inclusion of frequency channels in each needlet scale}

\texttt{pyilc} automatically drops frequency channels from the NILC in a given harmonic scale if their resolution is too low, based on a criterion which depends on the (user-described) resolution of the maps (this criterion is described in further detail in Ref.~\cite{2023arXiv230701043M}). The result is that the low-frequency maps (which also have lower resolution) are only included at low-resolution needlet scales (low $I$). We list explicitly in Table~\ref{tab:included_frequencies_CMB} the frequency channels included in each needlet scale.
\begin{table}[h!]
\begin{tabular}{|c |c |c|}\hline
Needlet scale & $\ell^{\mathrm{peak}}$&Frequencies included\\\hline\hline
0 &0&$\left\{30,44,70,100,143,217,353,545\right\}$ GHz\\\hline
1 &100&$\left\{30,44,70,100,143,217,353,545\right\}$ GHz\\\hline
2 &200&$\left\{30,44,70,100,143,217,353,545\right\}$ GHz\\\hline
3 &300&$\left\{30,44,70,100,143,217,353,545\right\}$ GHz\\\hline
4 &400&$\left\{30,44,70,100,143,217,353,545\right\}$ GHz\\\hline
5 &600&$\left\{30,44,70,100,143,217,353,545\right\}$ GHz\\\hline
6 &800&$\left\{44,70,100,143,217,353,545\right\}$ GHz\\\hline
7 &1000&$\left\{70,100,143,217,353,545\right\}$ GHz\\\hline
8 &1250&$\left\{70,100,143,217,353,545\right\}$ GHz\\\hline
9 &1400&$\left\{70,100,143,217,353,545\right\}$ GHz\\\hline
10 &1800&$\left\{70,100,143,217,353,545\right\}$ GHz\\\hline
11 &2200&$\left\{143,217,353,545\right\}$ GHz\\\hline
12 &4097&$\left\{143,217,353,545\right\}$ GHz\\\hline

\end{tabular}
\caption{The frequency channels included in each scale of the CMB NILC map}\label{tab:included_frequencies_CMB}
\end{table}

\end{document}